\documentclass[11pt]{article}
\usepackage{array}
\usepackage{amsfonts}
\usepackage{fullpage}
\usepackage{graphicx}
\usepackage{amsmath}
\usepackage{amsbsy}
\usepackage{amssymb}
\usepackage{mathtools}
\usepackage{mathrsfs}
\usepackage{tabularx}
\usepackage{indentfirst}
\usepackage{ifpdf}
\usepackage{subcaption,graphicx}
\usepackage{algorithm} 
\usepackage[noend]{algpseudocode}
\usepackage{algorithmicx}
\usepackage{algpseudocode}
\usepackage{setspace}
\usepackage{lscape}
\usepackage{pdflscape}
\usepackage{rotating}
\usepackage{enumitem}
\usepackage{empheq}
\usepackage[title]{appendix}
\algdef{SE}[DOWHILE]{Do}{doWhile}{\algorithmicdo}[1]{\algorithmicwhile\ #1}%

\usepackage{color}
\RequirePackage{xcolor}
\usepackage{mathtools}
\usepackage[utf8]{inputenc}
\usepackage{geometry}
\usepackage[export]{adjustbox}
\usepackage{float}
\usepackage[english]{babel}
\usepackage{xr}
\usepackage{booktabs}
\usepackage{multirow}
\usepackage{rotating}
\usepackage{alphalph,etoolbox}
\usepackage[english]{babel}
\usepackage[utf8]{inputenc}
\usepackage[T1]{fontenc}
\usepackage{mathtools}
\usepackage{mathrsfs}
\usepackage[official]{eurosym}

\newcounter{relctr} 
\everydisplay\expandafter{\the\everydisplay\setcounter{relctr}{0}} 

\AtBeginDocument{} 

\addto\captionsenglish{
	\renewcommand{\contentsname}%
	{Appendices}%
}
\patchcmd{\subequations}{\alph{equation}}{\alphalph{\value{equation}}}{}{}

\DeclareMathOperator*{\argmax}{argmax} 
\algcblockdefx[Then]{If}{Then}{EndThen}{$)$ $\{$}{$\}$}
\algcblockdefx[Else]{Then}{Else}{EndElse}{$\}$ \textbf{else} $\{$}{$\}$}

\date{}
\begin{document}
\newgeometry{top=20mm, bottom=20mm, left=20mm, right=20mm}
\renewcommand{\thepage}{\roman{page}}
\setcounter{page}{0}

\title{\textbf{Optimizing Freight Rail Electrification: A Framework for Charge Station Selection and Battery Charge/Swap Scheduling}}
\author{Jia Guo\footnote{jia\_guo.guo@unsw.edu.au} \\
	Elnaz Irannezhad\footnote{Corresponding author: e.irannezhad@unsw.edu.au}}
\maketitle
\thispagestyle{empty}

\begin{center}
	\text{Research Centre for Integrated Transport Innovation} \\
	\text{School of Civil and Environmental Engineering}\\
	\text{The University of New South Wales} \\
	\text{Sydney, New South Wales, Australia}\\
\end{center}
\vspace{0.5in}
\begin{center}
	Submitted: June 2025
\end{center}

\maketitle
\thispagestyle{empty}

\newgeometry{top=20mm, bottom=20mm, left=20mm, right=20mm}

\newpage
\begin{center}\textbf{Optimizing Freight Rail Electrification: A Framework for Charge Station Selection and Battery Charge/Swap Scheduling}\\
\end{center}

\begin{abstract}
	Battery electric freight trains are crucial for decarbonization by providing zero-emission transportation alternatives. The proper adoption of battery electric freight trains depends on an efficient battery electrification strategy, involving both infrastructure setup and charge scheduling. The study presents a comprehensive model for the optimal design of charging infrastructure and charge scheduling for each train. To provide more refueling flexibility, we allow batteries to be either charged or swapped in a deployed station, and each train can carry multiple batteries. This problem is formulated as a mixed integer linear programming model. To obtain real-time solutions for a large scale network, we develop three algorithms to solve the optimization problem: (1) a Rectangle Piecewise Linear Approximation technique, (2) a Fixed Algorithm heuristic, and (3) Benders Decomposition algorithm. In computational experiments, we use the three proposed algorithms to solve instances with up to 25 stations. Statistical analysis verifies that Benders Decomposition outperforms the other two algorithms with respect to the objective function value, closely followed by the Rectangle Piecewise Linear Approximation technique, and the Fixed Algorithm provides the least optimal solution.
\end{abstract}

\renewcommand{\thepage}{\arabic{page}}
\setcounter{page}{1}
{\tiny }

\section{Introduction}

\noindent The global freight rail industry is undergoing a transformative shift as it transitions from traditional diesel-powered locomotives to greener, more sustainable alternatives. This electrification process is driven by the urgent need to reduce carbon emissions and enhance energy efficiency, aligning with international climate goals and the broader decarbonization of transportation systems. Freight rail plays a critical role in global supply chains, and its shift toward electrification promises a cleaner, more efficient future for freight transportation. Battery electric freight trains, in particular, offer a zero-emission alternative to diesel-powered trains, significantly lowering greenhouse gas emissions, improving air quality, and contributing to more sustainable logistics networks. A study by Aredah et al. (2024) highlights the superiority of electricity over diesel, biodiesel, diesel-hybrid, biodiesel-hybrid, and hydrogen fuel cell alternatives, revealing that electric trains reduce energy consumption by 56\% compared to traditional diesel trains, with the potential for zero CO2 emissions when powered by renewable energy sources.

\indent The adoption of battery electric trains is gaining momentum worldwide, particularly in regions such as North America and Europe. In the United States, electrified rail lines for freight are present in areas like the Northeast Corridor and certain industrial zones, though they constitute only about 1\% of total railroad mileage due to high costs and infrastructure challenges (Lu and Allen 2024).In the UK, Varamis Rail became the first electric-only freight train company in 2022, using converted electric passenger trains to transport parcels at speeds over four times faster than traditional diesel-powered trains (Hughes 2023). Germany also operates electric freight trains on key routes like the Rhine-Alpine Corridor, which connects Rotterdam to Genoa via major German cities (Cech 2024, Mazzone et al. 2018).These global developments underscore both the opportunities and challenges of battery electric freight trains, emphasizing the need for optimized, long-term implementation strategies.

\indent However, the path to widespread freight rail electrification is fraught with challenges. Installing and maintaining overhead catenary systems for freight is particularly complex due to three major hurdles: infrastructure and clearance difficulties, economic and operational constraints, and vulnerability and resilience issues. Freight trains, especially those carrying double-stacked containers or bulk materials, require significantly larger vertical clearances than passenger trains. Accommodating these heights often necessitates costly modifications to tunnels, bridges, and platforms. Additionally, the capital investment for comprehensive freight-line electrification is staggering, with the Federal Railroad Administration noting that U.S. Class I railroads have historically found it "prohibitively expensive" (FreightWaves, 2025). Operational challenges further complicate the picture, as the intermittent nature of freight traffic and the need for robust train-to-wire connectivity over long distances make electrification less attractive for low-density or irregular routes. Moreover, overhead power systems introduce single points of failure, where disruptions—such as downed wires or damaged masts—can halt entire corridors, unlike diesel locomotives, which operate independently of the grid.

\indent Battery electric trains, while promising, also face operational limitations, particularly for long-distance trips. Their large, heavy batteries provide only about 10\% of the travel distance of diesel-electric locomotives, necessitating frequent recharging (Barbosa, 2024). This underscores the critical need for optimized charge infrastructure and scheduling to ensure efficient operations. Without well-designed electrification strategies, freight networks risk unnecessary delays, low infrastructure utilization, and high operational costs.

\indent Given these challenges, optimizing charge infrastructure and scheduling is essential for the effective deployment of battery electric vehicles. While significant research exists on these optimization problems for battery-electric trucks, the literature on freight trains remains sparse. Biedenbach and Strunz (2024) optimized charge levels for electric trucks, incorporating bidirectional charging to maximize cash flow from arbitrage trading and minimize costs. Their mixed-integer linear programming (MILP) model demonstrated annual savings of up to €3,300 per vehicle in 2021 and €10,000 in 2022 with bidirectional charging. Wang et al. (2024) addressed charge scheduling for truck battery swap stations, using a two-layered optimization model to minimize operational costs and the investment payback period. Their hybrid heuristic algorithm reduced the payback period by 10.8\% and increased net profits by 4.07\%. Ye et al. (2024) developed a spatio-temporal optimization model to determine charge station locations, power capacities, and battery capacities for electric trucks, achieving full grid capacity allocation to meet decarbonization goals by 2030. Zähringer et al. (2024) focused on dynamic charging strategies, minimizing idle time for trucks while adhering to driver Hours of Service constraints, with their dynamic programming algorithm reducing time loss by up to 50\% compared to diesel trucks.

\indent Despite these advances, the literature has largely overlooked the electrification of freight trains, particularly the joint optimization of charge station selection and charge/swap scheduling for long-distance operations. This gap is critical given the unique operational constraints of freight rail—fixed routes, higher energy demands, and extensive network coordination—compared to trucks. Existing studies on trucks do not fully address the complexities of rail infrastructure, such as integrating charge/swap options and managing battery capacities across long-haul freight routes.

\indent This study seeks to bridge this gap by developing a comprehensive framework for charge station selection and charge/swap scheduling tailored to battery electric freight trains. By incorporating factors such as station locations, battery capacities, charge/swap options, demand requirements, charging times, and setup costs, this research extends the current literature and provides actionable solutions for decarbonizing freight rail. Building on the insights from truck electrification studies, we aim to address the specific needs of long-distance freight train operations, offering a novel contribution to the field.

\indent We formulate an MILP model that incorporates a piecewise linear approximation technique to capture the nonlinear relationship between the state of charge (SOC) and charging time. To solve the model efficiently for practical instances, we develop a Fixed Algorithm and apply the Benders Decomposition algorithm to enhance computational efficiency, particularly for large-scale problems. Through extensive computational experiments on small, medium, and large instances, we evaluate and compare the performance of these algorithms, demonstrating their practical applicability.

\indent The primary contributions of this study are as follows: (i) it provides a comprehensive framework for the charge station selection and scheduling of battery electric freight trains; (ii) an MILP model is formulated, incorporating a Rectangle Piecewise Linear Approximation technique to capture the nonlinear relationship between the SOC and charging time; (iii) a Fixed Algorithm is developed to efficiently solve the model for practical instances; (iv) Benders Decomposition algorithm is applied to further enhance computational efficiency, particularly for large problems; (v) extensive computational experiments on small, medium, and large instances are conducted to evaluate and compare the performance of three algorithms, demonstrating the practical applicability of the proposed approaches; (vi) sensitivity analysis is implemented to investigate the impact of delay penalty weight on the algorithm performance.

\section{Literature Review} \label{sec_sec_literature}

\noindent The transition to battery-electric vehicles is a pivotal strategy for decarbonizing transportation, with significant implications for both road and rail freight systems. Optimizing charge station locations and charge scheduling is essential to ensure operational efficiency and cost-effectiveness, particularly for battery-electric trucks and, by extension, freight trains. While extensive research has addressed these challenges in the context of electric trucks, the literature on freight rail electrification remains unexplored. This review synthesizes key studies on charge station location and charge scheduling for battery-electric trucks. It highlights their contributions, relevance to freight rail, and identifies a critical gap that this study aims to address.

\indent The literature review is separated into two sections. Section \ref{sec_literature_trucks} reviews the electrification problems for battery electric trucks. Section \ref{sec_literature_gap} clarifies the existing research gap and highlights the novel contributions of this study.

\subsection{Review of Charge Station Location and Charge Scheduling Problems for Battery-Electric Trucks} \label{sec_literature_trucks}
\noindent Electrifying heavy-duty trucks poses unique challenges compared to light-duty EVs or passenger rail. Truck applications must balance large energy demands, driver rest regulations, and widely varying route patterns, all while ensuring grid compatibility and minimizing both capital and operational costs. Over the past few years, researchers have proposed a range of analytical and simulation models—from mixed-integer programming to agent-based simulation—to tackle facets of this problem: where to site chargers or swap stations, how to schedule charging or swapping, and how to exploit bidirectional power flows or hybrid powertrains.

\indent Biedenbach and Strunz (2024) developed an MILP that optimizes each truck’s state-of-charge and allows bidirectional charging (vehicle-to-grid) to arbitrage spot-market prices. Their objective blends revenue from power sales, grid and levy costs, and battery-degradation opportunity costs, subject to energy–power balance, minimum battery SOC, and grid connection limits. Solving for 30 trucks, they show that bidirectional operation raises annual per-vehicle savings from about €1,500 (unidirectional) to €3,300 in 2021—and up to €10,000 in 2022—highlighting the financial upside of smart charging strategies.

\indent A few studies focused on battery-swap station and depot scheduling. Wang et al. (2024) applied MILP for fleets using battery-swap stations where trucks got out of charge batteries swapped by new batteries, and the out of charge batteries would then be charged and used by future trucks. Decision variables included two layers: (1) the number of batteries in the swap station, number of chargers and photovoltaic capacity installed; (2) the number of chargers activated in each time interval and low-demand periods. The objective function was to minimize the operational costs (electricity costs, wait-time penalties and equipment maintenance) and the dynamic investment payback period which reflected the time required to recover the initial investment. Constraints involved the upper limit on the number of batteries, chargers, capacity and truck waiting time. The authors proposed a two-layered MILP optimization model, which was solved by a multi-stage hybrid heuristic algorithm based on Particle Swarm Optimization and Genetic Algorithm. In computational experiments with instances of 12 candidate batteries, the proposed algorithm was able to reduce the dynamic investment payback period by 10.8\%, compared to a base configuration. Furthermore, net profits were increased by 4.07\% and electricity costs were reduced by 5.7\%.

\indent Ye et al. (2024) likewise focused on locational and sizing decisions—but still without swapping—by jointly optimizing charger siting, station capacities, per-truck charging power, and truck battery sizes over space and time. The objective was to maximize the number of electric trucks, and to minimize the costs on infrastructure, power delivery and vehicle operational expenses. Constraints included supply capacity and demand requirement. In computational experiments, the spacio-temporal optimization model was able to solve problems with 344 potential charging locations and 255 electrical substations. The algorithm allocated 100\% of the remaining grid capacity to the drayage truck sector which met the 2030 goal, while a 50\% allocation might reach the goal around 2029.

\indent Facility location and network-sizing was studied by a few researchers. Whitehead et al. (2022) apply a coverage-based, capacity-constrained facility-location framework to Queensland’s short-haul sector. By capping service radius and facility count, they show that as few as ten well-placed public chargers can meet most regional fleet needs. Speth et al. (2025) extend this with an MILP integrated with queuing model to plan a fast-charging network for heavy trucks across Germany, explicitly capturing service delays under stochastic truck arrivals. Their results guide how many chargers each site needs to maintain acceptable wait times. 

\indent A few studies focused on dynamic charging. For example, Zähringer et al. (2024) shifted focus to en-route “dynamic” charging, where trucks receive power at roadside infrastructure rather than battery swapping. Through a dynamic-programming algorithm that embeds driver-Hours-of-Service regulations, they optimized charging durations, power levels, and rest breaks for a sequence of 100 transport tasks, cutting total idle time by up to 50\% relative to diesel trucks. The authors did not study the charging station location problem, and set the objective to minimize the idle time (waiting time, charging time and rest time) the truck spent in all stations. One novel contribution from this study was that it considered Hours of Service constraints for drivers. For example, drivers must rest for 45 minutes after every 4.5 hours of driving. This rest could be split into 15 and 30-minute intervals. Bai et al. (2023) investigated truck charging and rest decisions along pre-planned routes under regulatory rest rules. They formulated a dual-objective MILP (minimizing the total costs associated with charging and rest decisions) and proposed a rollout-based approximate solver that runs in near real-time on Swedish road‐network data.

\indent A few other studies applied simulation approaches to explore charging infrastructure scenarios for electric trucks. Karlsson and Grauers (2023) and Adegbohum et al. (2023) developed agent-based simulation to model heterogeneous truck flows. Karlsson and Grauers (2023) developed an agent-based simulation model to investigate the performance of a charging infrastructure for long-haul trucks between the Swedish cities of Helsingborg and Stockholm. The simulation model estimated about 140 stations with 900 kW chargers would be required to serve the truck flow in this route. Adegbohum et al. (2023) developed a discrete-time agent-based model to simulate charging operations of heavy-duty electric and autonomous truck between three major cities in Texas, and quantified cost of charging infrastructure requirements, using synthetic data. 

\subsection{Research Gap} \label{sec_literature_gap}
\noindent While the reviewed studies provide robust frameworks for optimizing charge station location and charge scheduling for battery-electric trucks, they do not address the specific challenges of freight rail electrification. Freight trains operate on fixed routes with higher energy demands and require extensive network coordination—factors not fully captured in truck-focused research. For instance, battery swapping could reduce rail downtime, but their implementation along tracks or at sidings remains unexplored. Similarly, facility location models must account for rail infrastructure, such as electrification gaps or the amount of energy required to charge multiple battery consists. 

\indent More importantly, freight rail systems can accommodate multiple battery consists and even specialized “battery cars”, unlike trucks, which are restricted to a single set of batteries due to spatial and weight limitations. This distinction enables rail systems to utilize one or multiple battery swapping and charging options at stations, adding complexity that demands sophisticated optimization models. This multi‐module architecture dramatically increases operational complexity: planners must decide not only where and when to charge or swap each individual consist, but also how many battery modules to allocate per train, how to sequence simultaneous charging and swapping activities, and how to stage spare battery consists at intermediate yards. Current research falls short in providing models that fully capture the dynamics and management of these multiple battery and charging/swapping combinations, highlighting a significant gap in advancing rail electrification.

\indent This study fills this gap by developing a tailored framework for charge station selection and charge/swap scheduling for battery-electric freight trains. Drawing on the methodologies of truck electrification, we address rail-specific constraints, including station locations, battery capacities, charging times, and setup costs. Specifically, factors such as station locations, number of batteries on each train, capacity of each candidate station (the number of fully charged batteries and the number of chargers), charge/swap options, demand requirement, charging time and SOC level in each station, waiting time and setup cost are all taken into account. As we discussed in Section \ref{sec_literature_trucks}, none of the previous work has combined those factors in one study. Table \ref{tbl_literature} displays the differences and similarities between this paper and prior studies in Section \ref{sec_literature_trucks}

\begin{table}[htbp]
	\centering
	\caption{Summary of differences and similarities between this study and prior work on truck electrification}
	\resizebox{\textwidth}{!}{%
	\begin{tabular}{|p{8cm}|p{2cm}|p{2cm}|p{2cm}|p{2cm}|p{2cm}|}
		\toprule
		Components & This study & \multicolumn{1}{p{4.215em}|}{Biedenbach \newline{} \& Strunz\newline{}(2024)} & \multicolumn{1}{p{4.215em}|}{Wang \newline{} et al. \newline{}(2024)} & \multicolumn{1}{p{4.215em}|}{Ye et al.\newline{}(2024)} & \multicolumn{1}{p{4.215em}|}{Zähringer et al.\newline{}(2024)} \\
		\midrule
		Station location & \textcolor[rgb]{ .278,  .278,  .278}{\checkmark} &       &       & \textcolor[rgb]{ .278,  .278,  .278}{\checkmark} &  \\
		No.  batteries on the train/truck & \textcolor[rgb]{ .278,  .278,  .278}{\checkmark} &       &       & \textcolor[rgb]{ .278,  .278,  .278}{\checkmark} &  \\
		No.  fully-charged batteries in each station & \textcolor[rgb]{ .278,  .278,  .278}{\checkmark} &       & \textcolor[rgb]{ .278,  .278,  .278}{\checkmark} &       &  \\
		No. chargers in each station & \textcolor[rgb]{ .278,  .278,  .278}{\checkmark} & \multicolumn{1}{l|}{\textcolor[rgb]{ .278,  .278,  .278}{\checkmark}} & \textcolor[rgb]{ .278,  .278,  .278}{\checkmark} & \textcolor[rgb]{ .278,  .278,  .278}{\checkmark} &  \\
		Demand requirement & \textcolor[rgb]{ .278,  .278,  .278}{\checkmark} & \multicolumn{1}{l|}{\textcolor[rgb]{ .278,  .278,  .278}{\checkmark}} & \textcolor[rgb]{ .278,  .278,  .278}{\checkmark} & \textcolor[rgb]{ .278,  .278,  .278}{\checkmark} & \multicolumn{1}{l|}{\textcolor[rgb]{ .278,  .278,  .278}{\checkmark}} \\
		Swap/charge decision & \textcolor[rgb]{ .278,  .278,  .278}{\checkmark} &       & \textcolor[rgb]{ .278,  .278,  .278}{\checkmark} &       &  \\
		Charging time for each battery in each station & \textcolor[rgb]{ .278,  .278,  .278}{\checkmark} & \multicolumn{1}{l|}{\textcolor[rgb]{ .278,  .278,  .278}{\checkmark}} & \textcolor[rgb]{ .278,  .278,  .278}{\checkmark} & \textcolor[rgb]{ .278,  .278,  .278}{\checkmark} & \multicolumn{1}{l|}{\textcolor[rgb]{ .278,  .278,  .278}{\checkmark}} \\
		Charging level for each battery in each station & \textcolor[rgb]{ .278,  .278,  .278}{\checkmark} & \multicolumn{1}{l|}{\textcolor[rgb]{ .278,  .278,  .278}{\checkmark}} & \textcolor[rgb]{ .278,  .278,  .278}{\checkmark} & \textcolor[rgb]{ .278,  .278,  .278}{\checkmark} & \multicolumn{1}{l|}{\textcolor[rgb]{ .278,  .278,  .278}{\checkmark}} \\
		Waiting time of each train/truck & \textcolor[rgb]{ .278,  .278,  .278}{\checkmark} &       & \textcolor[rgb]{ .278,  .278,  .278}{\checkmark} &       & \multicolumn{1}{l|}{\textcolor[rgb]{ .278,  .278,  .278}{\checkmark}} \\
		Infrastructure cost & \textcolor[rgb]{ .278,  .278,  .278}{\checkmark} &       & \textcolor[rgb]{ .278,  .278,  .278}{\checkmark} & \textcolor[rgb]{ .278,  .278,  .278}{\checkmark} &  \\
		\bottomrule
	\end{tabular}
	}
	\label{tbl_literature}%
\end{table}%

\section{Problem Description} \label{sec_probdes}

\noindent Each battery electric train departs the origin, drives along the route with a number of potential stations, and arrives at the destination. When a train leaves the origin, all its batteries are fully charged (SOC = 100\%). The SOC decreases as the train travels forward, and hence charge stations need to be deployed along the route for refuel purpose. In a deployed charge station, one or more batteries on the train can be either charged or swapped. To swap a battery means to replace a used battery on the train with a fully charged one in the station. If we choose to charge a battery, the SOC after charging depends on the charging time. The time required to swap a battery is usually much shorter than that to charge a battery to 100\%. The reason we might choose to charge a battery could be: (1) there are no remaining fully charged batteries in the station for swap; and (2) even though it takes much longer for a battery to be fully charged, we may just charge the SOC to a certain level (far below 100\%) to support the following trip, which requires shorter than the swapping time. In conclusion, when a train arrives at a deployed station, for each battery on the train, we have three options: swap, charge, and do nothing. Note that for each train, we cannot swap a battery and charge a second battery in the same station, but we can always neither charge nor swap a battery regardless the operation on others.

\indent In this study, two sets of decisions are to be made. First, we need to decide whether to deploy a charge station in each potential location. Second, for each train and deployed station, we will construct a charge/swap schedule for all batteries on the train. The schedule includes the number of batteries on a train, whether to charge or swap batteries in a station, the amount of charging time, and the SOC level at departure.

\indent We would like to minimize the total cost including the fixed cost of stations and the sum of delay of trains in all stations. Delay refers to the total amount of time trains spend in charging/swapping at all stations, minus the planned waiting time. The planned waiting time is common for freight trains. For example, in big cities such as Melbourne or Adelaide, tracks could be utilized to their full capacity due to a large number of trains, so freight trains would give way to passenger trains and be scheduled to wait for a certain amount of time. We can take advantage of the planned waiting time to charge or swap batteries. This can significantly reduce delay.

\indent The difficulty in finding an optimal solution to this problem is twofold. First, there is a tradeoff between the two objective terms: station fixed cost and delay. Setting up more stations leads to a higher fixed cost, but can reduce the delay time. For instance, if a train is scheduled to wait an hour in station $i$, and $i$ is deployed, then we may have the trains' batteries swapped or charged in station $i$ so as to take advantage of the planned waiting time and reduce delay. Second, the charge/swap schedule is for each battery of the train. Since a train can carry multiple batteries, the problem size grows large, making it even harder to find an optimal solution.

\indent The remainder of this section discusses components of the problem including decision variables, constraints, objective, assumptions and the model formulation.

\subsection{Decision Variables} \label{sec_probdes_var}
\begin{enumerate}
	\item Decisions on charge stations
	\begin{enumerate}
		\item Whether to deploy potential stations along the route
	\end{enumerate}
	\item Decisions on battery-consist assignment
	\begin{enumerate}
		\item The number of consists carrying batteries.
	\end{enumerate}
	\item Decisions on charge/swap scheduling
	\begin{enumerate}
		\item Whether to charge a consist with battery in each station
		\item The charge start time if a consist with battery is to be charged in a certain station
		\item The state-of-charge (SOC) of a battery when a train leaves a station
		\item Whether to swap a consist with battery in each station
		\item The swap start time if a consist with battery is to be charged in a certain station
	\end{enumerate}
\end{enumerate}

\subsection{Constraints} \label{sec_probdes_cons}
\begin{enumerate}
	\item Constraints on power support for each train
	\begin{enumerate}
		\item Charge stations must provide sufficient power for trains to travel from the origin to the destination.
	\end{enumerate}
	\item Requirement on minimum carrying capacity for each train
	\begin{enumerate}
		\item A lower limit on the capacity of each train should be added to ensure the cargoes can be transported. Otherwise, the consists might all be filled with batteries.
	\end{enumerate}
	\item Capacity of each charge station
	\begin{enumerate}
		\item An upper limit on the number of chargers 
		\item An upper limit on the number of fully-charged batteries
	\end{enumerate}
\end{enumerate}

\subsection{Objective} \label{sec_probdes_obj}
\begin{enumerate}
	\item Minimize the total station setup cost
	\item Minimize the total delay due to charging or swapping
\end{enumerate}

\subsection{Assumptions} \label{sec_probdes_assum}
\begin{enumerate}
	\item Charging a battery takes much longer than swapping a battery.
	\item In each charge station, multiple batteries can be charged at the same time. Similarly, multiple batteries can be swapped at the same time.
	\item For each train, we cannot both charge and swap batteries in a station. In other words, when a train arrives at a station, we must choose from exactly one options from the three: (1) charge some or all the batteries; (2) swap some or all the batteries; (3) do nothing.
	\item The charging time varies based on the remaining power in the battery.
	\item In each deployed station, if we choose to charge, we do not necessarily need to make SOC reach 100\%. Instead, battery SOC can be any value between 0\% and 100\%.
	\item If we choose to swap batteries in a charge station, we assume the new batteries are fully charged (SOC = $100\%$).
	\item When a train has batteries charged or swapped in a station, the operations take place in a rest area, which will not impact the schedule of succeeding trains if the succeeding ones do not charge or swap batteries in that station.
	\item Charging or swapping can be implemented during waiting period of the train. Waiting period refers to the time during which a freight train gives way to passenger trains. We may take advantage of this period to charge or swap the batteries so that delay could be minimized.
	\item When a train leaves the origin, all its batteries are fully charged (SOC = 100\%).
	\item When a train is traveling with multiple batteries, we assume there is a sequential battery management system, where one battery is used up first, then another takes over. 
\end{enumerate}

\subsection{Model} \label{sec_probdes_model}
\noindent The following gives notation and formulation of the problem. \\
\\
\begin{minipage}{\columnwidth}
	\noindent \textit{Sets and indices} \\
	\setlength{\parindent}{0em}
	\indent
	\begin{tabularx}{\textwidth}{p{2.2 cm}X}
		$i$		& index for potential charge station  \\
		$I$		& set of potential charge stations  \\
		$j$		& index for train \\
		$J$		& set of trains \\
		$k$		& index for consist \\
		$K_j$	& set of consists in train $j$ \\
		$o, \;d$  & indices of origin and destination 
	\end{tabularx}
\end{minipage}
\\
\\
\\
\begin{minipage}{\columnwidth}
	\noindent \textit{Parameters} \\
	\setlength{\parindent}{0em}
	\indent
	\begin{tabularx}{\textwidth}{p{2.2 cm}X}
		$\alpha^{\text{F}}$, $\alpha^{\text{D}}$	& penalty weight on one unit of fixed cost and  delay \\
		$a_{i_1,i_2}$ & 1 if station $i_2$ is immediately after station $i_1$ along the route, 0 otherwise \\
		$b^{\text{train}}_j$	& upper bound on the number of batteries that train $j$ can carry \\
		$b^{\text{station}}_i$	& the number of fully charged batteries that are available in station $i$ \\
		$c_i$   & the number of available chargers in station $i$, which is the maximum number of batteries that can be charged simultaneously in station $i$ \\
		$e_{i_1,i_2,j}$  & amount of power required for train $j$ to travel from charge station $i_1$ to charge station $i_2$ \\
		$\epsilon$  & a sufficiently small positive constant \\
		$f_i$	& fixed cost to establish charge station $i$ \\
		$l^{\text{c}}_{s_1, s_2}$  & the amount of time required to charge a battery from SOC $s_1$ to $s_2$ ($0\% \leq s_1 \leq s_2 \leq 100\%$) \\
		$l^{\text{s}}$  & the amount of time required to swap a battery \\
		$M$	& a sufficiently large constant \\
		$r^{\text{c}}$  & charge rate for each battery \\
		$t_{i_1,i_2,j}$	& travel time from stations $i_1$ to $i_2$ for train $j$ \\
		$w_{ij}$ & the amount of time train $j$ has to wait at station $i$ (to give way to passenger trains) \\
	\end{tabularx}
\end{minipage}
\\
\\
\\
\begin{minipage}{\columnwidth}
	\noindent \textit{Decision variables} \\
	\setlength{\parindent}{0em}
	\indent
	\begin{tabularx}{\textwidth}{p{2.2 cm}X}
		$D_{ij}$	& (nonnegative) amount of delay time for train $j$ in station $i$ \\
		$S^{\text{arrive}}_{ijk}$	& (nonnegative) SOC of the battery in consist $k$ train $j$  when $j$ arrives at station $i$. \\ 
		& $0 \leq S^{\text{arrive}}_{ijk} \leq 1$. \\
		$S^{\text{depart}}_{ijk}$	& (nonnegative) SOC of the battery in consist $k$ train $j$ when $j$ leaves station $i$. \\
		& $0 \leq S^{\text{arrive}}_{ijk} \leq S^{\text{depart}}_{ijk} \leq 1$. \\
		$T^{\text{arrive}}_{ij}$  & (nonnegative) the arrival time of train $j$ at station $i$ \\
		$T^{\text{depart}}_{ij}$  & (nonnegative) the departure time of train $j$ from station $i$ \\
		$X_i$		& (binary) 1 if potential charge station $i$ is established, 0 otherwise.  \\
		$Y_{jk}$   & (binary) 1 if consist $k$ in train $j$ carries a battery, 0 otherwise. \\
		$Z^{\text{c}}_{ijk}$		& (binary) 1 if the battery in consist $k$ train $j$ is charged in station $i$, 0 otherwise.  \\
		$Z^{\text{s}}_{ijk}$		& (binary) 1 if the battery in consist $k$ train $j$ is swapped in station $i$, 0 otherwise.  \\
	\end{tabularx}
\end{minipage}
\\
\\
\\
\noindent \textit{Model} \\
\begin{subequations} \label{model}
	\begin{alignat}{3}
	\text{Minimize} \qquad \alpha^{\text{F}} \sum\limits_{i \in I} f_i X_i  +  \alpha^{\text{D}} \sum\limits_{i \in I} \sum\limits_{j \in J} (D_{ij} - w_{ij}) \label{eq:model_obj}
	\end{alignat}
	Subject to 
	\begin{enumerate}
		\item Define delay time for train $j$ in station $i$.
		\begin{alignat}{3}
		& D_{ij} \geq T^{\text{depart}}_{ij} - T^{\text{arrive}}_{ij} &    \qquad \qquad \qquad \qquad \forall \; i \in I, \; j \in J \label{eq:model_delay} 
		\end{alignat}	
		
		\item If station $i$ is not deployed, then batteries can be neither swapped nor charged in station $i$.
		\begin{alignat}{3}
		& \sum\limits_{j \in J} \sum\limits_{k \in K_j} (Z^{\text{c}}_{ijk} + Z^{\text{s}}_{ijk}) \leq 2 |J| |K_j| \cdot X_i   & \qquad \qquad \qquad \forall \; i \in I \label{eq:model_station_chargeswap}
		\end{alignat}
		
		\item If station $i$ is deployed, then at least one battery must be swapped or charged in station $i$.
		\begin{alignat}{3}
			& \sum\limits_{j \in J} \sum\limits_{k \in K_j} (Z^{\text{c}}_{ijk} + Z^{\text{s}}_{ijk}) \geq  X_i   & \qquad \qquad \qquad \forall \; i \in I \label{eq:model_station_chargeswap_must}
		\end{alignat}
		
		\item Train $j$ cannot both swap and charge batteries in station $i$.
		\begin{alignat}{3}
		&  Z^{\text{s}}_{i,j,k_1} + Z^{\text{c}}_{i,j,k_2} \leq 1  & \qquad \qquad \qquad \forall \; i \; \in I, \; j \in J, \; k_1, k_2 \in K_j \label{eq:model_no_chargeswap}
		\end{alignat}
			
		\item Train $j$ cannot depart station $i$ until passenger trains have passed.
		\begin{alignat}{3}
		& T^{\text{depart}}_{ij} - T^{\text{arrive}}_{ij} \geq w_{ij}  & \qquad \qquad \qquad \forall \; i \; \in I, \; j \in J \label{eq:model_wait}
		\end{alignat}
		
		\item Upper bound on the number of batteries that train $j$ can carry.
		\begin{alignat}{3} 
		& \sum\limits_{k \in K_j} Y_{jk} \leq b^{\text{train}}_j & \qquad \qquad \qquad \forall \; j \in J \label{eq:model_max_batteries_train}
		\end{alignat}
		
		\item In train $j$, if consist $k$ does not carry a battery, then all the consists behind $k$ do not carry batteries, either. In other words, for each train, batteries can only be stored in the first few \textit{consecutive} consists. Assume the consists closer to the locomotive have smaller indices.
		\begin{alignat}{3}
			\sum\limits_{k' \in \{k'| k' \geq k+1, k' \in K_j \} } Y_{jk'} \leq M \cdot Y_{jk}  & \qquad \qquad \qquad \forall \; j \in J, \; k \in K_j \label{eq:model_consecutive_battery}
		\end{alignat}
		
		\item If train $j$ chooses to swap batteries in station $i$, the number of swapped batteries cannot exceed the number of available batteries in $i$.
		\begin{alignat}{3}
		& \sum\limits_{k \in K_j} Z^{\text{s}}_{ijk} \leq b^{\text{station}}_i  & \qquad \qquad \qquad \forall \; i \in I, \; j \in J \label{eq:model_max_batteries_station}
		\end{alignat} 
		
		\item If train $j$ chooses to charge batteries in station $i$, the number of charged batteries cannot exceed the number of available chargers in $i$.
		\begin{alignat}{3}
		& \sum\limits_{k \in K_j} Z^{\text{c}}_{ijk} \leq c_i  & \qquad \qquad \qquad \forall \; i \in I, \; j \in J \label{eq:model_max_charger_station}
		\end{alignat}
		
		\item If station $i_2$ is immediately after station $i_1$ along the route, then for train $j$, the SOC at its arrival at $i_2$ must equal the SOC at its departure from $i_1$ minus the amount of power required for train $j$ to travel from stations $i_1$ to $i_2$.
		\begin{alignat}{3}
		& \sum\limits_{k \in K_j} S^{\text{depart}}_{i_1,j,k} - \sum\limits_{k \in K_j} S^{\text{arrive}}_{i_2,j,k} = e_{i_1, i_2, j}  & \qquad \qquad \forall \; (i_1, i_2) \in \{(i_1, i_2) | i_1, i_2 \in I, a_{i_1,i_2} = 1\}, \; j \in J \label{eq:model_powersupport}
		\end{alignat}
		
		\item For each train $j$, electricity in each battery is consumed sequentially, starting from the battery closest to the locomotive. In other words, the battery in $k+1$ will not be used before the battery in $k$ is run out (assume the smaller the consist index is, the closer the consist is to the locomotive).
		\begin{alignat}{3}
			S^{\text{arrive}}_{ijk} \times (S^{\text{arrive}}_{i,j,k+1} - 1) = 0  \qquad \qquad \quad \forall \; i \in I, \; j \in J, \; k \in K_j  \label{eq:model_battery_sequential} 
		\end{alignat}
		
		\indent The purpose of constraints \eqref{eq:model_battery_sequential} is to force $S^{\text{arrive}}_{i,j,k+1}$ to equal 1 when $S^{\text{arrive}}_{ijk}>0$, and $S^{\text{arrive}}_{i,j,k+1}$ can equal any value when $S^{\text{arrive}}_{ijk}=0$. To linearize constraints \eqref{eq:model_battery_sequential}, we introduce new binary variables $B_{ijk}$, and hence constraints \eqref{eq:model_battery_sequential} can be replaced by following equations.		
		\begin{alignat}{3}
			S^{\text{arrive}}_{ijk} - M \cdot B_{ijk} \leq 0 & \qquad \qquad \qquad \forall \; i \in I, \; j \in J, \; k \in K_j  \label{eq:model_battery_sequential_linear_a} \\
			 (1-S^{\text{arrive}}_{i,j,k+1}) + M \cdot B_{ijk} \leq M & \qquad \qquad \qquad \forall \; i \in I, \; j \in J, \; k \in K_j  \label{eq:model_battery_sequential_linear_b} 
		\end{alignat}
		
		\item When train $j$ departs station $i$, the SOC of battery in consist $k$ of train $j$ cannot be lower than the SOC at its arrival.
		\begin{alignat}{3}
			& S^{\text{depart}}_{ijk} - S^{\text{arrive}}_{ijk} \geq 0 & \qquad \quad \qquad \forall \; i \in I, \; j \in J, \; k \in K_j \label{eq:model_soc_depart_arrival}   
		\end{alignat}
		
		\item If the battery in consist $k$ of train $j$ is swapped in station $i$, then the SOC is 100\% when train $j$ leaves station $i$.
		\begin{alignat}{3}
		& S^{\text{depart}}_{ijk} \geq Z^{\text{s}}_{ijk} & \qquad \qquad \qquad \forall \; i \in I, \; j \in J, \; k \in K_j \label{eq:model_swap_soc_depart}
		\end{alignat}
		
		\item If station $i_2$ is immediately after $i_1$, then for each battery in consist $k$ of train $j$, the SOC at its arrival in station $i_2$ must be no higher than that at its departure from station $i_1$.
		\begin{alignat}{3}
			S^{\text{arrive}}_{i_2,j,k} \leq S^{\text{depart}}_{i_1,j,k} & \qquad \qquad \forall \; (i_1,i_2) \in \{(i_1, i_2) | i_1, i_2 \in I, a_{i_1,i_2}=1 \}, \; j \in J, \; k \in K_j \label{eq:model_swap_soc_arrive_leq_depart}
		\end{alignat}

		\item If the battery in consist $k$ of train $j$ is swapped in station $i$, then train $j$ must stay in $i$ for at least $l^{\text{s}}$.
		\begin{alignat}{3}
			& T^{\text{depart}}_{ij} - T^{\text{arrive}}_{ij} \geq l^{\text{s}} Z^{\text{s}}_{ijk} & \qquad \qquad \qquad \forall \; i \in I, \; j \in J, \; k \in K_j \label{eq:model_time_swap}
		\end{alignat}
		
		\item If the battery in consist $k$ of train $j$ is charged in station $i$, then when train $j$ leaves station $i$, the SOC of consist $k$ is a function of the SOC when $j$ arrives at $i$, and charging time.
		\begin{alignat}{3}
			& S^{\text{depart}}_{ijk} = S^{\text{arrive}}_{ijk} +  r^{\text{c}} \cdot \min \{ \frac{1-S^{\text{arrive}}_{ijk}}{r^{\text{c}}}, \; T^{\text{depart}}_{ij} - T^{\text{arrive}}_{ij} \}  & \qquad \qquad \forall \; i \in I, \; j \in J, \; k \in K_j \label{eq:model_charge_soc_depart} 
		\end{alignat}
		To linearize constraints \eqref{eq:model_charge_soc_depart}, and to incorporate variables $Z^{\text{c}}_{ijk}$ and $Z^{\text{s}}_{ijk}$, we introduce a new variable $T^{\text{c}}_{ijk} = \min \{ \frac{1-S^{\text{arrive}}_{ijk}}{r^{\text{c}}}, \; T^{\text{depart}}_{ij} - T^{\text{arrive}}_{ij} \}$. Constraints \eqref{eq:model_charge_soc_depart} can be replaced by the constraints below.
		 \begin{alignat}{3}
		 	& S^{\text{depart}}_{ijk} \leq S^{\text{arrive}}_{ijk} + r^{\text{c}} \cdot T^{\text{c}}_{ijk} + M \cdot Z^{\text{s}}_{ijk}  & \qquad \qquad \forall \; i \in I, \; j \in J, \; k \in K_j \label{eq:model_charge_soc_depart_linear_1_Zs} \\
		 	& S^{\text{depart}}_{ijk} \geq S^{\text{arrive}}_{ijk} + r^{\text{c}} \cdot T^{\text{c}}_{ijk} - \epsilon \cdot Z^{\text{c}}_{ijk}  & \qquad \qquad \forall \; i \in I, \; j \in J, \; k \in K_j \label{eq:model_charge_soc_depart_linear_1_Zc} \\
		 	& T^{\text{c}}_{ijk} \leq \frac{1-S^{\text{arrive}}_{ijk}}{r^{\text{c}}} & \qquad \quad \qquad \forall \; i \in I, \; j \in J, \; k \in K_j \label{eq:model_charge_soc_depart_linear_2} \\
		 	& T^{\text{c}}_{ijk} \leq T^{\text{depart}}_{ij} - T^{\text{arrive}}_{ij} & \qquad \quad \qquad \forall \; i \in I, \; j \in J, \; k \in K_j \label{eq:model_charge_soc_depart_linear_3} \\
		 	& 0 \leq T^{\text{c}}_{ijk} \leq  M \cdot Z^{\text{c}}_{ijk} & \qquad \quad \qquad \forall \; i \in I, \; j \in J, \; k \in K_j \label{eq:model_charge_soc_depart_linear_4}  \\
		 	& 0 \leq Z^{\text{c}}_{ijk} \leq  M \cdot T^{\text{c}}_{ijk} & \qquad \quad \qquad \forall \; i \in I, \; j \in J, \; k \in K_j \label{eq:model_charge_soc_depart_linear_5}  
		 \end{alignat}
		 \begin{enumerate}
		 	\item When $Z^{\text{c}}_{ijk} = Z^{\text{s}}_{ijk} = 0$, constraints \eqref{eq:model_charge_soc_depart_linear_1_Zs} - \eqref{eq:model_charge_soc_depart_linear_4}, together with constraints \eqref{eq:model_soc_depart_arrival}, force $S^{\text{depart}}_{ijk} = S^{\text{arrive}}_{ijk} $. 
		 	\item When $Z^{\text{c}}_{ijk} = 1$, constraints \eqref{eq:model_no_chargeswap} force $Z^{\text{s}}_{ijk} = 0$, and hence we have $S^{\text{depart}}_{ijk} = S^{\text{arrive}}_{ijk} + r^{\text{c}} \cdot T^{\text{c}}_{ijk}$ from constraints \eqref{eq:model_charge_soc_depart_linear_1_Zs} and \eqref{eq:model_charge_soc_depart_linear_1_Zc}.
		 	\item  When $Z^{\text{s}}_{ijk} = 1$, constraints \eqref{eq:model_no_chargeswap} force $Z^{\text{c}}_{ijk} = 0$, so we have $T^{\text{c}}_{ijk}=0$ from constraints \eqref{eq:model_charge_soc_depart_linear_4}. Constraints \eqref{eq:model_charge_soc_depart_linear_1_Zs} become redundant, and constraints \eqref{eq:model_charge_soc_depart_linear_1_Zc} become equivalent to constraints \eqref{eq:model_soc_depart_arrival}. Constraints \eqref{eq:model_swap_soc_depart} make $S^{\text{depart}}_{ijk} = 1$.
		 \end{enumerate}
		
		\item If the battery in consist $k$ of train $j$ is neither charged nor swapped in station $i$, then the SOC at its departure and arrival should be same.
		\begin{alignat}{3}
			& S^{\text{depart}}_{ijk} - S^{\text{arrive}}_{ijk} \leq Z^{\text{c}}_{ijk} + Z^{\text{s}}_{ijk}  & \qquad \quad \qquad \forall \; i \in I, \; j \in J, \; k \in K_j \label{eq:model_charge_swap_soc}   
		\end{alignat}
		
		\item If consist $k$ of train $j$ carries a battery, then it is assumed to be fully charged when the train departs the origin. Also, to save computational time, the SOC of the battery is assumed to be 100\% when it arrives at the origin. 
		\begin{alignat}{3}
			& S^{\text{depart}}_{ojk} \geq Y_{jk}  & \qquad \qquad \qquad \qquad \qquad \forall \; j \in J, \; k \in K_j \label{eq:model_origin_SOC_depart} \\
			& S^{\text{arrive}}_{ojk} \geq Y_{jk}  & \qquad \qquad \qquad \qquad \qquad \forall \; j \in J, \; k \in K_j \label{eq:model_origin_SOC_arrive} 
		\end{alignat}
		
		\item If consist $k$ of train $j$ carries a battery, then the battery is neither charged nor swapped at the origin.
		\begin{alignat}{3}
			& Z^{\text{c}}_{ojk} + Z^{\text{s}}_{ojk} \leq 2 Y_{jk}  & \qquad \qquad \qquad \qquad \forall \; j \in J, \; k \in K_j \label{eq:model_origin_no swapcharge} 
		\end{alignat}
		
		\item For train $j$, the arrival and departure time at the origin equal 0.
		\begin{alignat}{3}
			& T^{\text{depart}}_{oj} + T^{\text{arrive}}_{oj} = 0 & \qquad \qquad \qquad \forall\; j \in J \label{eq:model_T_origin_zero}
		\end{alignat}
		
		\item For train $j$, if station $i_2$ is immediately after station $i_1$ on the route, then the arrival time at $i_2$ should equal the departure time at $i_1$ plus the travel time from stations $i_1$ to $i_2$.
		\begin{alignat}{3}
			T^{\text{arrive}}_{i_2,j} = T^{\text{depart}}_{i_1,j} + t_{i_1, i_2, j} & \qquad  \qquad \qquad \forall \; (i_1, i_2) \in \{(i_1, i_2) | i_1, i_2 \in I, \; a_{i_1,i_2}=1\}, \; j \in J \label{eq:model_T_traveltime}
		\end{alignat}
		
		\item If consist $k$ in train $j$ does not carry a battery, then we can neither charge nor swap the battery of consist $k$ in train $j$ at any stations.
		\begin{alignat}{3}
			& \sum\limits_{i \in I} Z^{\text{c}}_{ijk} + \sum\limits_{i \in I} Z^{\text{s}}_{ijk} \leq 2 |I| \cdot Y_{jk}  & \qquad \qquad \qquad \qquad \forall \; j \in J, \; k \in K_j \label{eq:model_nobattery_noswapcharge}
		\end{alignat}
		
		\item If consist $k$ of train $j$ does not carry a battery, then the SOC of the "dummy" battery in consist $k$ of train $j$ must equal zero at all stations.
		\begin{alignat}{3}
		& \sum\limits_{i \in I} S^{\text{arrive}}_{ijk} + \sum\limits_{i \in I} S^{\text{depart}}_{ijk} \leq 2|I| \cdot Y_{jk}   & \qquad \qquad \qquad \qquad \forall \; j \in J, \; k \in K_j \label{eq:model_nobattery_zerosoc}
		\end{alignat}
		
		\item Variables definition
		\begin{alignat}{3}
			& D_{ij}, \; T^{\text{depart}}_{ij}, \; T^{\text{arrive}}_{ij} \geq 0 & \qquad \quad \qquad \forall \; i \in I, \; j \in J \label{eq:model_var_DT}    \\
			& 0 \leq S^{\text{arrive}}_{ijk} \leq S^{\text{depart}}_{ijk} \leq 1  & \qquad \quad \qquad \forall \; i \in I, \; j \in J, \; k \in K_j \label{eq:model_var_S}   \\
			& X_i, \; Y_{jk}, \; Z^{\text{c}}_{ijk}, \; Z^{\text{s}}_{ijk} \in \{0,1\}& \qquad \quad \qquad \forall \; i \in I, \; j \in J, \; k \in K_j \label{eq:model_var_XYZ}
		\end{alignat}
	\end{enumerate}
	
	\indent Note that in constraints \eqref{eq:model_charge_soc_depart_linear_1_Zs} and \eqref{eq:model_charge_soc_depart_linear_1_Zc}, we assume the charge rate $r^{\text{c}}$ is a constant. However, in practice, the charge rate reduces as SOC grows (Wang et al. 2016). For simplicity, we assume the charge rate $r$ is linearly correlated to the SOC $s$. Let $r_0$ ($r_0$ is a constant) be the charge rate when the battery is empty ($s=0$). Furthermore, when the battery is fully charge ($s=1$), the charge rate $r$ becomes 0. Therefore, points $(0, r_0)$ and $(1,0)$ are on the curve of linear function $r = f(s)$. By plugging the two points in the linear equation with parameters, we can get function $f(s)$ and its inverse function $f^{-1}(r)$:
	\begin{alignat}{3}
		r = f(s) = -r_0 \cdot s + r_0  \label{eq:r_fs}\\
		s = f^{-1}(r) = -\frac{1}{r_0} \cdot r + 1   \label{eq:s_f1r}
	\end{alignat} 
	
	\indent Let $s(t|S^{\text{A}})$ and $h(t|S^{\text{A}})$ respectively be the SOC and charge rate of a battery when it has been charged $t$ units of time given an initial SOC of $S^{\text{A}}$ (which is the SOC of a battery when it arrives at a station). Actually, the value of $h(t|S^{\text{A}})$ does not depend on $t$ but $s(t|S^{\text{A}})$, but since $s(t|S^{\text{A}})$ depends on $t$, mathematically, we would like to express the charge rate $h(t|S^{\text{A}})$ as a function of $t$, with a given initial SOC of $S^{\text{A}}$. In other words, $h(t|S^{\text{A}})$ is the charge rate when the SOC equals $s(t|S^{\text{A}})$. For simplicity, during each single time unit, let us assume the charge rate remains same from the beginning to the end of the unit. We also assume the length of each time unit is 1 and the charge rate is the amount of power increase during each time unit.
	
	\indent To find the expression of $h(t|S^{\text{A}})$, let us start with the relationship between $h(t|S^{\text{A}})$ and $h(t+1|S^{\text{A}})$. After $t$ time units of charging with initial SOC being $S^{\text{A}}$, the charge rate is $h(t|S^{\text{A}})$ and SOC is $s(t|S^{\text{A}})$. With additional one time unit ($t+1$ time units since the start of charging), the SOC will increase by $h(t|S^{\text{A}})$. From equation \eqref{eq:r_fs}, we can indicate that the charge rate will decrease by $r_0 \cdot h(t|S^{\text{A}})$, based on $h(t|S^{\text{A}})$. Consequently, we have $h(t+1|S^{\text{A}}) = h(t|S^{\text{A}}) - r_0 \cdot h(t|S^{\text{A}}) = (1-r_0) \cdot h(t|S^{\text{A}})$. Therefore, $h(t|S^{\text{A}})$ is a geometric sequence with a common ratio of $1-r_0$. The first term of $h(t|S^{\text{A}})$ is $h(0|S^{\text{A}}) = f(S^{\text{A}}) = -r_0 \cdot S^{\text{A}} + r_0$. Hence, the equation of $h(t|S^{\text{A}})$ is:
	\begin{alignat}{3}
		h(t|S^{\text{A}}) = r_0 \cdot (1-S^{\text{A}}) \cdot (1-r_0)^t \qquad \qquad \qquad t = 0, 1, 2, 3, ... \label{eq:ht}
	\end{alignat}
	
	\indent It can be indicated that the SOC by the end of $t$ time units ($s(t|S^{\text{A}})$) equals the initial SOC ($S^{\text{A}}$) plus the sum of the products of unit time length and the charge rate after each time unit since the start of charging. Together with constraints \eqref{eq:model_charge_soc_depart_linear_1_Zs} and \eqref{eq:model_charge_soc_depart_linear_1_Zc} that accommodate the interaction between swap decision and SOC, this can be mathematically formulated as the following equation. 
	\begin{alignat}{3}
		s(t|S^{\text{A}}) \leq S^{\text{A}} + h(0|S^{\text{A}}) + h(1|S^{\text{A}}) + h(2|S^{\text{A}}) + ... + h(t-1|S^{\text{A}}) + M \cdot Z^{\text{s}} & \nonumber \\
		= S^{\text{A}} + \sum\limits_{i=0}^{t-1} h(i|S^{\text{A}}) + M \cdot Z^{\text{s}}  \quad \quad \qquad \qquad \qquad \qquad \qquad \qquad \qquad  & \nonumber \\
		= 1 - (1 - S^{\text{A}}) \cdot (1-r_0)^t + M \cdot Z^{\text{s}} \; \; \quad \qquad \qquad \qquad \qquad \qquad \qquad&  t = 0, 1, 2, 3, ... \label{eq:St_SA_t_Zs} \\
		s(t|S^{\text{A}}) \geq S^{\text{A}} + h(0|S^{\text{A}}) + h(1|S^{\text{A}}) + h(2|S^{\text{A}}) + ... + h(t-1|S^{\text{A}}) - \epsilon \cdot Z^{\text{c}} & \nonumber \\
		= S^{\text{A}} + \sum\limits_{i=0}^{t-1} h(i|S^{\text{A}}) - \epsilon \cdot Z^{\text{c}}  \quad \quad \qquad \qquad \qquad \qquad \qquad \qquad \qquad  & \nonumber \\
		= 1 - (1 - S^{\text{A}}) \cdot (1-r_0)^t - \epsilon \cdot Z^{\text{c}} \; \; \quad \qquad \qquad \qquad \qquad \qquad \qquad&  t = 0, 1, 2, 3, ... \label{eq:St_SA_t_Zc} 
	\end{alignat}
	
	\indent Let $S^{\text{arrive}}$, $S^{\text{depart}}$ and $T^{\text{c}}$ be the SOC at arrival, SOC at departure and charging time of a battery in a station, respectively. We may combine equations \eqref{eq:model_charge_soc_depart_linear_1_Zs}, \eqref{eq:model_charge_soc_depart_linear_1_Zc}, \eqref{eq:St_SA_t_Zs} and \eqref{eq:St_SA_t_Zc}, and convert to the following constraints.
	\begin{alignat}{3}
		S^{\text{depart}}_{ijk} \leq 1 - (1 - S^{\text{arrive}}_{ijk}) \cdot (1-r_0)^{T^{\text{c}}_{ijk}} + M \cdot Z^{\text{s}}_{ijk}  \qquad \qquad \forall \; i \in I, \; j \in J, \; k \in K_j \label{eq:model_charge_soc_depart_exp_Zs} \\
		S^{\text{depart}}_{ijk} \geq 1 - (1 - S^{\text{arrive}}_{ijk}) \cdot (1-r_0)^{T^{\text{c}}_{ijk}} - \epsilon \cdot Z^{\text{c}}_{ijk}  \qquad \qquad \; \; \forall \; i \in I, \; j \in J, \; k \in K_j \label{eq:model_charge_soc_depart_exp_Zc} 
	\end{alignat}
	\indent For computational efficiency, we should keep the value of $M$ as small as possible. From observation of Constraints \eqref{eq:model_charge_soc_depart_exp_Zs}, we fix the value of $M$ at 2.
	
	\indent The next step is to linearize constraints \eqref{eq:model_charge_soc_depart_exp_Zs} and \eqref{eq:model_charge_soc_depart_exp_Zc}. We use Piecewise Linear Approximation Rectangle Method (D'Ambrosio et al (2010)). Assume $s_{ijku}$ ($u=0,1,...,n$) be the breakpoints over the domain of variable $S^{\text{arrive}}_{ijk}$, where $s_{ijk0} = 0$, $s_{ijkn}=1$, and $n$ is a constant selected by us (for example, $n=10$). Similarly, let $t_{ijkv}$ ($v=0,1,...,m$) be the breakpoints over the domain of variable $T^{\text{c}}_{ijk}$, where $t_{ijk0} = 0$, $t_{ijkm}$ is the maximum number of consecutive charging hours of a battery (such as 12 hours), and $m$ is a constant selected by us (for example, $m=12$). Let function $g(s_{ijku}, t_{ijkv}) = (1 - s_{ijku}) \cdot (1-r_0)^{t_{ijkv}}$. In addition, we have constants $w_{ijkuv} = \min \{g(s_{ijku}, t_{ijk,v+1}) - g(s_{ijku}, t_{ijkv}), g(s_{ijk,u+1}, t_{ijk,v+1}) - g(s_{ijk,u+1}, t_{ijkv}) \}$. \\
	\indent To conduct Piecewise Linear Approximation with Rectangle Method, we introduce new continuous variables $F_{ijk} = (1 - S^{\text{arrive}}_{ijk}) \cdot (1-r_0)^{T^{\text{c}}_{ijk}}$. Then we have binary variables $\beta_{ijku}$ ($u = 0,1,...,n-1$) and continuous variables $\gamma_{ijku}$ ($u=0,1,...,n$) for variables $S^{\text{arrive}}_{ijk}$. Similarly, we have binary variables $\tau_{ijkv}$ ($v = 0,1,..., m-1$) and continuous variables $\eta_{ijkv}$ ($v=0,1,...,m$) for variables $T^{\text{c}}_{ijk}$. \\
	\indent Therefore, we can replace constraints \eqref{eq:model_charge_soc_depart_exp_Zs} and \eqref{eq:model_charge_soc_depart_exp_Zc} with the following linear constraints.
	\begin{alignat}{2}
		&S^{\text{depart}}_{ijk} \leq 1 - F_{ijk} + M \cdot Z^{\text{s}}_{ijk}  & \forall \; i \in I, \; j \in J, \; k \in K_j \label{eq:model_charge_soc_depart_exp_Zs_PLA} \\
		&S^{\text{depart}}_{ijk} \geq 1 - F_{ijk} - \epsilon \cdot Z^{\text{c}}_{ijk} &  \forall \; i \in I, \; j \in J, \; k \in K_j \label{eq:model_charge_soc_depart_exp_Zc_PLA} \\
		&\sum\limits_{u=0}^{n-1} \beta_{ijku} = 1  & \forall \; i \in I, \; j \in J, \; k \in K_j \label{eq:model_charge_soc_depart_exp_PLA_beta1} \\
		&\sum\limits_{u=0}^{n} \gamma_{ijku} = 1  & \forall \; i \in I, \; j \in J, \; k \in K_j \label{eq:model_charge_soc_depart_exp_PLA_gamma1} \\
		&\sum\limits_{v=0}^{m-1} \tau_{ijkv} = 1  & \forall \; i \in I, \; j \in J, \; k \in K_j \label{eq:model_charge_soc_depart_exp_PLA_tau1} \\
		&\sum\limits_{u=0}^{n} (s_{ijku} \cdot \gamma_{ijku}) = S^{\text{arrive}}_{ijk}  & \forall \; i \in I, \; j \in J, \; k \in K_j \label{eq:model_charge_soc_depart_exp_PLA_gamma_S} \\
		&\sum\limits_{v=0}^{m-1} [t_{ijkv} \cdot \tau_{ijkv} + (t_{ijk,v+1}-t_{ijkv}) \cdot \eta_{ijkv}] = T^{\text{c}}_{ijk}  & \forall \; i \in I, \; j \in J, \; k \in K_j \label{eq:model_charge_soc_depart_exp_PLA_tau_eta_T} \\
		&\gamma_{ijku} \leq \beta_{ijk,u-1} + \beta_{ijku}  & \forall \; i \in I, \; j \in J, \; k \in K_j, \; u=1, 2,...,n-1 \label{eq:model_charge_soc_depart_exp_PLA_gamma_beta} \\
		&\eta_{ijkv} \leq \tau_{ijk,v-1} + \tau_{ijkv}  & \forall \; i \in I, \; j \in J, \; k \in K_j, \; v=1, 2,...,m-1 \label{eq:model_charge_soc_depart_exp_PLA_eta_tau} \\
		&\eta_{ijkv} \leq \tau_{ijkv}  & \forall \; i \in I, \; j \in J, \; k \in K_j, \; v=1, 2,...,m-1 \label{eq:model_charge_soc_depart_exp_PLA_eta_tau1} \\
		&\gamma_{ijk0} \leq \beta_{ijk0}  & \forall \; i \in I, \; j \in J, \; k \in K_j \label{eq:model_charge_soc_depart_exp_PLA_gamma_beta_0} \\
		&\eta_{ijk0} \leq \tau_{ijk0}  & \forall \; i \in I, \; j \in J, \; k \in K_j \label{eq:model_charge_soc_depart_exp_PLA_eta_tau_0} \\
		&\gamma_{ijkn} \leq \beta_{ijk,n-1}  & \forall \; i \in I, \; j \in J, \; k \in K_j \label{eq:model_charge_soc_depart_exp_PLA_gamma_beta_n} \\
		&\eta_{ijkm} \leq \tau_{ijk,m-1}  & \forall \; i \in I, \; j \in J, \; k \in K_j \label{eq:model_charge_soc_depart_exp_PLA_eta_tau_m}  
	\end{alignat}
	\begin{alignat}{3}
		&F_{ijk} \leq \sum\limits_{w=0}^n  (\gamma_{ijkw} \cdot g(s_{ijkw},t_{ijkv})) + \eta_{ijkv} w_{ijkuv} + M \cdot (2 - \tau_{ijkv} - \beta_{ijku}) \nonumber \\
		& \qquad \qquad \qquad \qquad \qquad \qquad  \qquad \qquad \forall \; i \in I, \; j \in J, \; k \in K_j, \; u = 0,1,...,n-1, \; v=0,1,...,m-1 \label{eq:model_charge_soc_depart_exp_PLA_F_leq} \\
		&F_{ijk} \geq \sum\limits_{w=0}^n  (\gamma_{ijkw} \cdot g(s_{ijkw},t_{ijkv})) + \eta_{ijkv} w_{ijkuv} - M \cdot (2 - \tau_{ijkv} - \beta_{ijku})  \nonumber \\
		& \qquad \qquad \qquad \qquad \qquad \qquad  \qquad \qquad \forall \; i \in I, \; j \in J, \; k \in K_j, \; u = 0,1,...,n-1, \; v=0,1,...,m-1 \label{eq:model_charge_soc_depart_exp_PLA_F_geq} \\
		&\beta_{ijku}, \; \tau_{ijkv} \in \{0,1\} \qquad \qquad \qquad \quad \; \; \forall \; i \in I, \; j \in J, \; k \in K_j, \; u = 0,1,...,n-1, \; v=0,1,...,m-1 \label{eq:model_charge_soc_depart_exp_PLA_vardef_beta_tau} \\
		&0 \leq \gamma_{ijku} \leq 1, \; 0 \leq \eta_{ijkv} \leq 1  \qquad \qquad \qquad \quad  \forall \; i \in I, \; j \in J, \; k \in K_j, \; u = 0,1,...,n, \; v=0,1,...,m \label{eq:model_charge_soc_depart_exp_PLA_vardef_gamma_eta} \\
		&F_{ijk} \geq 0 \qquad \qquad \qquad \qquad \qquad \qquad \qquad \qquad \qquad\qquad \qquad \qquad \qquad \qquad \quad  \forall \; i \in I, \; j \in J, \; k \in K_j \label{eq:model_charge_soc_depart_exp_PLA_vardef_F} 
	\end{alignat}

\indent In summary, the Integrated Model with Rectangle Piecewise Linear Approximation ($\mathcal{IMRPLA}$) is as follows.
\begin{alignat}{2}
	\text{Minimize} \; \; \qquad & \eqref{eq:model_obj} \nonumber \\
	\text{Subject to} \qquad & \eqref{eq:model_delay} - \eqref{eq:model_powersupport}, \; \eqref{eq:model_battery_sequential_linear_a} - \eqref{eq:model_time_swap}, \; \eqref{eq:model_charge_soc_depart_linear_3} - \eqref{eq:model_var_XYZ}, \; \eqref{eq:model_charge_soc_depart_exp_Zs_PLA} - \eqref{eq:model_charge_soc_depart_exp_PLA_vardef_F} \nonumber 
\end{alignat}

\end{subequations}

\section{Fix Algorithm} \label{sec_fix}
\noindent Since we are given the power demand between each potential station and we aim to minimize the number of stations, the basic idea of this algorithm is to deploy stations sequentially in the order of the "benefit" they provide. Assume we already have some deployed stations. The benefit of a potential but undeployed station $i$ includes the following aspects. 
\begin{enumerate}
	\item The maximum amount of power it can provide: $\bar{E}_i = b_i^{\text{station}} + |J| \cdot c_i$. \\
	If station $i$ is deployed, we assume (1) all batteries available in station $i$ are swapped with some depleted batteries on trains, and (2) all trains are fully charged in station $i$. \\[-20pt]
	\item The total amount of power required from station $i$ to the first downstream deployed station: $e_i^{\text{down}}$. \\[-20pt]
	\item The total amount of power required from the last upstream deployed station to station $i$: $e_i^{\text{up}}$. \\[-20pt]
	\item The negative value of the fixed cost of station $i$: $-f_i$. \\[-20pt]
	\item The total amount of planned waiting time in station $i$ for all trains: $\sum_{j \in J}w_{ij}$. 
\end{enumerate}

\noindent \noindent \textit{\textbf{Initialization.}} Let $I^{\text{deploy}}$ and $I^{\text{undeploy}}$ represent the set of deployed and undeployed stations, respectively, with $I^{\text{deploy}} \cup I^{\text{undeploy}} = I$. In the entire system, the total power demand is $\sum_{j \in J} \sum_{i=o}^{|I|-1} e_{i,i+1,j}$, and the total initial power supply at the origin is  $\sum_{j \in J} b_j^{\text{train}}$, so the total power provided by the stations along the route ($\Delta E$) should satisfy the following constraint. 
\begin{alignat}{3}
	\Delta E \geq \sum\limits_{j \in J} \sum\limits_{i=o}^{|I|-1} e_{i,i+1,j} - \sum\limits_{j \in J} b_j^{\text{train}}  \label{eq:Delta_E}
\end{alignat}

\indent Each potential station $i$ can provide maximum amount of power $\bar{E}_i = b_i^{\text{station}} + |J| \cdot c_i$. Let $\tilde{E}$ be the maximum amount of power from selected stations with initialization $\tilde{E}=0$. Then we (1) sort all potential stations in decreasing order of "benefit" value $B_i = \bar{E_i} + e_i^{\text{down}} + e_i^{\text{up}} - \alpha^{\text{F}} f_i + \alpha^{\text{D}} \sum_{j \in J}w_{ij}$, (2) deploy the one (station $i^*$) with the maximum "benefit" value: $i^* = \argmax_{i \in I^{\text{undeploy}}} \{B_i\}$, (3) fix $X_{i^*} = 1$, and (4) add $\bar{E}_i$ to $\tilde{E}$ until $\tilde{E} \geq \Delta E$. If, in step (2), multiple undeployed stations have the same maximum "benefit" value, then we randomly select one from those.

\indent Furthermore, since the objective is to minimize the number of deployed stations and delay hours due to charging or swapping, the optimal solution must have each train load the maximum number of batteries allowed. Therefore, each time we solve Model \ref{model}, the values of $Y$ variables are also fixed. 

\indent Even initialization fixes the value of $X$ and $Y$ variables in Model \ref{model} and significantly reduces the processing time, the termination condition $\tilde{E} \geq \Delta E$ cannot guarantee feasibility of Model \ref{model}, and hence we need to dig into iterations to deploy additional stations.

\noindent \textit{\textbf{Iterations.}} In each iteration of the Fix Algorithm, for all stations $i \in I^{\text{undeploy}}$, we update the "benefit" value $B_i$. Then we select the one (station $i^*$) with the maximum "benefit" value, and fix $X_{i^*} = 1$. If Model \ref{model} is infeasible due to an insufficient number of stations, we go to the next iteration and deploy another station based on "benefit" values of the remaining undeployed stations. The algorithm stops when Model \ref{model} becomes feasible. Algorithm \ref{alg_FixAlg} illustrates steps of the Fix Algorithm.

\begin{algorithm}
	\begin{algorithmic}
		\caption{Fix Algorithm} \label{alg_FixAlg}
		\State{\underline{\textbf{Initialization}}}
		\State{Calculate total power demand $E^{\text{demand}} \leftarrow \sum_{j \in J} \sum_{i=o}^{|I|-1} e_{i,i+1,j}$}
		\State{Calculate initial power supply in the origin $E^{\text{initial}} \leftarrow \sum_{j \in J} b_j^{\text{train}}$}
		\State{Calculate the minimum amount of power required from stations $\Delta E \leftarrow E^{\text{demand}} - E^{\text{initial}} $}
		\For{$i \in I$}
			\State Maximum amount of power provided from station $i$: $\bar{E}_i \leftarrow b_i^{\text{station}} + |J| \cdot c_i$. 
			\State Total amount of power required from station $i$ to the first downstream deployed station: $e_i^{\text{down}}$.
			\State Total amount of power required from the last upstream deployed station to station $i$: $e_i^{\text{up}}$. 
			\State Negative value of the fixed cost of station $i$: $-f_i$.
			\State Total amount of planned waiting time in station $i$ for all trains: $\sum_{j \in J}w_{ij}$. 
			\State $B_i \leftarrow \bar{E_i} + e_i^{\text{down}} + e_i^{\text{up}} - \alpha^{\text{F}} f_i + \alpha^{\text{D}} \sum_{j \in J}w_{ij}$
		\EndFor
		\State{Sort stations in decreasing order of $B_i$}
		\State{Initialize the total amount of power from deployed stations $\tilde{E} \leftarrow 0$}
		\State{Initialize the set of deployed stations $I^{\text{deployed}} \leftarrow \phi$}
		\State{Initialize the set of undeployed stations $I^{\text{undeploy}} \leftarrow I$, $X_i \leftarrow 0 \; (\forall \; i \in I)$ }
		\While{$\tilde{E} < \Delta E$}
			\For{$i \in I^{\text{undeploy}}$}
			\State Update $e_i^{\text{down}}$, $e_i^{\text{up}}$ \& $B_i$
			\EndFor
			\State{$i^* \leftarrow \argmax_{i \in I^{\text{undeploy}}} \{B_i\}$ }
			\State{Deploy station $i^*$ by fixing $X_{i^*}=1$ in Model \ref{model}}
			\State{$\tilde{E} \leftarrow \tilde{E} + \bar{E}_{i^*}$}
			\State{$I^{\text{deploy}} \leftarrow I^{\text{deploy}} \cup \{i^*\}$, $\; I^{\text{undeploy}} \leftarrow I^{\text{undeploy}} \backslash \{i^*\}$}
		\EndWhile
		\State{Fix $Y=1$ to make each train load the maximum number of batteries allowed}
		\State{\underline{\textbf{Iterations}}}
		\While{Model \ref{model} is infeasible}
			\For{$i \in I^{\text{undeploy}}$}
				\State Update $e_i^{\text{down}}$, $e_i^{\text{up}}$ \& $B_i$
			\EndFor
			\State{$i^* \leftarrow \argmax_{i \in I^{\text{undeploy}}} \{B_i\}$}
			\State{Deploy station $i^*$ by fixing $X_{i^*}=1$ in Model \ref{model}}
			\State{$I^{\text{deploy}} \leftarrow I^{\text{deploy}} \cup \{i^*\}$, $\; I^{\text{undeploy}} \leftarrow I^{\text{undeploy}} \backslash \{i^*\}$}
			\If{Model \ref{model} is feasible}
				\State{Solve Model \ref{model} and record the feasible solution}
				\State{Break}
			\Else
				\State{Go to the next iteration}
			\EndIf
		\EndWhile
	\end{algorithmic}
\end{algorithm}

\section{Benders Decomposition} \label{sec_benders}
\noindent In this section, we use Benders Decomposition to solve $\mathcal{IMRPLA}$. Section \ref{sec_benders_traditional} introduces the application of traditional Benders Decomposition. Section \ref{sec_benders_cuts} discusses extra feasibility cuts. 

\subsection{Traditional Benders Decomposition} \label{sec_benders_traditional}
\noindent The motivation behind Benders Decomposition is: we may divide variables into two categories: binary variables $  v =$ ($X$, $Y$, $Z^{\text{c}}$, $Z^{\text{s}}$, $\beta$, $\tau$) and continuous variables $u  = $ ($D$, $S^{\text{arrive}}$, $S^{\text{depart}}$, $T^{\text{arrive}}$, $T^{\text{depart}}$, $\gamma$, $\eta$, $F$). For simplicity, we formulate $\mathcal{IMRPLA}$ as the model below. Let's name this model $\mathcal{OP}$ (Original Problem).

\begin{subequations} \label{model_benderOP}
\begin{alignat}{3}
	\mathcal{OP}: \qquad  & \text{Minimize} \qquad  \;  c^\intercal u + f  v \label{eq:benderOP_obj} \\
	& \text{Subject to} \quad  \; \; \;   A u + D v \geq b \label{eq:benderOP_jointconstraints} \\
	& \qquad \qquad \qquad \; \; v \in V \label{eq:benderOP_v} \\
	& \qquad \qquad \qquad \; \; u \geq 0 \label{eq:benderOP_u} 
\end{alignat}
\end{subequations}

\indent $\mathcal{OP}$ (Model \eqref{model_benderOP}) is equivalent to $\mathcal{IMRPLA}$. In the objective function \eqref{eq:benderOP_obj}, coefficients $c$ and $f$ correspond to those of the continuous and binary variables in \eqref{eq:model_obj}, respectively. Constraints \eqref{eq:benderOP_jointconstraints} are equivalent to the $\mathcal{IMRPLA}$ constraints that involve at least one variable from $ u$ and $ v$, respectively. Constraints \eqref{eq:benderOP_v} represent the $\mathcal{IMRPLA}$ constraints that contain $ v$ variables only. Constraints \eqref{eq:benderOP_u} require all $ u$ variables to  be nonnegative.

\indent For fixed value of $ v$ variables ($\hat{ v}$), $\mathcal{OP}$ is given by Model \eqref{model_benderFixmu}.
\begin{subequations} \label{model_benderFixmu}
	\begin{alignat}{3}
	f \hat{ v} \qquad + \qquad \text{Minimize} \qquad  \; & c^\intercal u \label{eq:benderFixmu_obj} \\
	\text{Subject to} \quad  \; \; \;  & A u \geq b -  D\hat{ v} \label{eq:benderFixmu_jointconstraints} \\
	&  u \geq 0 \label{eq:benderFixmu_u} 
	\end{alignat}
\end{subequations}

\indent Therefore, the resulting model to solve is the following subproblem ($\mathcal{SP}$).
\begin{subequations} \label{model_benderSP}
	\begin{alignat}{3}
	\mathcal{SP}: \qquad & \text{Minimize} \qquad  \;  c^\intercal  u \label{eq:benderSP_obj} \\
	& \text{Subject to} \quad  \; \; \;   A u \geq b -  D\hat{ v} \label{eq:benderSP_jointconstraints} \\
	& \qquad \qquad \qquad \; \; u \geq 0 \label{eq:benderSP_u} 
	\end{alignat}
\end{subequations}

\indent Let $\pi$ be the dual variables of constraints \eqref{eq:benderSP_jointconstraints}. The dual of $\mathcal{SP}$ ($\mathcal{DSP}$) is as follows.
\begin{subequations} \label{model_benderDSP}
	\begin{alignat}{3}
	\mathcal{DSP}: \qquad  & \text{Maximize} \qquad \;  (b - D\hat{v})^\intercal \pi   \label{eq:benderDSP_obj} \\
	& \text{Subject to} \quad  \; \; \;   A^\intercal \pi \leq c \label{eq:benderDSP_jointconstraints} \\
	& \qquad \qquad \qquad \; \;  \pi \geq 0 \label{eq:benderDSP_pi} 
	\end{alignat}
\end{subequations}

Let $P$ be the total number of extreme points in $\mathcal{DSP}$, and $\pi_p$ be extreme point $p$. $\mathcal{DSP}$ can be reformulated as follows (Model \eqref{model_benderDSPEqui}).
\begin{subequations} \label{model_benderDSPEqui}
	\begin{alignat}{3}
	\text{Minimize} \qquad \; & w   \label{eq:benderDSPEqui_obj} \\
	\text{Subject to} \quad \; \; \;  & w \geq (b - D\hat{v})^\intercal \pi_p \qquad \forall \; p = 1,..., P \label{eq:benderDSPEqui_jointconstraints} \\
	&  w \text{ free} \label{eq:benderDSPEqui_w} 
	\end{alignat}
\end{subequations}

\indent Consequently, $\mathcal{OP}$ can be reformulated as the following $\mathcal{MP}$ (Master Problem) with $v$ and $w$ variables.
\begin{subequations} \label{model_benderMP}
	\begin{alignat}{3}
	\mathcal{MP}: \qquad & \text{Minimize} \qquad  \;  fv + w   \label{eq:benderMP_obj} \\
	& \text{Subject to} \quad  \; \; \;   w \geq (b - Dv)^\intercal \hat{\pi}_p \qquad \forall \; p = 1,..., P \label{eq:benderMP_jointconstraints} \\
	& \qquad \qquad \qquad \; \; v \in V \label{eq:benderMP_v}  \\
	& \qquad \qquad \qquad \; \;  w \text{ free} \label{eq:benderMP_w} 
	\end{alignat}
\end{subequations}

\indent For large problems where we cannot enumerate all extreme points of $\mathcal{DSP}$, let $Q<P$, $R$ be the number of extreme rays we have found for $\mathcal{DSP}$, and $\hat{\rho}_r$ is extreme ray $r$ of $\mathcal{DSP}$.  The Relaxed $\mathcal{MP}$ ($\mathcal{RMP}$) with fewer constraints is given by Model \eqref{model_benderRMP}.
\begin{subequations} \label{model_benderRMP}
	\begin{alignat}{3}
	\mathcal{RMP}: \qquad & \text{Minimize} \qquad  \;  fv + w   \label{eq:benderRMP_obj} \\
	& \text{Subject to} \quad  \; \; \;   w \geq (b - Dv)^\intercal \hat{\pi}_p \qquad \forall \; p = 1,..., Q \label{eq:benderRMP_optimality_cut} \\
	& \qquad \qquad \qquad \; \; 0 \geq (b - Dv)^\intercal \hat{\rho}_r \qquad \forall \; r = 1,..., R \label{eq:benderRMP_feasibility_cut} \\
	& \qquad \qquad \qquad \; \; v \in V \label{eq:benderRMP_v}  \\
	& \qquad \qquad \qquad \; \; w \text{ free} \label{eq:benderRMP_w} 
	\end{alignat}
\end{subequations}

\indent Algorithm \ref{alg_Bender} illustrates steps of the Benders Decomposition. In initialization, we find an initial feasible solution by fixing all $X$ and $Y$ variables at one, then use Gurobi to solve the $\mathcal{IMRPLA}$. We take the first feasible solution found by CPLEX  and use it as an initial feasible solution. In each iteration, we solve $\mathcal{RMP}$ to get a lower bound LB, then solve $\mathcal{DSP}$ to add cuts. Specifically, there are two cases: (1) If $\mathcal{DSP}$ is bounded, i.e., $\mathcal{SP}$ is feasible, then we get an extreme point of $\mathcal{DSP}$ and update the upper bound UB. If the UB and LB are close enough, then the current solution is roughly optimal and the algorithm stops. Otherwise, we add an optimality cut to $\mathcal{RMP}$ and go to the next iteration. (2) If $\mathcal{DSP}$ is unbounded, i.e., $\mathcal{SP}$ is infeasible, then we get an extreme ray of $\mathcal{DSP}$, add a feasibility cut to $\mathcal{RMP}$ and go to the next iteration. The algorithm stops when $\frac{\text{UB} - \text{LB}}{\text{UB} }$ falls below $\text{gap}_{\epsilon}$.

\begin{algorithm}
	\begin{algorithmic}
		\caption{Benders Decomposition Algorithm} \label{alg_Bender}
		\State{\underline{\textbf{Initialization}}}
		\State{Find an initial feasible solution $(\hat{u}, \hat{v})$ to $\mathcal{OP}$. }
		\State{Solve $\mathcal{DSP}$: $\text{Max}_{\pi} \{f\hat{v} + (b-D\hat{v})^\intercal \pi | A^\intercal \pi \leq c, \pi \geq 0 \} $, get extreme point $\hat{\pi}_1$}
		\State{$Q \leftarrow 1$, $R \leftarrow 0$}
		\State{UB = Upper bound $\leftarrow c^\intercal \hat{u} + f \hat{v}$}
		\State{LB = Lower bound $\leftarrow -\infty$}
		\State{\underline{\textbf{Iterations}}}
		\While{$\frac{\text{UB} - \text{LB}}{\text{UB} } > \text{gap}_{\epsilon}$}
		\State{\underline{Step 1}}
		\State{Solve $\mathcal{RMP}$: $\text{Min}_{v,w} \{fv+w | w \geq (b-Dv)^\intercal \hat{\pi}_p \; \forall \;  p=1,...,Q, \; v \in V, \; w \text{ free} \}$}
		\If{$\mathcal{RMP}$ is infeasibe} Stop
		\Else 
			\State{Let $(\hat{v}, \hat{w})$ be the optimal solution to $\mathcal{RMP}$}
			\State{$\text{LB} \leftarrow f\hat{v} + \hat{w}$}
			\State{Go to Step 2}
		\EndIf
		\State{\underline{Step 2}}
		\State{Solve $\mathcal{DSP}$: $\text{Max}_{\pi} \{f\hat{v} + (b-D\hat{v})^\intercal \pi | A^\intercal \pi \leq c, \pi \geq 0 \} $}
		\If{$\mathcal{DSP}$ is bounded} 
			\State{Get and extreme point $\hat{\pi}$ of $\mathcal{DSP}$}
			\State{$\text{UB} \leftarrow f\hat{v} + (b-D\hat{v})^\intercal \hat{\pi}$}
			\If{$\frac{\text{UB} - \text{LB}}{\text{UB}} \leq \text{gap}_{\epsilon}$}
				\State{The current solution is optimal, stop}
			\Else 
				\State{Add optimality cut $w \geq (b-Dv)^\intercal \hat{\pi}$ to $\mathcal{RMP}$, $\; Q \leftarrow Q+1$}
				\State{Go the the next iteration}
			\EndIf
		\Else 
			\State{Get an extreme ray $\hat{\rho}$ of $\mathcal{DSP}$}
			\State{Add feasibility cut $0 \geq (b-Dv)^\intercal \hat{\rho}$ to $\mathcal{RMP}$, $\; R \leftarrow R+1$}
			\State{Go to the next iteration}
		\EndIf

		\EndWhile
	\end{algorithmic}
\end{algorithm}

\subsection{Extra Feasibility Cuts}  \label{sec_benders_cuts}
\noindent To fasten the Benders Decomposition procedure, we add following extra feasibility cuts to $\mathcal{RMP}$. The purpose of adding these feasibility cuts is to prevent $\mathcal{RMP}$ from giving integer solutions that are feasible to $\mathcal{RMP}$ but infeasible to $\mathcal{SP}$.

\begin{enumerate}
	\item Feasibility cuts for $B$ variables \\
	From constraints \eqref{eq:model_battery_sequential_linear_a} and \eqref{eq:model_battery_sequential_linear_b}, we can conclude that $B_{ijk}=0$ indicates the power in the battery of consist $k$ train $j$ is used up (SOC=0) when train $j$ arrives at station $i$. Variable $B_{ijk}=1$ indicates the power in the battery of consist $k$ train $j$ is not used up (SOC>0) when train $j$ arrives at station $i$. Therefore, when $B_{ijk}=0$, the power in batteries of consists 1, ..., $k-1$ must have been used up, i.e., $\sum_{k'=1}^{k-1} B_{ijk'}=0$. Similarly, when $B_{ijk}=1$, the SOC in batteries of consists $k+1$, ..., $|K_j|$ must be $100\%$, i.e., $\sum_{k'=k+1}^{|K_j|} B_{ijk'} = |K_j|-k$. The feasibility cuts are as follows.
	\begin{subequations}
		\begin{alignat}{3}
		\sum\limits_{k'=1}^{k-1} B_{ijk'} \leq (k-1) B_{ijk}  & \qquad \qquad \qquad \forall \; i \in I, \; j \in J, \; k \in K_j  \label{eq:model_battery_sequential_linear_a_feas} \\
		\sum\limits_{k'=k+1}^{|K_j|} B_{ijk'} \geq (|K_j|-k) B_{ijk}  & \qquad \qquad \qquad \forall \; i \in I, \; j \in J, \; k \in K_j  \label{eq:model_battery_sequential_linear_b_feas} \\
		B_{i,j,k+1} \geq B_{ijk}   & \qquad \qquad \qquad \forall \; i \in I, \; j \in J, \; k \in K_j  \label{eq:model_battery_sequential_linear_c_feas} 
		\end{alignat}
	\item Feasibility cuts for the connection between $B$ and $X$ variables \\
	Variable $B_{ijk}=0$ indicates the SOC of battery $k$ in train $j$ equals 0 when it arrives at location $i$. Let $i'$ be the location immediately succeeding station $i$. When arriving at location $i$, if all batteries in train $j$ have a SOC of zero, and the power required from location $i$ to $i'$ is greater than zero ($e_{i,i',j}>0$), then a charging/swapping station must be deployed in location $i$. In other words, if $\sum_{k \in K_j} B_{ijk}=0$ and $e_{i,i',j}>0$, then $X_i$ must equal 1. The feasibility constraint can be formulated as follows.
	\begin{alignat}{3}
		M X_i \geq e_{i,i',j} - \sum\limits_{k \in K_j} B_{ijk} & \qquad \qquad \qquad \forall \; i \in I, \; j \in J  \label{eq:model_B_X_a_feas} 
	\end{alignat}
	\indent Also, when arriving at location $i$, if all batteries in train $j$ have a SOC of zero, and the power required from location $i$ to $i'$ is greater than zero ($e_{i,i',j}>0$), then a station must be deployed  in location $i$. In other words, if $\sum_{k \in K_j} B_{ijk}=0$ and $e_{i,i',j}>0$, then $X_i$ must equal 1. The feasibility constraint can be formulated as follows.
	\begin{alignat}{3}
		M X_i  \geq e_{i,i',j} (1 - \sum\limits_{k \in K_j} B_{ijk}) & \qquad \qquad \qquad \forall \; i \in I, \; j \in J \label{eq:model_B_X_b_feas} 
	\end{alignat}
	\item Feasibility cuts for the connection between $B$ and $Y$ variables \\
	If consist $k$ of train $j$ carries a battery ($Y_{jk}=1$), then its SOC is assumed to be 100\% when the train arrives the origin, and hence $B_{o,j,k}=1$. Otherwise, $B_{o,j,k}=0$. The feasibility cuts are as follows.
	\begin{alignat}{3}
		Y_{jk} = B_{o,j,k}	& \qquad \qquad \qquad \forall \; j \in J, \; k \in K_j  \label{eq:model_B_Y_feas} 
	\end{alignat}
	\item Feasibility cuts for the connection among $B$, $Z^{\text{c}}$ and $Z^{\text{s}}$ variables \\
	Variable $B_{ijk}=0$ indicates the SOC of battery $k$ in train $j$ equals 0 when it arrives at location $i$. The amount of power that train $j$ carries when it arrives at location $i$ ranges between $\min\{0, \sum_{k \in K_j} B_{ijk} - 1\}$ and $\sum_{k \in K_j} B_{ijk}$. Let $i'$ be the location immediately succeeding station $i$. If the maximum potential amount of power in train $j$ ($\sum_{k \in K_j} B_{ijk}$) is smaller than the power required from location $i$ to $i'$ ($e_{i,i',j}$), then at least one battery of train $j$ must be charged or swapped in location $i$.
	\begin{alignat}{3}
		M \sum\limits_{k \in K_j} (Z^{\text{c}}_{ijk} + Z^{\text{s}}_{ijk})  \geq e_{i,i',j} - \sum\limits_{k \in K_j} B_{ijk} & \qquad \qquad \qquad \forall \; i \in I, \; j \in J \label{eq:model_B_Z_a_feas} 
	\end{alignat}
	\indent Also, When arriving at location $i$, if all batteries in train $j$ have a SOC of zero, and the power required from location $i$ to $i'$ is greater than zero ($e_{i,i',j}>0$), then at least one battery must be charged or swapped in location $i$. In other words, if $\sum_{k \in K_j} B_{ijk}=0$ and $e_{i,i',j}>0$, then $\sum_{k \in K_j} (Z^{\text{c}}_{ijk} + Z^{\text{s}}_{ijk})$ must be greater than or equal to 1. The feasibility constraint can be formulated as follows.
	\begin{alignat}{3}
			M \sum\limits_{k \in K_j} (Z^{\text{c}}_{ijk} + Z^{\text{s}}_{ijk})  \geq e_{i,i',j} (1 - \sum\limits_{k \in K_j} B_{ijk}) & \qquad \qquad \qquad \forall \; i \in I, \; j \in J \label{eq:model_B_Z_b_feas} 
	\end{alignat}
	
	\item Feasibility cuts for the connection between $B$ and $\beta$ variables \\
	We can make two indications from \eqref{eq:model_battery_sequential_linear_a} and \eqref{eq:model_battery_sequential_linear_b}: (1) If $B_{ijk}=0$, then $\sum_{k'=1}^{k} S^{\text{arrive}}_{ijk'}=0$, and we have $\sum_{k'=1}^{k} \beta_{ijk'0}=k$. (2) If $B_{ijk}=1$, then $\sum_{k'=k+1}^{|K_j|} S^{\text{arrive}}_{ijk'} = |K_j|-k$, and we have $\sum_{k'=k+1}^{|K_j|} \beta_{ijk'n} = |K_j|-k$. The indications can be formulated as the following constraints.
	\begin{alignat}{3}
		\sum\limits_{k'=1}^{k} \beta_{ijk'0} \geq k(1 - B_{ijk}) & \qquad \qquad \qquad \forall \; i \in I, \; j \in J, \; k \in K_j  \label{eq:model_B_beta_a_feas} \\
		\sum\limits_{k'=k+1}^{|K_j|} \beta_{ijk'n} \geq (|K_j|-k) B_{ijk} & \qquad \qquad \qquad \forall \; i \in I, \; j \in J, \; k \in K_j  \label{eq:model_B_beta_b_feas} \\
	 	\beta_{ijk0} \geq 1 - B_{ijk} & \qquad \qquad \qquad \forall \; i \in I, \; j \in J, \; k \in K_j  \label{eq:model_B_beta_c_feas} \\
		\beta_{i,j,k+1,n} \geq B_{ijk} & \qquad \qquad \qquad \forall \; i \in I, \; j \in J, \; k \in K_j  \label{eq:model_B_beta_d_feas} 
	\end{alignat}
	\end{subequations}
\end{enumerate}

\section{Computational Experiments} \label{sec_comp}

\noindent The three algorithms - (1) integrated model with Piecewise Linear Approximation (PLA), (2) Fix Algorithm (FA), and (3) Benders Decomposition (BD) -  have been tested to determine the quality of charge location selection and charge schedules provided. Section \ref{sec_comp_dataparam} clarifies the data and parameters. Section \ref{sec_comp_result} presents computational results.  Section \ref{sec_comp_sensitivity} conducts the sensitivity analysis on the impact of delay penalty weight ($\alpha^{\text{D}}$).

\subsection{Data and Parameters}  \label{sec_comp_dataparam}

\noindent In this subsection, we introduce the data sets and parameters used in the computational experiments. 

\subsubsection{Data sets}
\noindent We generate three groups of random instances: small, medium and large. The difference among the three groups lies in the number of stations along the route, including the origin and destination. There are 6, 15 and 25 stations in small, medium and large instances, respectively. 

\indent In the instances among all groups, the number of trains is 2 and each train can carry at most 3 batteries for energy storage. We assume each train drives at 100 km/hr, each fully-charged battery can provide 7000 kWh energy, and a train consumes 30 kWh per kilometer. The distance between any two neighboring nodes follows normal distribution with a mean of 373 km and standard deviation of 146 km. These values are obtained from Google Maps, and calculated based on the distances between neighboring large cities from Sydney NSW to Perth WA. The fixed cost to deploy a station is assumed to be normally distributed and have a mean, standard deviation, lower and upper limit of 22, 3, 15 and 30 million dollars, respectively. Each potential station has 5 chargers and 10 fully-charged batteries on average, with standard deviations being 1 and 2. The ranges for the number of chargers and fully-charged batteries in each station are [1, 10] and [5, 15], respectively. We assume a train has 50\% of chance to wait in a station, so the average waiting time is 0 hour. The standard deviation is 0.3 hour. In addition, we assume the charge rate for an empty battery is 40\% per hour. The charge rate decreases approximately linearly as SOC grows. The swapping time (i.e., time to replace a used battery with a fully-charge battery on a train) is 2 hours.

\subsubsection{Parameters}
\noindent In the testing, we make the penalty weight for delay ($\alpha^{\text{D}}$) equal three times of the penalty weight for facility setup cost ($\alpha^{\text{F}}$). For PLA, we assume all trips can be completed within 100 hours, and each battery can be charged to 100\% within 10 hours, and hence we divide the entire planning horizon into 100 1-hour segments, and divide the charging period of a battery at each station into 10 1-hour segments.

\indent In PLA and FA tests, the tolerance gap is 1\% and time limit is 30 minutes. In BD tests, the tolerance gap and time limit for $\mathcal{RMP}$ are fixed at 0.5\% and 5 minutes, respectively. When the percentage gap between the upper and lower bound ($\text{gap}_{\epsilon}$) falls below 5\%, the algorithm stops.  Parameter $\epsilon$ is fixed at 0.000001
and $M$ is fixed at 1000. All algorithms are coded in Python, and Gurobi 12.0.1 is called to solve all optimization models.

\subsection{Computational Results}  \label{sec_comp_result}
\noindent This subsection discusses experimental results for small, medium and large instances, respectively. To evaluate the performance of the proposed algorithms, we record three sets of statistics. The first set is the cost measures including the objective function value, the number of deployed stations, infrastructure setup cost, the average number of delayed hours for each train (excluding the planned waiting time), and the average time in charging and swapping batteries for each train. The second set of statistics describes the utilization of deployed stations such as the average number of hours in charging and swapping batteries in each deployed station. The last set of statistics is processing times. Table \ref{tbl_output_delay3} in Appendix \ref{sec_append_output_result} presents the experimental results.

\subsubsection{Illustrative example} 
\noindent For illustrative purpose, we use a small instance (instance 1 in Table \ref{tbl_output_delay3}) to display the optimal schedule obtained from PLA. The parameters such as waiting times, power requirement and travel time between each potential stations are listed in Appendix \ref{sec_append_illustrative_data}.

\indent Table \ref{tble_illustrative_schedule} presents the optimal schedule of the two trains from PLA. Column 1 displays the train, consist and their corresponding output statistics names. Columns 2-7 illustrate the arrival/departure time of a train in a potential station, charge/swap decision and arrival/departure SOC of the battery in each potential station. Note that in row 1, an asterisk sign indicates the deployment of a potential station. Vertically, rows 2-13 are the results for Train 1 and rows 14-25 are results for Train 2. Rows 2-4 are the arrival/departure time and delay of Train 1. Rows 5-7, 8-10, 11-13 display the battery status, charge/swap decision, and arrival/departure SOC of the battery in Train 1 Consists 1, 2 and 3, respectively. Rows 14-25 are the schedule for Train 2, and the structure is similar to that of rows 2-13. For simplicity, we only discuss the schedule for Train 1 in this subsection.

\indent From the first row of Table \ref{tble_illustrative_schedule}, we can observe that Stations 1, 2 and 4 are deployed. Rows 5, 8 and 11 in column 1 indicate that all three consists contain a battery. Assume the time horizon starts from hour 0, and Train 1 leaves the origin at hour 0 with all batteries having SOC=100\%. There is no delay in origin by default. From Table \ref{tbl_illustrative_waittime} in Appendix \ref{sec_append_illustrative_data}, we know that the it takes 7 hours and 1.73 fully-charged batteries for Train 1 to travel from the origin to Station 1, so the arrival time at Station 1 is hour 7. Due to the sequential battery management system, only batteries in Consists 1 and 2 contribute to the trip from the origin to Station 1. Specifically, the batteries in Consists 1 and 2 are depleted to 0\% and 27.09\% respectively. In Station 1, batteries in Consists 1 and 2 are charged for 0.55 hours to 22.05\% and 44.73\%, respectively. Since there is no planned waiting time in Station 1, the delay is 0.55 hour. Train 1 departs Station 1 at hour 7.55, spends 2.18 hours and 1.67 fully-charged batteries on its way to Station 2. It arrives at Station 2 in hour 9.73 with batteries in all three consists totally depleted. The three batteries are swapped in Station 2 with departure SOC=100\%. Since the swapping takes 2 hours an the planned waiting time in Station 2 is 0.2 hour, the delay is 1.8 hours. Train leaves Station 2 in hour 11.73, spends 3.38 hours and 1-battery power, arrives at Station 3 in hour 15.12. The battery in Consist 1 is almost completely depleted and those in Consists 2 and 3 have SOC=100\%. Since Station 3 is not deployed, the battery can neither be swapped nor charged in Station 3. As planned, Train 1 waits in Station 3 for 0.14 hour and leaves in hour 15.25 with unchanged SOC of batteries. There is no delay in Station 3. After 3.60 hours, Train 1 arrives at Station 4 in hour 18.86. The SOCs of batteries in Consists 1, 2 and 3 are 0\%, 0\% and 16.94\%, respectively, which is a reflection of "1.83 fully-charged batteries" power requirement from Stations 3 to 4. All three batteries spend 2 hours getting swapped in Station 4 and Train 1 departs in hour 20.86. Since the planned waiting time in Station 4 is 0.24 hour, the delay is 1.76 hours. Then Train 1 travels 4.1 hours toward the destination. The SOCs of batteries in three Consists 1, 2 and 3 are 0\%, 0\% and 73.47\%, respectively, corresponding to the power requirement of 2.27 fully-charged batteries.

\begin{table}[H]
	\centering
	\caption{Schedule of the illustrative example obtained from PLA}
	\resizebox{\textwidth}{!}{
	\begin{tabular}{|l|l|l|l|l|l|l|}
		\toprule
		\textbf{Trains/Consists} & \textbf{Origin} & \textbf{Station 1 *} & \textbf{Station 2 *} & \textbf{Station 3} & \textbf{Station 4 *} & \textbf{Destination} \\
		\midrule
		\textbf{Train 1} &       &       &       &       &       &  \\
		(Arrival time, departure time) & (0.00, 0.00) & (7.00, 7.55) & (9.73, 11.73) & (15.12, 15.25) & (18.86, 20.86) & (24.96, 24.96) \\
		Delay & 0.00  & 0.55  & 1.80  & 0.00  & 1.76  & 0.00 \\
		\midrule
		Consist 1 in Train 1 (with battery) &       &       &       &       &       &  \\
		Charge? swap? &       & charge (0.55 hours) & swap  &       & swap  &  \\
		(Arrival SOC, departure SOC) & (100.00\%, 100.00\%) & (0.00\%, 22.05\%) & (0.00\%, 100.00\%) & (0.18\%, 0.18\%) & (0.00\%, 100.00\%) & (0.00\%, 0.00\%) \\
		\midrule
		Consist 2 in Train 1 (with battery) &       &       &       &       &       &  \\
		Charge? swap? &       & charge (0.55 hours) & swap  &       & swap  &  \\
		(Arrival SOC, departure SOC) & (100.00\%, 100.00\%) & (27.09\%, 44.73\%) & (0.00\%, 100.00\%) & (100.00\%, 100.00\%) & (0.00\%, 100.00\%) & (0.00\%, 0.00\%) \\
		\midrule
		Consist 3  in Train 1 (with battery) &       &       &       &       &       &  \\
		Charge? swap? &       &       & swap  &       & swap  &  \\
		(Arrival SOC, departure SOC) & (100.00\%, 100.00\%) & (100.00\%, 100.00\%) & (0.00\%, 100.00\%) & (100.00\%, 100.00\%) & (16.94\%, 100.00\%) & (73.47\%, 73.47\%) \\
		\midrule
		\textbf{Train 2} &       &       &       &       &       &  \\
		(Arrival time, departure time) & (0.00, 0.00) & (2.10, 2.38) & (6.44, 8.44) & (11.92, 11.92) & (15.42, 16.82) & (21.94, 21.94) \\
		Delay  & 0.00  & -0.00 & 1.70  & 0.00  & 1.39  & 0.00 \\
		\midrule
		Consist 1 in Train 2 (with battery) &       &       &       &       &       &  \\
		Charge? swap? &       & charge (0.28 hours) & swap  &       & charge (1.39 hours) &  \\
		(Arrival SOC, departure SOC) & (100.00\%, 100.00\%) & (0.00\%, 11.36\%) & (0.00\%, 100.00\%) & (34.91\%, 34.91\%) & (0.00\%, 49.46\%) & (0.00\%, 0.00\%) \\
		\midrule
		Consist 2 in Train 2 (with battery) &       &       &       &       &       &  \\
		Charge? swap? &       & charge (0.20 hours) & swap  &       & charge (1.39 hours) &  \\
		(Arrival SOC, departure SOC) & (100.00\%, 100.00\%) & (58.70\%, 62.77\%) & (0.00\%, 100.00\%) & (100.00\%, 100.00\%) & (0.00\%, 49.46\%) & (0.00\%, 0.00\%) \\
		\midrule
		Consist 3 in Train 2 (with battery) &       &       &       &       &       &  \\
		Charge? swap? &       & charge (0.00 hours) & swap  &       & charge (1.39 hours) &  \\
		(Arrival SOC, departure SOC) & (100.00\%, 100.00\%) & (100.00\%, 100.00\%) & (10.00\%, 100.00\%) & (100.00\%, 100.00\%) & (11.19\%, 55.22\%) & (0.00\%, 0.00\%) \\
		\bottomrule
	\end{tabular}
	}
	\label{tble_illustrative_schedule}%
\end{table}%

\subsubsection{Results analysis} 

\noindent Table \ref{tbl_output_delay3} in Appendix \ref{sec_append_output_result} presents the experimental results with $\alpha^{\text{D}}=3$. After implementing PLA, FA and BD algorithms on small, medium and large instances, we have summarized the result measures in eight aspects: (1) the objective function value, (2) setup cost of all deployed stations, (3) the number of deployed stations, (4) the average delay time for each train, (5) the average time in charging batteries of each train, (6) the average time in swapping batteries of each train, (7) the average time in charging batteries at each deployed station, and (8) the average time in swapping batteries at each deployed station. For each result measure, we conduct paired t-test to verify whether there is significant difference among the three algorithms. \\

\begin{figure}[H]
	\centering
	\includegraphics[width= .9\columnwidth]{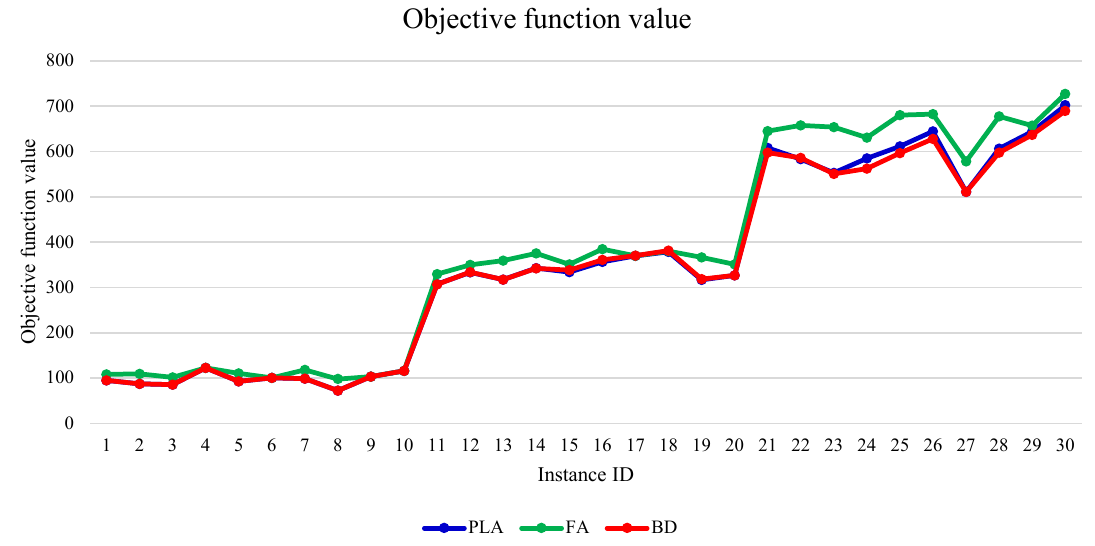}
	\caption{The objective function values from PLA, FA and BD algorithms in each instance with $\alpha^{\text{D}}=3$.}
	\label{fig_scatter_obj_delay3}
\end{figure}

\noindent \textit{\textbf{Objective function value.}} Figure \ref{fig_scatter_obj_delay3} depicts the objective function value of the 30 instances from PLA, FA and BD algorithms. One obvious observation is that the curves of PLA (blue) and BD (red) almost overlap except for several large instances (instances 21-30), indicating the PLA and BD provide similar objective function values in small and medium instances. In large instances, BD offers a slightly lower objective function value than PLA. The paired t-test between PLA and BD has a p-value of 0.0346. Compared with PLA and BD, FA gives solutions with higher objective function values, and the difference becomes even more apparent as the problem size grows. The difference between BD and FA is significant with a p-value of $9.35 \times 10^{-7}$. Therefore, BD has been verified to provide the lowest objective function value among the three algorithms. \\

\begin{figure}[H]
	\centering
	\includegraphics[width= 1.0\columnwidth]{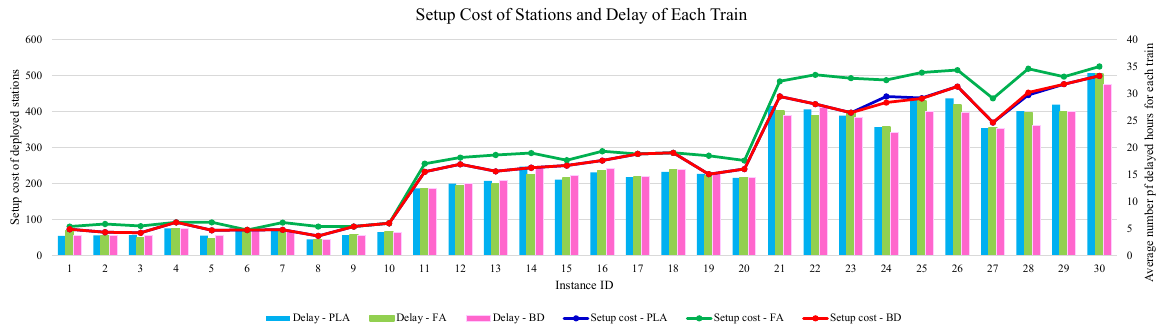}
	\caption{Setup cost of all stations and delay for each train from PLA, FA and BD algorithms in each instance with $\alpha^{\text{D}}=3$.}
	\label{fig_scatter_fixcostdelay_delay3}
\end{figure}

\noindent \textit{\textbf{Setup cost of deployed stations.}} Figure \ref{fig_scatter_fixcostdelay_delay3} illustrates the setup cost of all deployed stations and average delay time of each train. One observation which is similar to that from Figure \ref{fig_scatter_obj_delay3} is that BD outperforms PLA and FA regarding the setup cost, despite the slight difference between PLA and BD. The p-values from the paired t-tests for FA-BD and PLA-BD pairs are $4.42 \times 10^{-7}$ and 0.5487, respectively, indicating the difference between FA and BD is significant, while that between PLA and BD is trivial.  \\

\noindent \textit{\textbf{Average delay time for each train.}} The bars in Figure \ref{fig_scatter_fixcostdelay_delay3} illustrate the average delay time for each train. Note that in Figure \ref{fig_scatter_fixcostdelay_delay3}, for each algorithm, the colors for the setup cost and average delay time are similar with different shades. For example, for FA, the curve for setup cost is dark green, and the bar for delay time is light green. From the bars, we may conclude that for small instances, FA gives longer delay times than PLA and BD. In large instances, BD tends to provide shorter delay times, compared with PLA and FA. The p-values from paired t-tests for PLA-FA, PLA-BD and FA-BD pairs are 0.0456, 0.0382 and 0.3599, respectively. The test results indicate (i) PLA provides significantly longer delay times (15.32 hours) than FA and BD (15.10 and 14.94 hours), and (ii) the average delay times from FA and BD are similar.  \\

\begin{figure}[H]
	\centering
	\includegraphics[width= 1.0\columnwidth]{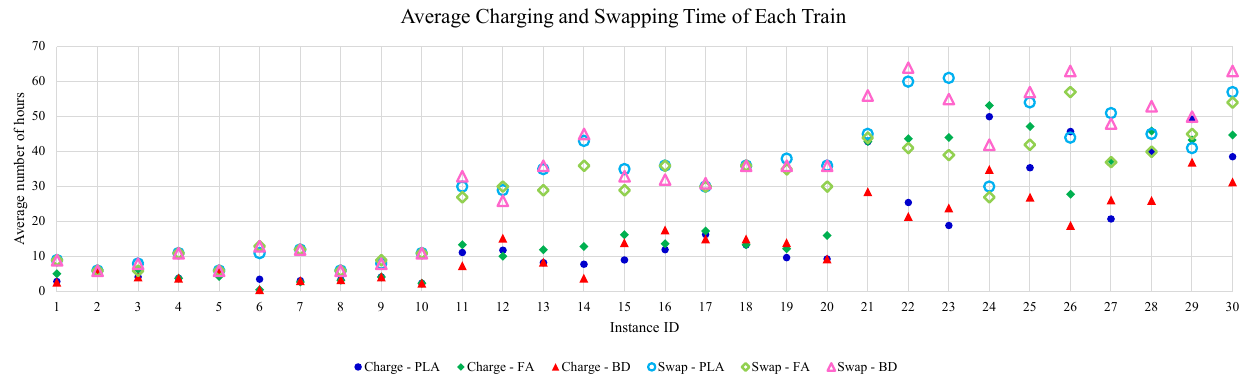}
	\caption{The average time in charging and swapping for each train from PLA, FA and BD algorithms in each instance with $\alpha^{\text{D}}=3$.}
	\label{fig_scatter_chargeswap_train_delay3}
\end{figure}

\noindent \textit{\textbf{Average time in battery charging and swapping for each train.}} Figure \ref{fig_scatter_chargeswap_train_delay3} depicts the average time in battery charging and swapping for each train. Note that in Figure \ref{fig_scatter_chargeswap_train_delay3}, for each algorithm, the markers for charging and swapping have the same color (with different shades) and same shape (with and without solid fill). For example, the PLA algorithm is denoted by blue rounds/circles. The marker is a dark blue round (with solid fill) for battery charging, and a light blue circle (without solid fill) for battery swapping. From Figure \ref{fig_scatter_chargeswap_train_delay3} and paired t-tests, we have following conclusions. \\[-20pt]
\begin{enumerate}
	\item In most instances, the three algorithms tend to refuel trains by battery swapping instead of battery charging, i.e., markers with lighter colors are above those with darker colors. This can be explained by the fact that battery swapping takes much shorter (2 hours) than charging battery to 100\% (about 10 hours). Therefore, whenever the deployed station has fully-charged batteries available, the algorithms would select the swapping option for trains. \\[-20pt]
	\item FA provides a longer battery charging time for each train (20.20 hours) than PLA (17.18 hours) and BD (14.37 hours), i.e., the dark green diamonds tend to appear above dark blue rounds and red triangles. From the paired t-tests, there has been verified significant difference among the three algorithms. The p-values for PLA-FA, PLA-BD and FA-BD pairs are 0.0404, 0.0460 and 0.0005, respectively. \\[-20pt]
	\item FA offers schedules with a shorter swapping time for each train (27.77 hours) than PLA (30.80 hours) and BD (32.83 hours), i.e., the light green diamonds are below light blue circles and pink triangles. This complements our second observation of longest charging time from FA. The p-values for PLA-FA, PLA-BD and FA-BD pairs are 0.0201, 0.0432 and 0.0003, respectively, indicating the three algorithms provide significantly different battery swapping times for each train. 
\end{enumerate}
  
\begin{figure}[H]
	\centering
	\includegraphics[width= 1.0\columnwidth]{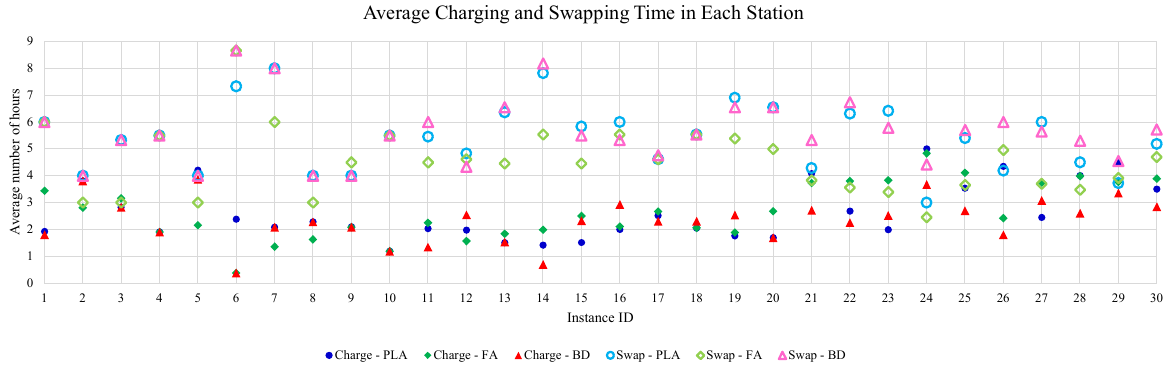}
	\caption{The average time in charging and swapping at each station from PLA, FA and BD algorithms in each instance with $\alpha^{\text{D}}=3$.}
	\label{fig_scatter_chargeswap_station_delay3}
\end{figure}

\noindent \textit{\textbf{Average time in battery charging and swapping at each deployed station.}} Figure \ref{fig_scatter_chargeswap_station_delay3} plots the average time in battery charging and swapping at each deployed station from the three algorithms. We use the markers in the same manner as that in Figure \ref{fig_scatter_chargeswap_train_delay3}, i.e., for each algorithm, the markers for charging and swapping have the same color (with different shades) and same shape (with and without solid fill). From Figure \ref{fig_scatter_chargeswap_station_delay3} and paired t-tests, we can draw following conclusions. \\[-20pt]

\begin{enumerate}
	\item The three algorithms tend to select the battery swapping option instead of charging option in deployed stations, i.e., markers with lighter colors are mostly above those with darker colors. This observation is same as that of refueling times for each train, and it perfectly matches the first conclusion from Figure \ref{fig_scatter_chargeswap_train_delay3}.\\[-20pt]
	\item BD, PLA and FA algorithms provide battery charging times at each station in increasing order (2.33, 2.64 and 2.66 hours, respectively). However, paired t-tests have shown that the difference is insignificant with p-values of PLA-FA, PLA-BD and FA-BD pairs being 0.9213, 0.0533 and 0.0546, respectively. \\[-20pt]
	\item FA gives shortest battery swapping times at each station (4.52 hours), followed by PLA (5.42 hours), and BD offers longest battery swapping times (5.65 hours). The difference is significant with verification from paired t-tests. The p-values of PLA-FA, PLA-BD and FA-BD pairs are $8.18 \times 10^{-5}$, 0.0445 and $7.45 \times 10^{-7}$, respectively. 
\end{enumerate}

\subsection{Sensitivity Analysis on the Impact of Delay Penalty Weight $\alpha^{\text{D}}$}  \label{sec_comp_sensitivity}

\noindent In this subsection, we investigate the impact of the delay penalty weight ($\alpha^{\text{D}}$) on the station deployment and schedules produced by the three algorithms. The parameter $\alpha^{\text{D}}$ is used to penalize the objective function for each hour of extra delay. A smaller $\alpha^{\text{D}}$ can lead to lower facility setup cost but a larger amount of delay. On the other hand, an unreasonably high $\alpha^{\text{D}}$ value would result in minimum delay but unnecessarily high facility setup cost. Therefore, it is essential to figure out how each cost measure varies as the $\alpha^{\text{D}}$ value changes. In Section \ref{sec_comp_result}, the computational results are obtained with $\alpha^{\text{D}}=3$. In this subsection, we fix $\alpha^{\text{D}}=5$ and compare the results with those from $\alpha^{\text{D}}=3$. 

\indent To investigate how $\alpha^{\text{D}}$ impacts cost measures under each algorithm, we record the cost measures obtained from each algorithm (PLA, FA and BD) with both $\alpha^{\text{D}}=3$ and $\alpha^{\text{D}}=5$. Then we compare the two costs from $\alpha^{\text{D}}=3$ and $\alpha^{\text{D}}=5$, conduct paired t-test, and calculate the difference ($\Delta$) by subtracting the cost of $\alpha^{\text{D}}=3$ from $\alpha^{\text{D}}=5$. All costs and test statistics are available in Tables \ref{tbl_sensitivity_output1} and \ref{tbl_sensitivity_output2} in Appendix \ref{sec_append_sensitivity_result}. 

\begin{figure}[H]
	\centering
	\includegraphics[width= 1.0\columnwidth]{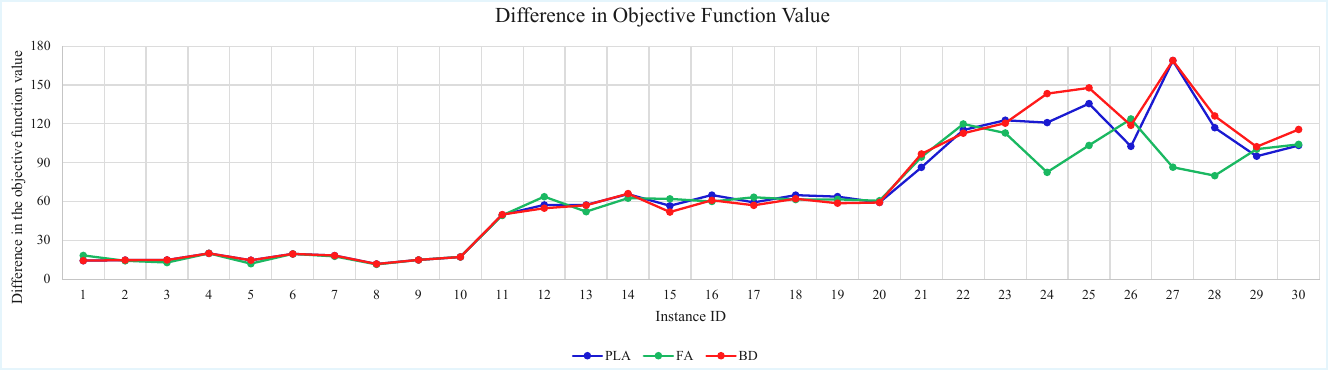}
	\caption{The difference in the objective function value from PLA, FA and BD algorithms in each instance between $\alpha^{\text{D}}=3$ and $\alpha^{\text{D}}=5$.}
	\label{fig_scatter_obj_compare}
\end{figure}

\begin{figure}[H]
	\centering
	\includegraphics[width= 1.0\columnwidth]{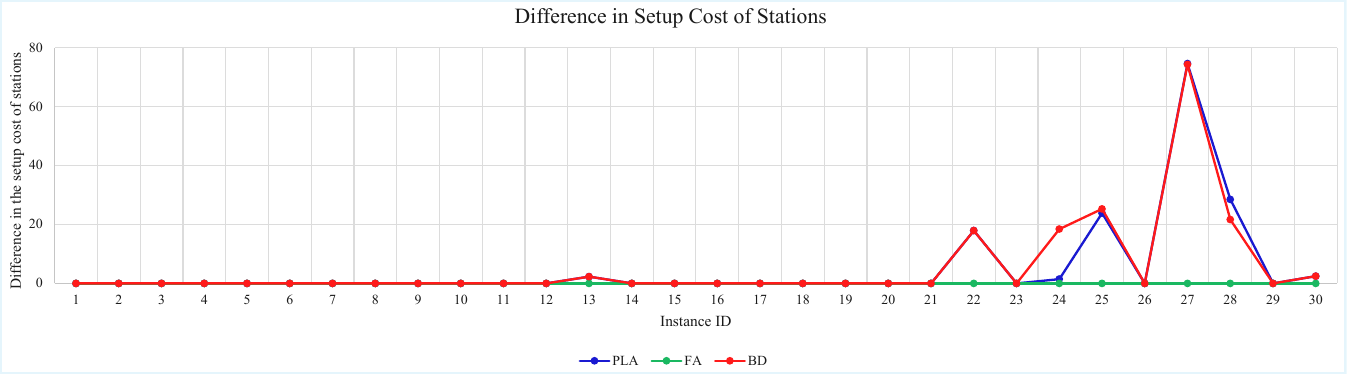}
	\caption{The difference in the setup cost of all deployed stations from PLA, FA and BD algorithms in each instance between $\alpha^{\text{D}}=3$ and $\alpha^{\text{D}}=5$.}
	\label{fig_scatter_fixcost_compare}
\end{figure}

\begin{figure}[H]
	\centering
	\includegraphics[width= 1.0\columnwidth]{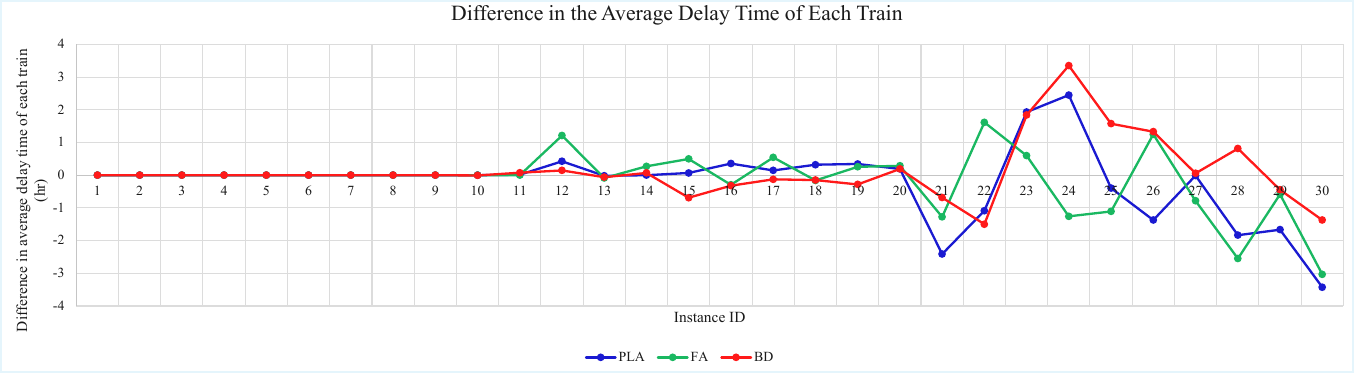}
	\caption{The difference in the average delay time of each train from PLA, FA and BD algorithms in each instance between $\alpha^{\text{D}}=3$ and $\alpha^{\text{D}}=5$.}
	\label{fig_scatter_delay_compare}
\end{figure}

\begin{figure}[H]
	\centering
	\includegraphics[width= 1.0\columnwidth]{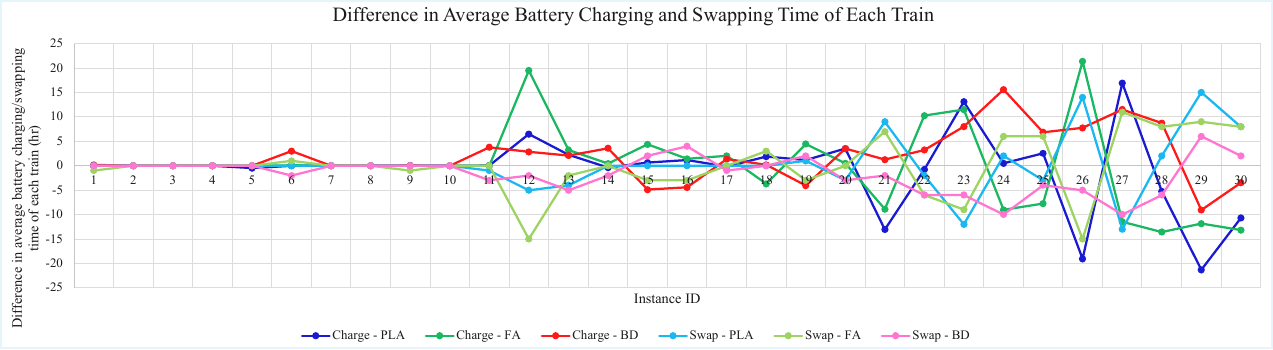}
	\caption{The difference in the average battery charging and swapping time of each train from PLA, FA and BD algorithms in each instance between $\alpha^{\text{D}}=3$ and $\alpha^{\text{D}}=5$.}
	\label{fig_scatter_chargeswap_train_compare}
\end{figure}

\begin{figure}[H]
	\centering
	\includegraphics[width= 1.0\columnwidth]{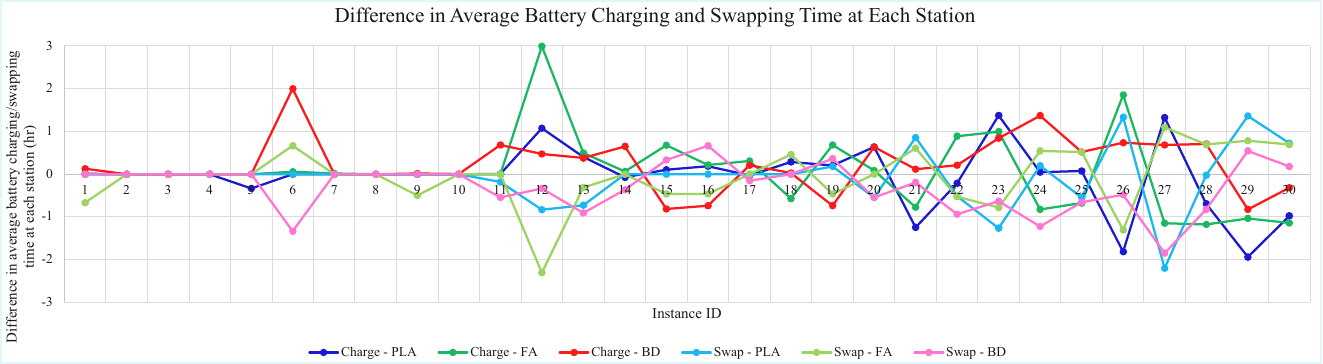}
	\caption{The difference in the average battery charging and swapping time at each deployed station from PLA, FA and BD algorithms in each instance between $\alpha^{\text{D}}=3$ and $\alpha^{\text{D}}=5$.}
	\label{fig_scatter_chargeswap_station_compare}
\end{figure}

\indent Figures \ref{fig_scatter_obj_compare}-\ref{fig_scatter_chargeswap_station_compare} plot the difference in each of the cost measures from the three algorithms between $\alpha^{\text{D}}=3$ and $\alpha^{\text{D}}=5$. The horizontal axis is the instance ID, and the vertical axis is the difference in cost measure. Each curve color represents an algorithm. The colors for PLA, FA and BD are blue, green and red, respectively, which are similar to figures in Section \ref{sec_comp_result}. \\

\noindent \textit{\textbf{Objective function value.}} Figure \ref{fig_scatter_obj_compare} plots the difference in the objective function value from $\alpha^{\text{D}}=3$ and $\alpha^{\text{D}}=5$. All three curves are above the 0-horizontal axis, indicating a larger $\alpha^{\text{D}}$ would result in a higher objective function value. The difference is significant with p-values of $7.95 \times 10^{-9}$, $9.72 \times 10^{-10}$ and $1.90 \times 10^{-8}$ for PLA, FA and BD algorithms, respectively. The significant difference might be due to a higher setup cost, or simply a larger constant ($\alpha^{\text{D}}$) value when we calculate the total cost. Further verification on the setup and delay costs is necessary and will be discussed below. The other observation from Figure \ref{fig_scatter_obj_compare} is that in small and medium instances, the three algorithms give similar objective function values. In large instances, however, the deviation increases. BD provides solutions with higher objective function values than PLA and FA, indicating that BD and large instances are more sensitive to $\alpha^{\text{D}}$.

\noindent \textit{\textbf{Setup cost of deployed stations.}} Figure \ref{fig_scatter_fixcost_compare} illustrates the difference in setup cost of deployed stations. The first observation is FA gives the same setup cost when $\alpha^{\text{D}}=3$ and $\alpha^{\text{D}}=5$ in all instances, so FA is not sensitive to $\alpha^{\text{D}}$ regarding setup cost. For PLA and BD, the difference is almost zero in small and medium instances, but in large instances, they provide higher setup costs when $\alpha^{\text{D}}=5$. The difference from BD is even more tremendous than that from PLA, so BD is the most sensitive algorithm among the three. The second observation is the differences in setup costs from the three algorithms are all non-negative. This can be explained by the fact that a higher $\alpha^{\text{D}}$ value encourages the algorithm to deploy more stations so that delay could be further reduced, and hence the setup cost increases. We conduct paired t-tests for PLA and BD, and the p-values are 0.0760 and 0.0559, respectively, indicating insignificant difference.

\noindent \textit{\textbf{Average delay time for each train.}} Figure \ref{fig_scatter_delay_compare} depicts the difference in average delay time of each train from the three algorithms between $\alpha^{\text{D}}=3$ and $\alpha^{\text{D}}=5$. As we can observe, the impact of $\alpha^{\text{D}}$ varies with the problem size. In small instances, $\alpha^{\text{D}}$ has little impact on the average delay. In medium instances, FA fluctuates most severely among the three algorithms. In large instances, the difference values from BD are higher than those from PLA and FA. Furthermore, it is apparent that for each algorithm, the difference becomes more obvious as the problem size grows. The reason is: in small instances, we have fewer potential stations, so each algorithm is more likely to find the same optimal solution for $\alpha^{\text{D}}=3$ and $\alpha^{\text{D}}=5$. When the problem size grows, the number of feasible solutions increases, and hence the possibility of having the same optimal solution for different $\alpha^{\text{D}}$ values decreases, making the curve fluctuates more sharply.

\indent It is surprising that not all differences are non-positive. We expected the difference in delay to be smaller than or equal to zero because a higher $\alpha^{\text{D}}$ value would result in shorter delay. The result, however, contradicts our expectation. This is because in medium and large instances with time limit, the three algorithms cannot guarantee optimal solutions, so the solutions behind Figure \ref{fig_scatter_delay_compare} are not necessarily optimal, and the difference in delay can be positive. To further verify the impact of $\alpha^{\text{D}}$ on delay, we conduct paired t-tests for all three algorithms. The p-values are 0.3346, 0.3922 and 0.4566 for PLA, FA and BD, respectively, indicating the impact of $\alpha^{\text{D}}$ is trivial on the average delay time of each train.

\noindent \textit{\textbf{Average time in battery charging and swapping for each train.}} Figure \ref{fig_scatter_chargeswap_train_compare} illustrates the difference in average time on battery charging and swapping for each train. The curve colors are defined in the same manner as that of Figures \ref{fig_scatter_chargeswap_train_delay3} and \ref{fig_scatter_chargeswap_station_delay3} in Section \ref{sec_comp_result}, i.e., for each algorithm, the curves for charging and swapping have the same color but different shades. Specifically, the charging curves for PLA, FA and BD are dark blue, dark green and red, respectively. Similarly, the swapping curves for PLA, FA and BD are light blue, light green and pink, respectively. From Figure \ref{fig_scatter_chargeswap_train_compare} and paired t-tests, we can draw the following conclusions.\\[-20pt]

\begin{enumerate}
	\item For each algorithm, the impact of $\alpha^{\text{D}}$ on the difference in charging and swapping time for each train becomes more obvious as the problem size increases. In small instances, the charging and swapping times of each train are hardly impacted by $\alpha^{\text{D}}$. In medium and large instances, however, the charging and swapping time of each train fluctuates with the $\alpha^{\text{D}}$ value, and the fluctuation becomes sharper as the problem size grows. This matches the observation of setup cost and delay in Figures \ref{fig_scatter_fixcost_compare} and \ref{fig_scatter_delay_compare}. \\[-20pt]
	\item For each algorithm, the sum of difference in charging time and swapping time is close to zero, i.e., the curves with the same (but different shades) are roughly symmetric with respect to the horizontal axis. Let's take BD for example: for each instance, if the red marker is above (below) the horizontal axis, then the pink marker is below (above) the horizontal axis, and the red and pink markers have similar absolute distances to the horizontal axis. This can be explained by the fact that to support the train to travel from the origin to destination, the total amount of time in battery refueling should not be cut tremendously, and hence the reduction of time in charging (swapping) is compensated by the increase of time in swapping (charging). \\[-20pt]
	\item The paired t-tests have verified the significant impact of $\alpha^{\text{D}}$ on both battery charging and swapping times of each train from BD with p-values of 0.0468 and 0.0163, respectively. For PLA and FA, the p-values are greater than 0.05 with respect to both charging and swapping times, indicating insignificant impact of $\alpha^{\text{D}}$.
\end{enumerate}

\noindent \textit{\textbf{Average time in battery charging and swapping at each deployed station.}} Figure \ref{fig_scatter_chargeswap_station_compare} plots the difference in average battery charging and swapping time at each deployed station from three algorithms. The curve colors are used in the same manner as that in Figure \ref{fig_scatter_chargeswap_train_compare}. Following conclusions can be drawn from Figure \ref{fig_scatter_chargeswap_station_compare} and paired t-tests. \\[-20pt]

\begin{enumerate}
	\item By comparing Figures \ref{fig_scatter_chargeswap_train_compare} and \ref{fig_scatter_chargeswap_station_compare}, we observe that despite the different values over the vertical axis, the curves in the two figures have roughly the same patterns and shapes within each group of instance size. This is because the total amount of time in charging/swapping are same from Figures \ref{fig_scatter_chargeswap_train_compare} and \ref{fig_scatter_chargeswap_station_compare}, and if we divide the total time by the number of deployed stations which is also identical for instances in the same group, then the pattern keeps unchanged within that group. Therefore, we can draw similar conclusions to those from Figure \ref{fig_scatter_chargeswap_train_compare}: (i) for each algorithm, the impact of $\alpha^{\text{D}}$ on the difference in charging and swapping time at each deployed station becomes more obvious as the problem size increases, (ii) for each algorithm, the sum of difference in charging time and swapping time at each deployed station is close to zero. \\[-20pt]
	\item When we solve the problem by BD, $\alpha^{\text{D}}$ is proved to have significant impact on the charging and swapping time at each deployed station with p-values of 0.0481 and 0.0081, respectively. In PLA and FA, the impact of $\alpha^{\text{D}}$ is insignificant with p-values greater than 0.05. 
\end{enumerate}

\section{Conclusions and Discussions} \label{sec_conclusion}

\noindent In this study, we investigated an optimization challenges of electrifying battery electric freight trains. By developing a tailored framework for charge station selection and charge/swap scheduling, we tackle the unique complexities of rail systems, which, unlike trucks, can accommodate multiple battery consists and specialized battery cars. Our model integrates key decisions, including the strategic placement of charge stations, the scheduling of charging and swapping activities for each battery consist, and the allocation of battery modules across trains. The decision variables include the location selection of charge stations and the scheduling of charging/swapping for each train battery in deployed stations. The objective function balances government budget constraints and freight train efficiency by minimizing total facility setup costs and delays caused by charging or swapping operations. Constraints such as station capacity, vehicle battery capacity, charge/swap decisions, energy demand, and sequential battery management are incorporated to reflect the operational realities of freight rail.

\indent To solve this multifaceted facility location and scheduling problem, we proposed three algorithms to solve the optimization problem: (i) a MILP model linearized by a Rectangle Piecewise Linear Approximation method (D'Ambrosio et al. 2010), (ii) a Fixed Algorithm, and (iii) Benders Decomposition. Computational experiments across small, medium, and large instances demonstrate that BD consistently delivers the lowest objective function values, closely followed by the PLA method, while FA yields the highest values. Sensitivity analysis further reveals that the BD algorithm significantly impacts refueling times for individual trains and stations, particularly when adjusting delay penalty weights, highlighting its robustness for practical applications.

\indent This research fills a critical gap in the literature, as prior studies on battery-electric vehicle electrification have primarily focused on trucks, neglecting the unique challenges of freight rail, such as managing multiple battery consists and coordinating charging/swapping across extensive rail networks. By drawing on methodologies from truck electrification and adapting them to rail-specific constraints—such as station locations, battery capacities, and charge/swap options—this study provides a novel framework for advancing sustainable freight rail systems.

\indent However, this study has limitations that warrant further exploration. First, it does not account for hours-of-service constraints for train drivers, such as mandatory breaks after extended driving periods, which could enhance crew welfare and operational realism. Second, the model assumes that charging and swapping cannot occur simultaneously at a station, a simplification that limits flexibility compared to real-world practices where such hybrid approaches could improve efficiency. Finally, the current framework focuses on single-trip scenarios, whereas freight trains often operate round trips, requiring models that account for return journeys with or without cargo.

\indent Future research should address these limitations by incorporating driver scheduling constraints, enabling simultaneous charging and swapping, and extending the model to round-trip operations. Additionally, exploring dynamic energy pricing and vehicle-to-grid integration could further optimize costs and enhance the resilience of electrified freight rail networks. By addressing these challenges, this study lays a foundation for transformative advancements in sustainable freight transportation, paving the way for more efficient and environmentally friendly freight rail systems. 

\vspace{10mm}

\newpage

\begin{appendices}
	
\section{Data for the Illustrative Example (Instance 1)} \label{sec_append_illustrative_data}
\noindent This appendix presents the data of the illustrative example. Table \ref{tbl_illustrative_waittime} shows the waiting time of trains at each station. Tables \ref{tbl_illustrative_power_train1} and \ref{tbl_illustrative_traveltime_train1} display the number of fully charged batteries and travel time (in hours) of train 1 between any two potential stations. Similarly, Tables \ref{tbl_illustrative_power_train2} and \ref{tbl_illustrative_traveltime_train2} introduce the number of fully charged batteries and travel time of train 2 between any two potential stations. Table \ref{tbl_illustrative_station} clarifies the maximum number of chargers and fully charged batteries at each potential station.

\begin{table}[H]
	\centering
	\caption{Waiting time (in hours) of trains at each station in the illustrative example}
	\begin{tabular}{|l|rrrrrr|}
		\toprule
		(Train, Station) & \multicolumn{1}{l}{Origin} & \multicolumn{1}{l}{Station 1} & \multicolumn{1}{l}{Station 2} & \multicolumn{1}{l}{Station 3} & \multicolumn{1}{l}{Station 4} & \multicolumn{1}{l|}{Destination} \\
		\midrule
		Train 1 & 0.00  & 0.00  & 0.20  & 0.14  & 0.24  & 0.00 \\
		Train 2 & 0.00  & 0.28  & 0.30  & 0.00  & 0.00  & 0.00 \\
		\bottomrule
	\end{tabular}%
	\label{tbl_illustrative_waittime}%
\end{table}%

\begin{table}[H]
	\centering
	\caption{Number of fully charged batteries required from Train 1 to travel between any two potential stations in the illustrative example}
	\begin{tabular}{|l|rrrrrr|}
		\toprule
		No. full batteries & \multicolumn{1}{l}{Origin} & \multicolumn{1}{l}{Station 1} & \multicolumn{1}{l}{Station 2} & \multicolumn{1}{l}{Station 3} & \multicolumn{1}{l}{Station 4} & \multicolumn{1}{l|}{Destination} \\
		\midrule
		Origin & 0.00  & 1.73  & 3.40  & 4.40  & 6.23  & 8.49 \\
		Station 1 & 1.73  & 0.00  & 1.67  & 2.67  & 4.50  & 6.76 \\
		Station 2 & 3.40  & 1.67  & 0.00  & 1.00  & 2.83  & 5.10 \\
		Station 3 & 4.40  & 2.67  & 1.00  & 0.00  & 1.83  & 4.10 \\
		Station 4 & 6.23  & 4.50  & 2.83  & 1.83  & 0.00  & 2.27 \\
		Destination & 8.49  & 6.76  & 5.10  & 4.10  & 2.27  & 0.00 \\
		\bottomrule
	\end{tabular}%
	\label{tbl_illustrative_power_train1}%
\end{table}%

\begin{table}[H]
	\centering
	\caption{Travel time required from Train 1 to travel between any two potential stations in the illustrative example}
	\begin{tabular}{|l|rrrrrr|}
		\toprule
		Travel Time (hr) & \multicolumn{1}{l}{Origin} & \multicolumn{1}{l}{Station 1} & \multicolumn{1}{l}{Station 2} & \multicolumn{1}{l}{Station 3} & \multicolumn{1}{l}{Station 4} & \multicolumn{1}{l|}{Destination} \\
		\midrule
		Origin & 0.00  & 7.00  & 9.18  & 12.57 & 16.17 & 20.27 \\
		Station 1 & 7.00  & 0.00  & 2.18  & 5.57  & 9.17  & 13.27 \\
		Station 2 & 9.18  & 2.18  & 0.00  & 3.38  & 6.99  & 11.09 \\
		Station 3 & 12.57 & 5.57  & 3.38  & 0.00  & 3.60  & 7.71 \\
		Station 4 & 16.17 & 9.17  & 6.99  & 3.60  & 0.00  & 4.10 \\
		Destination & 20.27 & 13.27 & 11.09 & 7.71  & 4.10  & 0.00 \\
		\bottomrule
	\end{tabular}%
	\label{tbl_illustrative_traveltime_train1}%
\end{table}%

\begin{table}[H]
	\centering
	\caption{Number of fully charged batteries required from Train 2 to travel between any two potential stations in the illustrative example}
	\begin{tabular}{|l|rrrrrr|}
		\toprule
		No. full batteries  & \multicolumn{1}{l}{Origin} & \multicolumn{1}{l}{Station 1} & \multicolumn{1}{l}{Station 2} & \multicolumn{1}{l}{Station 3} & \multicolumn{1}{l}{Station 4} & \multicolumn{1}{l|}{Destination} \\
		\midrule
		Origin & 0.00  & 1.41  & 3.05  & 3.71  & 5.94  & 7.48 \\
		Station 1 & 1.41  & 0.00  & 1.64  & 2.29  & 4.53  & 6.07 \\
		Station 2 & 3.05  & 1.64  & 0.00  & 0.65  & 2.89  & 4.43 \\
		Station 3 & 3.71  & 2.29  & 0.65  & 0.00  & 2.24  & 3.78 \\
		Station 4 & 5.94  & 4.53  & 2.89  & 2.24  & 0.00  & 1.54 \\
		Destination & 7.48  & 6.07  & 4.43  & 3.78  & 1.54  & 0.00 \\
		\bottomrule
	\end{tabular}%
	\label{tbl_illustrative_power_train2}%
\end{table}%

\begin{table}[H]
	\centering
	\caption{Travel time required from Train 2 to travel between any two potential stations in the illustrative example}
	\begin{tabular}{|l|rrrrrr|}
		\toprule
		Travel Time (hr) & \multicolumn{1}{l}{Origin} & \multicolumn{1}{l}{Station 1} & \multicolumn{1}{l}{Station 2} & \multicolumn{1}{l}{Station 3} & \multicolumn{1}{l}{Station 4} & \multicolumn{1}{l|}{Destination} \\
		\midrule
		Origin & 0.00  & 2.10  & 6.16  & 9.64  & 13.14 & 18.26 \\
		Station 1 & 2.10  & 0.00  & 4.06  & 7.54  & 11.04 & 16.16 \\
		Station 2 & 6.16  & 4.06  & 0.00  & 3.48  & 6.98  & 12.10 \\
		Station 3 & 9.64  & 7.54  & 3.48  & 0.00  & 3.50  & 8.62 \\
		Station 4 & 13.14 & 11.04 & 6.98  & 3.50  & 0.00  & 5.12 \\
		Destination & 18.26 & 16.16 & 12.10 & 8.62  & 5.12  & 0.00 \\
		\bottomrule
	\end{tabular}%
	\label{tbl_illustrative_traveltime_train2}%
\end{table}%

\begin{table}[H]
	\centering
	\caption{The maximum number of chargers and fully charged batteries at each potential station in the illustrative example}
	\begin{tabular}{|l|rrr|}
		\toprule
		Stations & \multicolumn{1}{l}{Fixed Cost} & \multicolumn{1}{l}{No. chargers} & \multicolumn{1}{l|}{No. batteries} \\
		\midrule
		Station 1 & 21.72 & 7     & 8 \\
		Station 2 & 21.47 & 5     & 8 \\
		Station 3 & 29.02 & 6     & 9 \\
		Station 4 & 30.00 & 3     & 13 \\
		\bottomrule
	\end{tabular}%
	\label{tbl_illustrative_station}%
\end{table}%

\begin{landscape}
\section{Output of Computational Experiments for $\alpha^{\text{D}}=3$} \label{sec_append_output_result}
\noindent Table \ref{tbl_output_delay3} presents the experimental results of small, medium and large instances from PLA, FA and BD algorithms. The first and second columns identify the instance group and ID, respectively. Columns 3-26 display 8 result measures from PLA, FA and BD algorithms. Columns 3-5 list the objective function value. Columns 6-8, 9-11 and 12-14 discuss the number of deployed stations, total setup cost and the average number of delayed hours for each train, respectively. Columns 15-17 and 18-20 present the average time each train spends in charging and swapping batteries, respectively. Columns 21-23 and 24-26 show the average time each deployed station spends in charging and swapping batteries, respectively.

\begin{table}[H]
	\centering
	\caption{Results output from PLA, FA and BD algorithms for small, medium and large instances with $\alpha^{\text{D}}=3$.}
	\resizebox{\textwidth}{!}{%
	\begin{tabular}{|l|c|ccc|ccc|ccc|ccc|ccc|ccc|ccc|ccc|}
		\toprule
		\multirow{2}{*}{\textbf{Instance Group}} & \multirow{2}{*}{\textbf{Instance ID}} & \multicolumn{3}{p{12.705em}|}{\textbf{Objective function value}} & \multicolumn{3}{p{10.14em}|}{\textbf{Number of deployed stations}} & \multicolumn{3}{p{12.705em}|}{\textbf{Fixed cost of deployed stations}} & \multicolumn{3}{p{12.705em}|}{\textbf{Average number of delayed hours for each train}} & \multicolumn{3}{p{12.705em}|}{\textbf{Average number of hours in charging for each train}} & \multicolumn{3}{p{12.36em}|}{\textbf{Average number of hours in swapping for each train}} & \multicolumn{3}{p{14.64em}|}{\textbf{Average number of hours in charging for each deployed station}} & \multicolumn{3}{p{13.14em}|}{\textbf{Average number of hours in swapping for each deployed station}} \\
		\cmidrule{3-26}      &     & \multicolumn{1}{p{4.235em}}{\textbf{PLA}} & \multicolumn{1}{p{4.235em}}{\textbf{FA}} & \multicolumn{1}{p{4.235em}|}{\textbf{BD}} & \multicolumn{1}{p{3.38em}}{\textbf{PLA}} & \multicolumn{1}{p{3.38em}}{\textbf{FA}} & \multicolumn{1}{p{3.38em}|}{\textbf{BD}} & \multicolumn{1}{p{4.235em}}{\textbf{PLA}} & \multicolumn{1}{p{4.235em}}{\textbf{FA}} & \multicolumn{1}{p{4.235em}|}{\textbf{BD}} & \multicolumn{1}{p{4.235em}}{\textbf{PLA}} & \multicolumn{1}{p{4.235em}}{\textbf{FA}} & \multicolumn{1}{p{4.235em}|}{\textbf{BD}} & \multicolumn{1}{p{4.235em}}{\textbf{PLA}} & \multicolumn{1}{p{4.235em}}{\textbf{FA}} & \multicolumn{1}{p{4.235em}|}{\textbf{BD}} & \multicolumn{1}{p{4.12em}}{\textbf{PLA}} & \multicolumn{1}{p{4.12em}}{\textbf{FA}} & \multicolumn{1}{p{4.12em}|}{\textbf{BD}} & \multicolumn{1}{p{4.88em}}{\textbf{PLA}} & \multicolumn{1}{p{4.88em}}{\textbf{FA}} & \multicolumn{1}{p{4.88em}|}{\textbf{BD}} & \multicolumn{1}{p{4.38em}}{\textbf{PLA}} & \multicolumn{1}{p{4.38em}}{\textbf{FA}} & \multicolumn{1}{p{4.38em}|}{\textbf{BD}} \\
		\midrule
		\multirow{10}{*}{Small} & 1     & 94.81 & 108.45 & 94.81 & 3     & 3     & 3     & 73.19 & 80.74 & 73.19 & 3.60  & 4.62  & 3.60  & 2.89  & 5.16  & 2.71  & 9     & 9.00  & 9     & 1.92  & 3.44  & 1.81  & 6.00  & 6.00  & 6.00 \\
		& 2     & 87.05 & 109.24 & 87.05 & 3     & 4     & 3     & 64.81 & 87.77 & 64.81 & 3.71  & 3.58  & 3.71  & 5.71  & 5.62  & 5.71  & 6     & 6.00  & 6     & 3.81  & 2.81  & 3.81  & 4.00  & 3.00  & 4.00 \\
		& 3     & 85.71 & 101.56 & 85.71 & 3     & 4     & 3     & 63.31 & 82.11 & 63.31 & 3.73  & 3.24  & 3.73  & 4.24  & 6.32  & 4.24  & 8     & 6.00  & 8     & 2.83  & 3.16  & 2.83  & 5.33  & 3.00  & 5.33 \\
		& 4     & 122.59 & 122.59 & 122.59 & 4     & 4     & 4     & 92.47 & 92.47 & 92.47 & 5.02  & 5.02  & 5.02  & 3.80  & 3.80  & 3.80  & 11    & 11.00 & 11    & 1.90  & 1.90  & 1.90  & 5.50  & 5.50  & 5.50 \\
		& 5     & 92.52 & 110.55 & 92.52 & 3     & 4     & 3     & 70.40 & 92.37 & 70.40 & 3.69  & 3.03  & 3.69  & 6.31  & 4.31  & 5.81  & 6     & 6.00  & 6     & 4.21  & 2.16  & 3.87  & 4.00  & 3.00  & 4.00 \\
		& 6     & 100.40 & 100.40 & 100.40 & 3     & 3     & 3     & 70.95 & 70.95 & 70.95 & 4.91  & 4.91  & 4.91  & 3.57  & 0.56  & 0.57  & 11    & 13.00 & 13    & 2.38  & 0.37  & 0.38  & 7.33  & 8.67  & 8.67 \\
		& 7     & 98.98 & 117.99 & 98.98 & 3     & 4     & 3     & 71.43 & 91.36 & 71.43 & 4.59  & 4.44  & 4.59  & 3.13  & 2.72  & 3.13  & 12    & 12.00 & 12    & 2.09  & 1.36  & 2.09  & 8.00  & 6.00  & 8.00 \\
		& 8     & 72.41 & 98.11 & 72.41 & 3     & 4     & 3     & 54.62 & 80.76 & 54.62 & 2.97  & 2.89  & 2.97  & 3.42  & 3.26  & 3.42  & 6     & 6.00  & 6     & 2.28  & 1.63  & 2.28  & 4.00  & 3.00  & 4.00 \\
		& 9     & 103.01 & 103.01 & 103.01 & 4     & 4     & 4     & 80.53 & 80.53 & 80.53 & 3.75  & 3.75  & 3.75  & 4.20  & 4.16  & 4.16  & 8     & 9.00  & 8     & 2.10  & 2.08  & 2.08  & 4.00  & 4.50  & 4.00 \\
		& 10    & 115.76 & 115.75 & 115.80 & 4     & 4     & 4     & 89.83 & 89.83 & 89.83 & 4.32  & 4.32  & 4.33  & 2.39  & 2.38  & 2.37  & 11    & 11.00 & 11    & 1.19  & 1.19  & 1.19  & 5.50  & 5.50  & 5.50 \\
		\midrule
		\multirow{10}{*}{Medium }& 11    & 307.39 & 329.44 & 306.94 & 11    & 12    & 11    & 233.01 & 255.39 & 233.01 & 12.40 & 12.34 & 12.32 & 11.16 & 13.45 & 7.40  & 30    & 27.00 & 33    & 2.03  & 2.24  & 1.34  & 5.45  & 4.50  & 6.00 \\
		& 12    & 333.34 & 349.77 & 333.68 & 12    & 13    & 12    & 253.54 & 272.24 & 253.54 & 13.30 & 12.92 & 13.36 & 11.85 & 10.17 & 15.29 & 29    & 30.00 & 26    & 1.97  & 1.56  & 2.55  & 4.83  & 4.62  & 4.33 \\
		& 13    & 317.25 & 359.38 & 317.50 & 11    & 13    & 11    & 234.22 & 279.69 & 234.22 & 13.84 & 13.28 & 13.88 & 8.29  & 11.97 & 8.42  & 35    & 29.00 & 36    & 1.51  & 1.84  & 1.53  & 6.36  & 4.46  & 6.55 \\
		& 14    & 342.55 & 374.94 & 342.14 & 11    & 13    & 11    & 243.88 & 285.01 & 243.88 & 16.45 & 14.99 & 16.38 & 7.80  & 12.92 & 3.82  & 43    & 36.00 & 45    & 1.42  & 1.99  & 0.70  & 7.82  & 5.54  & 8.18 \\
		& 15    & 334.01 & 350.65 & 338.15 & 12    & 13    & 12    & 249.85 & 264.85 & 249.85 & 14.03 & 14.30 & 14.72 & 9.08  & 16.27 & 13.97 & 35    & 29.00 & 33    & 1.51  & 2.50  & 2.33  & 5.83  & 4.46  & 5.50 \\
		& 16    & 356.57 & 384.44 & 360.63 & 12    & 13    & 12    & 264.24 & 289.87 & 264.24 & 15.39 & 15.76 & 16.07 & 12.01 & 13.70 & 17.58 & 36    & 36.00 & 32    & 2.00  & 2.11  & 2.93  & 6.00  & 5.54  & 5.33 \\
		& 17    & 369.52 & 369.52 & 370.29 & 13    & 13    & 13    & 282.61 & 282.61 & 282.61 & 14.48 & 14.48 & 14.61 & 16.33 & 17.36 & 15.01 & 30    & 30.00 & 31    & 2.51  & 2.67  & 2.31  & 4.62  & 4.62  & 4.77 \\
		& 18    & 378.25 & 380.30 & 381.05 & 13    & 13    & 13    & 285.46 & 285.46 & 285.46 & 15.46 & 15.81 & 15.93 & 13.34 & 13.40 & 15.03 & 36    & 36.00 & 36    & 2.05  & 2.06  & 2.31  & 5.54  & 5.54  & 5.54 \\
		& 19    & 316.65 & 366.24 & 318.35 & 11    & 13    & 11    & 225.96 & 277.47 & 225.96 & 15.11 & 14.79 & 15.40 & 9.71  & 12.25 & 13.96 & 38    & 35.00 & 36    & 1.76  & 1.89  & 2.54  & 6.91  & 5.38  & 6.55 \\
		& 20    & 326.40 & 350.70 & 326.40 & 11    & 12    & 11    & 240.29 & 263.96 & 240.29 & 14.35 & 14.46 & 14.35 & 9.30  & 16.04 & 9.30  & 36    & 30.00 & 36    & 1.69  & 2.67  & 1.69  & 6.55  & 5.00  & 6.55 \\
		\midrule
		\multirow{10}{*}{Large} & 21    & 608.06 & 645.09 & 597.69 & 21    & 23    & 21    & 442.16 & 484.28 & 442.16 & 27.65 & 26.80 & 25.92 & 42.84 & 42.88 & 28.56 & 45    & 44.00 & 56    & 4.08  & 3.73  & 2.72  & 4.29  & 3.83  & 5.33 \\
		& 22    & 583.34 & 657.84 & 585.82 & 19    & 23    & 19    & 420.99 & 502.20 & 420.99 & 27.06 & 25.94 & 27.47 & 25.48 & 43.69 & 21.47 & 60    & 41.00 & 64    & 2.68  & 3.80  & 2.26  & 6.32  & 3.57  & 6.74 \\
		& 23    & 552.54 & 653.48 & 550.45 & 19    & 23    & 19    & 397.20 & 492.89 & 397.20 & 25.89 & 26.77 & 25.54 & 18.86 & 44.08 & 23.93 & 61    & 39.00 & 55    & 1.98  & 3.83  & 2.52  & 6.42  & 3.39  & 5.79 \\
		& 24    & 584.53 & 630.47 & 562.13 & 20    & 22    & 19    & 441.92 & 487.61 & 424.95 & 23.77 & 23.81 & 22.86 & 50.00 & 53.17 & 34.91 & 30    & 27.00 & 42    & 5.00  & 4.83  & 3.67  & 3.00  & 2.45  & 4.42 \\
		& 25    & 611.34 & 679.97 & 596.44 & 20    & 23    & 20    & 437.69 & 508.27 & 436.25 & 28.94 & 28.62 & 26.70 & 35.36 & 47.21 & 27.00 & 54    & 42.00 & 57    & 3.54  & 4.11  & 2.70  & 5.40  & 3.65  & 5.70 \\
		& 26    & 644.18 & 682.38 & 627.96 & 21    & 23    & 21    & 469.56 & 515.46 & 469.56 & 29.10 & 27.82 & 26.40 & 45.73 & 27.84 & 18.93 & 44    & 57.00 & 63    & 4.36  & 2.42  & 1.80  & 4.19  & 4.96  & 6.00 \\
		& 27    & 510.93 & 578.18 & 510.80 & 17    & 20    & 17    & 369.43 & 436.62 & 369.73 & 23.58 & 23.59 & 23.51 & 20.76 & 37.51 & 26.19 & 51    & 37.00 & 48    & 2.44  & 3.75  & 3.08  & 6.00  & 3.70  & 5.65 \\
		& 28    & 606.32 & 677.49 & 597.26 & 20    & 23    & 20    & 446.02 & 519.16 & 452.87 & 26.72 & 26.39 & 24.06 & 40.03 & 45.86 & 26.05 & 45    & 40.00 & 53    & 4.00  & 3.99  & 2.61  & 4.50  & 3.48  & 5.30 \\
		& 29    & 643.77 & 656.81 & 636.46 & 22    & 23    & 22    & 476.19 & 497.27 & 476.19 & 27.93 & 26.59 & 26.71 & 49.24 & 43.33 & 36.98 & 41    & 45.00 & 50    & 4.48  & 3.77  & 3.36  & 3.73  & 3.91  & 4.55 \\
		& 30    & 702.07 & 727.35 & 689.73 & 22    & 23    & 22    & 499.34 & 525.53 & 499.34 & 33.79 & 33.64 & 31.73 & 38.54 & 44.74 & 31.33 & 57    & 54.00 & 63    & 3.50  & 3.89  & 2.85  & 5.18  & 4.70  & 5.73 \\
		\midrule
		\multicolumn{2}{|p{10.705em}|}{Average} & 346.74 & 376.40 & 344.11 & 11.70 & 13.07 & 11.67 & 254.84 & 285.82 & 254.46 & 15.32 & 15.10 & 14.94 & 17.18 & 20.20 & 14.37 & 30.80 & 27.77 & 32.83 & 2.64  & 2.66  & 2.33  & 5.42  & 4.52  & 5.65 \\
		\bottomrule
	\end{tabular}%
	\label{tbl_output_delay3}
	}
\end{table}%

\section{Output of Results Comparison between $\alpha^{\text{D}}=3$ and $\alpha^{\text{D}}=5$} \label{sec_append_sensitivity_result}

\noindent Tables \ref{tbl_sensitivity_output1} and \ref{tbl_sensitivity_output2} present the sensitivity analysis results on each cost measure from PLA, FA and BD algorithms between $\alpha^{\text{D}}=3$ and $\alpha^{\text{D}}=5$. In Table \ref{tbl_sensitivity_output1}, we discuss the impact of $\alpha^{\text{D}}$ on the objective function value, the number of deployed stations, the setup cost, and the average delay time of each train. Table \ref{tbl_sensitivity_output2} displays the sensitivity analysis on the average battery charging/swapping time of each train, and the average battery charging/swapping time at each deployed station. Specifically, in both Tables \ref{tbl_sensitivity_output1} and \ref{tbl_sensitivity_output2}, the first two columns tell the instance group and instance ID. Then we divide the remaining columns into four groups, each group representing a cost measure with nine columns. The nine columns are further split into three sub-groups, and each sub-group discusses the impact of $\alpha^{\text{D}}$ on the cost measure from a particular algorithm (PLA, FA or BD). For example, the first 9-column group is columns 3-11, which discusses the objective function value. Columns 3-5, 6-8 and 9-11 are for PLA, FA and BD algorithms, respectively. Columns 3, 6 and 9 are the objective function values from the three algorithms when $\alpha^{\text{D}}=3$. Columns 4, 7 and 10 are the objective function values from the three algorithms when $\alpha^{\text{D}}=5$. Columns 5, 8 and 11 are the difference in the objective function value between $\alpha^{\text{D}}=3$ and $\alpha^{\text{D}}=5$. The $\Delta$ values in these columns are calculated by subtracting the objective function value when $\alpha^{\text{D}}=3$ from that when $\alpha^{\text{D}}=5$.

\begin{table}[H]
	\centering
	\caption{Output of results comparison on the objective function value, the number of deployed stations, setup cost of deployed stations and average delay time for each train from PLA, FA and BD algorithms between $\alpha^{\text{D}}=3$ and $\alpha^{\text{D}}=5$.}
	\resizebox{\textwidth}{!}{%
	\begin{tabular}{|p{5em}c|ccc|ccc|ccc|ccc|ccc|ccc|ccc|ccc|ccc|ccc|ccc|ccc|}
		\toprule
		\multicolumn{1}{|c|}{\multirow{3}[6]{*}{\textbf{Instance Group}}} & \multicolumn{1}{c|}{\multirow{3}[6]{*}{\textbf{Instance ID}}} & \multicolumn{9}{p{37.935em}|}{\textbf{Objective function value}}      & \multicolumn{9}{p{37.935em}|}{\textbf{Number of deployed stations}}   & \multicolumn{9}{p{37.935em}|}{\textbf{Fixed cost of deployed stations}} & \multicolumn{9}{p{37.935em}|}{\textbf{Average number of delayed hours for each train}} \\
		\cmidrule{3-38}    \multicolumn{1}{|c|}{} &       & \multicolumn{3}{p{12.645em}|}{\textbf{PLA}} & \multicolumn{3}{p{12.645em}|}{\textbf{FA}} & \multicolumn{3}{p{12.645em}|}{\textbf{BD}} & \multicolumn{3}{p{12.645em}|}{\textbf{PLA}} & \multicolumn{3}{p{12.645em}|}{\textbf{FA}} & \multicolumn{3}{p{12.645em}|}{\textbf{BD}} & \multicolumn{3}{p{12.645em}|}{\textbf{PLA}} & \multicolumn{3}{p{12.645em}|}{\textbf{FA}} & \multicolumn{3}{p{12.645em}|}{\textbf{BD}} & \multicolumn{3}{p{12.645em}|}{\textbf{PLA}} & \multicolumn{3}{p{12.645em}|}{\textbf{FA}} & \multicolumn{3}{p{12.645em}|}{\textbf{BD}} \\
		\cmidrule{3-38}    \multicolumn{1}{|c|}{} &       & \multicolumn{1}{p{4.215em}}{\textbf{$\alpha^{\text{D}}$=3}} & \multicolumn{1}{p{4.215em}}{\textbf{$\alpha^{\text{D}}$=5}} & \multicolumn{1}{p{4.215em}|}{\textbf{$\Delta$}} & \multicolumn{1}{p{4.215em}}{\textbf{$\alpha^{\text{D}}$=3}} & \multicolumn{1}{p{4.215em}}{\textbf{$\alpha^{\text{D}}$=5}} & \multicolumn{1}{p{4.215em}|}{\textbf{$\Delta$}} & \multicolumn{1}{p{4.215em}}{\textbf{$\alpha^{\text{D}}$=3}} & \multicolumn{1}{p{4.215em}}{\textbf{$\alpha^{\text{D}}$=5}} & \multicolumn{1}{p{4.215em}|}{\textbf{$\Delta$}} & \multicolumn{1}{p{4.215em}}{\textbf{$\alpha^{\text{D}}$=3}} & \multicolumn{1}{p{4.215em}}{\textbf{$\alpha^{\text{D}}$=5}} & \multicolumn{1}{p{4.215em}|}{\textbf{$\Delta$}} & \multicolumn{1}{p{4.215em}}{\textbf{$\alpha^{\text{D}}$=3}} & \multicolumn{1}{p{4.215em}}{\textbf{$\alpha^{\text{D}}$=5}} & \multicolumn{1}{p{4.215em}|}{\textbf{$\Delta$}} & \multicolumn{1}{p{4.215em}}{\textbf{$\alpha^{\text{D}}$=3}} & \multicolumn{1}{p{4.215em}}{\textbf{$\alpha^{\text{D}}$=5}} & \multicolumn{1}{p{4.215em}|}{\textbf{$\Delta$}} & \multicolumn{1}{p{4.215em}}{\textbf{$\alpha^{\text{D}}$=3}} & \multicolumn{1}{p{4.215em}}{\textbf{$\alpha^{\text{D}}$=5}} & \multicolumn{1}{p{4.215em}|}{\textbf{$\Delta$}} & \multicolumn{1}{p{4.215em}}{\textbf{$\alpha^{\text{D}}$=3}} & \multicolumn{1}{p{4.215em}}{\textbf{$\alpha^{\text{D}}$=5}} & \multicolumn{1}{p{4.215em}|}{\textbf{$\Delta$}} & \multicolumn{1}{p{4.215em}}{\textbf{$\alpha^{\text{D}}$=3}} & \multicolumn{1}{p{4.215em}}{\textbf{$\alpha^{\text{D}}$=5}} & \multicolumn{1}{p{4.215em}|}{\textbf{$\Delta$}} & \multicolumn{1}{p{4.215em}}{\textbf{$\alpha^{\text{D}}$=3}} & \multicolumn{1}{p{4.215em}}{\textbf{$\alpha^{\text{D}}$=5}} & \multicolumn{1}{p{4.215em}|}{\textbf{$\Delta$}} & \multicolumn{1}{p{4.215em}}{\textbf{$\alpha^{\text{D}}$=3}} & \multicolumn{1}{p{4.215em}}{\textbf{$\alpha^{\text{D}}$=5}} & \multicolumn{1}{p{4.215em}|}{\textbf{$\Delta$}} & \multicolumn{1}{p{4.215em}}{\textbf{$\alpha^{\text{D}}$=3}} & \multicolumn{1}{p{4.215em}}{\textbf{$\alpha^{\text{D}}$=5}} & \multicolumn{1}{p{4.215em}|}{\textbf{$\Delta$}} \\
		\midrule
		\multicolumn{1}{|c|}{\multirow{10}[2]{*}{Small}} & 1     & 94.81 & 109.23 & 14.41 & 108.45 & 126.92 & 18.47 & 94.81 & 109.23 & 14.41 & 3     & 3     & 0.00  & 3     & 3     & 0.00  & 3     & 3     & 0.00  & 73.19 & 73.19 & 0.00  & 80.74 & 80.74 & 0.00  & 73.19 & 73.19 & 0.00  & 3.60  & 3.60  & 0.00  & 4.62  & 4.62  & 0.00  & 3.60  & 3.60  & 0.00 \\
		\multicolumn{1}{|c|}{} & 2     & 87.05 & 101.88 & 14.83 & 109.24 & 123.55 & 14.31 & 87.05 & 101.88 & 14.83 & 3     & 3     & 0.00  & 4     & 4     & 0.00  & 3     & 3     & 0.00  & 64.81 & 64.81 & 0.00  & 87.77 & 87.77 & 0.00  & 64.81 & 64.81 & 0.00  & 3.71  & 3.71  & 0.00  & 3.58  & 3.58  & 0.00  & 3.71  & 3.71  & 0.00 \\
		\multicolumn{1}{|c|}{} & 3     & 85.71 & 100.65 & 14.93 & 101.56 & 114.53 & 12.97 & 85.71 & 100.65 & 14.93 & 3     & 3     & 0.00  & 4     & 4     & 0.00  & 3     & 3     & 0.00  & 63.31 & 63.31 & 0.00  & 82.11 & 82.11 & 0.00  & 63.31 & 63.31 & 0.00  & 3.73  & 3.73  & 0.00  & 3.24  & 3.24  & 0.00  & 3.73  & 3.73  & 0.00 \\
		\multicolumn{1}{|c|}{} & 4     & 122.59 & 142.67 & 20.08 & 122.59 & 142.67 & 20.08 & 122.59 & 142.67 & 20.08 & 4     & 4     & 0.00  & 4     & 4     & 0.00  & 4     & 4     & 0.00  & 92.47 & 92.47 & 0.00  & 92.47 & 92.47 & 0.00  & 92.47 & 92.47 & 0.00  & 5.02  & 5.02  & 0.00  & 5.02  & 5.02  & 0.00  & 5.02  & 5.02  & 0.00 \\
		\multicolumn{1}{|c|}{} & 5     & 92.52 & 107.27 & 14.75 & 110.55 & 122.67 & 12.12 & 92.52 & 107.27 & 14.75 & 3     & 3     & 0.00  & 4     & 4     & 0.00  & 3     & 3     & 0.00  & 70.40 & 70.40 & 0.00  & 92.37 & 92.37 & 0.00  & 70.40 & 70.40 & 0.00  & 3.69  & 3.69  & 0.00  & 3.03  & 3.03  & 0.00  & 3.69  & 3.69  & 0.00 \\
		\multicolumn{1}{|c|}{} & 6     & 100.40 & 120.03 & 19.63 & 100.40 & 120.03 & 19.63 & 100.40 & 120.03 & 19.63 & 3     & 3     & 0.00  & 3     & 3     & 0.00  & 3     & 3     & 0.00  & 70.95 & 70.95 & 0.00  & 70.95 & 70.95 & 0.00  & 70.95 & 70.95 & 0.00  & 4.91  & 4.91  & 0.00  & 4.91  & 4.91  & 0.00  & 4.91  & 4.91  & 0.00 \\
		\multicolumn{1}{|c|}{} & 7     & 98.98 & 117.35 & 18.37 & 117.99 & 135.74 & 17.75 & 98.98 & 117.35 & 18.37 & 3     & 3     & 0.00  & 4     & 4     & 0.00  & 3     & 3     & 0.00  & 71.43 & 71.43 & 0.00  & 91.36 & 91.36 & 0.00  & 71.43 & 71.43 & 0.00  & 4.59  & 4.59  & 0.00  & 4.44  & 4.44  & 0.00  & 4.59  & 4.59  & 0.00 \\
		\multicolumn{1}{|c|}{} & 8     & 72.41 & 84.28 & 11.86 & 98.11 & 109.68 & 11.57 & 72.41 & 84.28 & 11.86 & 3     & 3     & 0.00  & 4     & 4     & 0.00  & 3     & 3     & 0.00  & 54.62 & 54.62 & 0.00  & 80.76 & 80.76 & 0.00  & 54.62 & 54.62 & 0.00  & 2.97  & 2.97  & 0.00  & 2.89  & 2.89  & 0.00  & 2.97  & 2.97  & 0.00 \\
		\multicolumn{1}{|c|}{} & 9     & 103.01 & 117.99 & 14.98 & 103.01 & 117.99 & 14.98 & 103.01 & 117.99 & 14.98 & 4     & 4     & 0.00  & 4     & 4     & 0.00  & 4     & 4     & 0.00  & 80.53 & 80.53 & 0.00  & 80.53 & 80.53 & 0.00  & 80.53 & 80.53 & 0.00  & 3.75  & 3.75  & 0.00  & 3.75  & 3.75  & 0.00  & 3.75  & 3.75  & 0.00 \\
		\multicolumn{1}{|c|}{} & 10    & 115.76 & 133.04 & 17.29 & 115.75 & 133.04 & 17.29 & 115.80 & 133.04 & 17.25 & 4     & 4     & 0.00  & 4     & 4     & 0.00  & 4     & 4     & 0.00  & 89.83 & 89.83 & 0.00  & 89.83 & 89.83 & 0.00  & 89.83 & 89.83 & 0.00  & 4.32  & 4.32  & 0.00  & 4.32  & 4.32  & 0.00  & 4.33  & 4.32  & -0.01 \\
		\midrule
		\multicolumn{1}{|c|}{\multirow{10}[2]{*}{Medium}} & 11    & 307.39 & 357.13 & 49.74 & 329.44 & 378.81 & 49.37 & 306.94 & 356.97 & 50.03 & 11    & 11    & 0.00  & 12    & 12    & 0.00  & 11    & 11    & 0.00  & 233.01 & 233.01 & 0.00  & 255.39 & 255.39 & 0.00  & 233.01 & 233.01 & 0.00  & 12.40 & 12.41 & 0.02  & 12.34 & 12.34 & 0.00  & 12.32 & 12.40 & 0.07 \\
		\multicolumn{1}{|c|}{} & 12    & 333.34 & 390.79 & 57.45 & 349.77 & 413.59 & 63.82 & 333.68 & 388.57 & 54.88 & 12    & 12    & 0.00  & 13    & 13    & 0.00  & 12    & 12    & 0.00  & 253.54 & 253.54 & 0.00  & 272.24 & 272.24 & 0.00  & 253.54 & 253.54 & 0.00  & 13.30 & 13.72 & 0.43  & 12.92 & 14.13 & 1.21  & 13.36 & 13.50 & 0.15 \\
		\multicolumn{1}{|c|}{} & 13    & 317.25 & 374.72 & 57.47 & 359.38 & 411.62 & 52.25 & 317.50 & 374.72 & 57.22 & 11    & 11    & 0.00  & 13    & 13    & 0.00  & 11    & 11    & 0.00  & 234.22 & 236.52 & 2.30  & 279.69 & 279.69 & 0.00  & 234.22 & 236.52 & 2.30  & 13.84 & 13.82 & -0.02 & 13.28 & 13.19 & -0.09 & 13.88 & 13.82 & -0.06 \\
		\multicolumn{1}{|c|}{} & 14    & 342.55 & 408.33 & 65.78 & 374.94 & 437.59 & 62.65 & 342.14 & 408.33 & 66.20 & 11    & 11    & 0.00  & 13    & 13    & 0.00  & 11    & 11    & 0.00  & 243.88 & 243.88 & 0.00  & 285.01 & 285.01 & 0.00  & 243.88 & 243.88 & 0.00  & 16.45 & 16.45 & 0.00  & 14.99 & 15.26 & 0.27  & 16.38 & 16.45 & 0.07 \\
		\multicolumn{1}{|c|}{} & 15    & 334.01 & 390.79 & 56.78 & 350.65 & 412.82 & 62.17 & 338.15 & 390.12 & 51.97 & 12    & 12    & 0.00  & 13    & 13    & 0.00  & 12    & 12    & 0.00  & 249.85 & 249.85 & 0.00  & 264.85 & 264.85 & 0.00  & 249.85 & 249.85 & 0.00  & 14.03 & 14.09 & 0.07  & 14.30 & 14.80 & 0.50  & 14.72 & 14.03 & -0.69 \\
		\multicolumn{1}{|c|}{} & 16    & 356.57 & 421.67 & 65.10 & 384.44 & 444.55 & 60.11 & 360.63 & 421.67 & 61.04 & 12    & 12    & 0.00  & 13    & 13    & 0.00  & 12    & 12    & 0.00  & 264.24 & 264.24 & 0.00  & 289.87 & 289.87 & 0.00  & 264.24 & 264.24 & 0.00  & 15.39 & 15.74 & 0.35  & 15.76 & 15.47 & -0.29 & 16.07 & 15.74 & -0.32 \\
		\multicolumn{1}{|c|}{} & 17    & 369.52 & 428.90 & 59.38 & 369.52 & 432.89 & 63.38 & 370.29 & 427.45 & 57.17 & 13    & 13    & 0.00  & 13    & 13    & 0.00  & 13    & 13    & 0.00  & 282.61 & 282.61 & 0.00  & 282.61 & 282.61 & 0.00  & 282.61 & 282.61 & 0.00  & 14.48 & 14.63 & 0.14  & 14.48 & 15.03 & 0.54  & 14.61 & 14.48 & -0.13 \\
		\multicolumn{1}{|c|}{} & 18    & 378.25 & 443.30 & 65.06 & 380.30 & 441.91 & 61.61 & 381.05 & 443.30 & 62.25 & 13    & 13    & 0.00  & 13    & 13    & 0.00  & 13    & 13    & 0.00  & 285.46 & 285.46 & 0.00  & 285.46 & 285.46 & 0.00  & 285.46 & 285.46 & 0.00  & 15.46 & 15.78 & 0.32  & 15.81 & 15.64 & -0.16 & 15.93 & 15.78 & -0.15 \\
		\multicolumn{1}{|c|}{} & 19    & 316.65 & 380.53 & 63.88 & 366.24 & 427.97 & 61.73 & 318.35 & 377.12 & 58.77 & 11    & 11    & 0.00  & 13    & 13    & 0.00  & 11    & 11    & 0.00  & 225.96 & 225.96 & 0.00  & 277.47 & 277.47 & 0.00  & 225.96 & 225.96 & 0.00  & 15.11 & 15.46 & 0.34  & 14.79 & 15.05 & 0.26  & 15.40 & 15.12 & -0.28 \\
		\multicolumn{1}{|c|}{} & 20    & 326.40 & 385.71 & 59.32 & 350.70 & 411.41 & 60.71 & 326.40 & 385.71 & 59.32 & 11    & 11    & 0.00  & 12    & 12    & 0.00  & 11    & 11    & 0.00  & 240.29 & 240.29 & 0.00  & 263.96 & 263.96 & 0.00  & 240.29 & 240.29 & 0.00  & 14.35 & 14.54 & 0.19  & 14.46 & 14.74 & 0.29  & 14.35 & 14.54 & 0.19 \\
		\midrule
		\multicolumn{1}{|c|}{\multirow{10}[2]{*}{Large}} & 21    & 608.06 & 694.53 & 86.47 & 645.09 & 739.53 & 94.43 & 597.69 & 694.53 & 96.83 & 21    & 21    & 0.00  & 23    & 23    & 0.00  & 21    & 21    & 0.00  & 442.16 & 442.16 & 0.00  & 484.28 & 484.28 & 0.00  & 442.16 & 442.16 & 0.00  & 27.65 & 25.24 & -2.41 & 26.80 & 25.52 & -1.28 & 25.92 & 25.24 & -0.69 \\
		\multicolumn{1}{|c|}{} & 22    & 583.34 & 698.65 & 115.30 & 657.84 & 777.77 & 119.92 & 585.82 & 698.65 & 112.83 & 19    & 20    & 1.00  & 23    & 23    & 0.00  & 19    & 20    & 1.00  & 420.99 & 438.94 & 17.94 & 502.20 & 502.20 & 0.00  & 420.99 & 438.94 & 17.94 & 27.06 & 25.97 & -1.09 & 25.94 & 27.56 & 1.62  & 27.47 & 25.97 & -1.50 \\
		\multicolumn{1}{|c|}{} & 23    & 552.54 & 675.41 & 122.87 & 653.48 & 766.51 & 113.03 & 550.45 & 671.02 & 120.58 & 19    & 19    & 0.00  & 23    & 23    & 0.00  & 19    & 19    & 0.00  & 397.20 & 397.20 & 0.00  & 492.89 & 492.89 & 0.00  & 397.20 & 397.20 & 0.00  & 25.89 & 27.82 & 1.93  & 26.77 & 27.36 & 0.60  & 25.54 & 27.38 & 1.84 \\
		\multicolumn{1}{|c|}{} & 24    & 584.53 & 705.55 & 121.03 & 630.47 & 713.11 & 82.64 & 562.13 & 705.55 & 143.42 & 20    & 20    & 0.00  & 22    & 22    & 0.00  & 19    & 20    & 1.00  & 441.92 & 443.39 & 1.47  & 487.61 & 487.61 & 0.00  & 424.95 & 443.39 & 18.44 & 23.77 & 26.22 & 2.45  & 23.81 & 22.55 & -1.26 & 22.86 & 26.22 & 3.35 \\
		\multicolumn{1}{|c|}{} & 25    & 611.34 & 747.01 & 135.67 & 679.97 & 783.36 & 103.38 & 596.44 & 744.29 & 147.85 & 20    & 21    & 1.00  & 23    & 23    & 0.00  & 20    & 21    & 1.00  & 437.69 & 461.56 & 23.87 & 508.27 & 508.27 & 0.00  & 436.25 & 461.56 & 25.31 & 28.94 & 28.55 & -0.40 & 28.62 & 27.51 & -1.11 & 26.70 & 28.27 & 1.57 \\
		\multicolumn{1}{|c|}{} & 26    & 644.18 & 746.87 & 102.69 & 682.38 & 806.16 & 123.77 & 627.96 & 746.87 & 118.91 & 21    & 21    & 0.00  & 23    & 23    & 0.00  & 21    & 21    & 0.00  & 469.56 & 469.56 & 0.00  & 515.46 & 515.46 & 0.00  & 469.56 & 469.56 & 0.00  & 29.10 & 27.73 & -1.37 & 27.82 & 29.07 & 1.25  & 26.40 & 27.73 & 1.33 \\
		\multicolumn{1}{|c|}{} & 27    & 510.93 & 679.76 & 168.84 & 578.18 & 664.71 & 86.54 & 510.80 & 679.76 & 168.96 & 17    & 20    & 3.00  & 20    & 20    & 0.00  & 17    & 20    & 3.00  & 369.43 & 444.08 & 74.65 & 436.62 & 436.62 & 0.00  & 369.73 & 444.08 & 74.35 & 23.58 & 23.57 & -0.01 & 23.59 & 22.81 & -0.78 & 23.51 & 23.57 & 0.06 \\
		\multicolumn{1}{|c|}{} & 28    & 606.32 & 723.35 & 117.03 & 677.49 & 757.55 & 80.05 & 597.26 & 723.35 & 126.09 & 20    & 21    & 1.00  & 23    & 23    & 0.00  & 20    & 21    & 1.00  & 446.02 & 474.54 & 28.53 & 519.16 & 519.16 & 0.00  & 452.87 & 474.54 & 21.68 & 26.72 & 24.88 & -1.84 & 26.39 & 23.84 & -2.55 & 24.06 & 24.88 & 0.82 \\
		\multicolumn{1}{|c|}{} & 29    & 643.77 & 738.83 & 95.06 & 656.81 & 757.30 & 100.49 & 636.46 & 738.83 & 102.37 & 22    & 22    & 0.00  & 23    & 23    & 0.00  & 22    & 22    & 0.00  & 476.19 & 476.19 & 0.00  & 497.27 & 497.27 & 0.00  & 476.19 & 476.19 & 0.00  & 27.93 & 26.26 & -1.67 & 26.59 & 26.00 & -0.59 & 26.71 & 26.26 & -0.45 \\
		\multicolumn{1}{|c|}{} & 30    & 702.07 & 805.40 & 103.33 & 727.35 & 831.55 & 104.20 & 689.73 & 805.40 & 115.67 & 22    & 22    & 0.00  & 23    & 23    & 0.00  & 22    & 22    & 0.00  & 499.34 & 501.80 & 2.46  & 525.53 & 525.53 & 0.00  & 499.34 & 501.80 & 2.46  & 33.79 & 30.36 & -3.43 & 33.64 & 30.60 & -3.03 & 31.73 & 30.36 & -1.37 \\
		\midrule
		\multicolumn{2}{|c|}{Average} & 346.74 & 411.05 & 64.31 & 376.40 & 435.25 & 58.85 & 344.11 & 410.55 & 66.45 & 11.70 & 11.90 & 0.20  & 13.07 & 13.07 & 0.00  & 11.67 & 11.90 & 0.23  & 254.84 & 259.88 & 5.04  & 285.82 & 285.82 & 0.00  & 254.46 & 259.88 & 5.42  & 15.32 & 15.12 & -0.20 & 15.10 & 14.94 & -0.15 & 14.94 & 15.07 & 0.13 \\
		\midrule
		\multicolumn{2}{|p{10em}|}{P-value from paired t-test} & \multicolumn{3}{c|}{$7.95 \times 10^{-9}$} & \multicolumn{3}{c|}{$9.72 \times 10^{-10}$} & \multicolumn{3}{c|}{$1.90 \times 10^{-8}$} & \multicolumn{3}{c|}{0.0831} & \multicolumn{3}{c|}{close to 1} & \multicolumn{3}{c|}{0.0504} & \multicolumn{3}{c|}{0.0760} & \multicolumn{3}{c|}{close to 1} & \multicolumn{3}{c|}{0.0559} & \multicolumn{3}{c|}{0.3346} & \multicolumn{3}{c|}{0.3922} & \multicolumn{3}{c|}{0.4566} \\
		\midrule
		\multicolumn{2}{|p{10em}|}{Significant difference} & \multicolumn{3}{c|}{\checkmark} & \multicolumn{3}{c|}{\checkmark} & \multicolumn{3}{c|}{\checkmark} & \multicolumn{3}{c|}{} & \multicolumn{3}{c|}{} & \multicolumn{3}{c|}{} & \multicolumn{3}{c|}{} & \multicolumn{3}{c|}{} & \multicolumn{3}{c|}{} & \multicolumn{3}{c|}{} & \multicolumn{3}{c|}{} & \multicolumn{3}{c|}{} \\
		\bottomrule
	\end{tabular}%
	}
	\label{tbl_sensitivity_output1}%
\end{table}%

\begin{table}[H]
	\centering
	\caption{Output of results comparison on the average battery charging/swapping time for each train, and average battery charging/swapping time at each deployed station from PLA, FA and BD algorithms between $\alpha^{\text{D}}=3$ and $\alpha^{\text{D}}=5$.}
	\resizebox{\textwidth}{!}{%
	\begin{tabular}{|p{5.145em}c|ccc|ccc|ccc|ccc|ccc|ccc|ccc|ccc|ccc|ccc|ccc|ccc|}
		\toprule
		\multicolumn{1}{|c|}{\multirow{3}[6]{*}{\textbf{Instance Group}}} & \multicolumn{1}{c|}{\multirow{3}[6]{*}{\textbf{Instance ID}}} & \multicolumn{9}{p{37.935em}|}{\textbf{Average number of hours in charging for each train}} & \multicolumn{9}{p{37.935em}|}{\textbf{Average number of hours in swapping for each train}} & \multicolumn{9}{p{37.935em}|}{\textbf{Average number of hours in charging for each deployed station}} & \multicolumn{9}{p{37.935em}|}{\textbf{Average number of hours in swapping for each deployed station}} \\
		\cmidrule{3-38}    \multicolumn{1}{|c|}{} &       & \multicolumn{3}{p{12.645em}|}{\textbf{PLA}} & \multicolumn{3}{p{12.645em}|}{\textbf{FA}} & \multicolumn{3}{p{12.645em}|}{\textbf{BD}} & \multicolumn{3}{p{12.645em}|}{\textbf{PLA}} & \multicolumn{3}{p{12.645em}|}{\textbf{FA}} & \multicolumn{3}{p{12.645em}|}{\textbf{BD}} & \multicolumn{3}{p{12.645em}|}{\textbf{PLA}} & \multicolumn{3}{p{12.645em}|}{\textbf{FA}} & \multicolumn{3}{p{12.645em}|}{\textbf{BD}} & \multicolumn{3}{p{12.645em}|}{\textbf{PLA}} & \multicolumn{3}{p{12.645em}|}{\textbf{FA}} & \multicolumn{3}{p{12.645em}|}{\textbf{BD}} \\
		\cmidrule{3-38}    \multicolumn{1}{|c|}{} &       & \multicolumn{1}{p{4.215em}}{\textbf{$\alpha^{\text{D}}$=3}} & \multicolumn{1}{p{4.215em}}{\textbf{$\alpha^{\text{D}}$=5}} & \multicolumn{1}{p{4.215em}|}{\textbf{$\Delta$}} & \multicolumn{1}{p{4.215em}}{\textbf{$\alpha^{\text{D}}$=3}} & \multicolumn{1}{p{4.215em}}{\textbf{$\alpha^{\text{D}}$=5}} & \multicolumn{1}{p{4.215em}|}{\textbf{$\Delta$}} & \multicolumn{1}{p{4.215em}}{\textbf{$\alpha^{\text{D}}$=3}} & \multicolumn{1}{p{4.215em}}{\textbf{$\alpha^{\text{D}}$=5}} & \multicolumn{1}{p{4.215em}|}{\textbf{$\Delta$}} & \multicolumn{1}{p{4.215em}}{\textbf{$\alpha^{\text{D}}$=3}} & \multicolumn{1}{p{4.215em}}{\textbf{$\alpha^{\text{D}}$=5}} & \multicolumn{1}{p{4.215em}|}{\textbf{$\Delta$}} & \multicolumn{1}{p{4.215em}}{\textbf{$\alpha^{\text{D}}$=3}} & \multicolumn{1}{p{4.215em}}{\textbf{$\alpha^{\text{D}}$=5}} & \multicolumn{1}{p{4.215em}|}{\textbf{$\Delta$}} & \multicolumn{1}{p{4.215em}}{\textbf{$\alpha^{\text{D}}$=3}} & \multicolumn{1}{p{4.215em}}{\textbf{$\alpha^{\text{D}}$=5}} & \multicolumn{1}{p{4.215em}|}{\textbf{$\Delta$}} & \multicolumn{1}{p{4.215em}}{\textbf{$\alpha^{\text{D}}$=3}} & \multicolumn{1}{p{4.215em}}{\textbf{$\alpha^{\text{D}}$=5}} & \multicolumn{1}{p{4.215em}|}{\textbf{$\Delta$}} & \multicolumn{1}{p{4.215em}}{\textbf{$\alpha^{\text{D}}$=3}} & \multicolumn{1}{p{4.215em}}{\textbf{$\alpha^{\text{D}}$=5}} & \multicolumn{1}{p{4.215em}|}{\textbf{$\Delta$}} & \multicolumn{1}{p{4.215em}}{\textbf{$\alpha^{\text{D}}$=3}} & \multicolumn{1}{p{4.215em}}{\textbf{$\alpha^{\text{D}}$=5}} & \multicolumn{1}{p{4.215em}|}{\textbf{$\Delta$}} & \multicolumn{1}{p{4.215em}}{\textbf{$\alpha^{\text{D}}$=3}} & \multicolumn{1}{p{4.215em}}{\textbf{$\alpha^{\text{D}}$=5}} & \multicolumn{1}{p{4.215em}|}{\textbf{$\Delta$}} & \multicolumn{1}{p{4.215em}}{\textbf{$\alpha^{\text{D}}$=3}} & \multicolumn{1}{p{4.215em}}{\textbf{$\alpha^{\text{D}}$=5}} & \multicolumn{1}{p{4.215em}|}{\textbf{$\Delta$}} & \multicolumn{1}{p{4.215em}}{\textbf{$\alpha^{\text{D}}$=3}} & \multicolumn{1}{p{4.215em}}{\textbf{$\alpha^{\text{D}}$=5}} & \multicolumn{1}{p{4.215em}|}{\textbf{$\Delta$}} \\
		\midrule
		\multicolumn{1}{|c|}{\multirow{10}[2]{*}{Small}} & 1     & 2.89  & 2.91  & 0.02  & 5.16  & 5.16  & 0.00  & 2.71  & 2.91  & 0.20  & 9     & 9.00  & 0.00  & 9.00  & 8.00  & -1.00 & 9.00  & 9.00  & 0.00  & 1.92  & 1.94  & 0.01  & 3.44  & 3.44  & 0.00  & 1.81  & 1.94  & 0.13  & 6.00  & 6.00  & 0.00  & 6.00  & 5.33  & -0.67 & 6.00  & 6.00  & 0.00 \\
		\multicolumn{1}{|c|}{} & 2     & 5.71  & 5.71  & 0.00  & 5.62  & 5.62  & 0.00  & 5.71  & 5.71  & 0.00  & 6     & 6.00  & 0.00  & 6.00  & 6.00  & 0.00  & 6.00  & 6.00  & 0.00  & 3.81  & 3.81  & 0.00  & 2.81  & 2.81  & 0.00  & 3.81  & 3.81  & 0.00  & 4.00  & 4.00  & 0.00  & 3.00  & 3.00  & 0.00  & 4.00  & 4.00  & 0.00 \\
		\multicolumn{1}{|c|}{} & 3     & 4.24  & 4.24  & 0.00  & 6.32  & 6.32  & 0.00  & 4.24  & 4.24  & 0.00  & 8     & 8.00  & 0.00  & 6.00  & 6.00  & 0.00  & 8.00  & 8.00  & 0.00  & 2.83  & 2.83  & 0.00  & 3.16  & 3.16  & 0.00  & 2.83  & 2.83  & 0.00  & 5.33  & 5.33  & 0.00  & 3.00  & 3.00  & 0.00  & 5.33  & 5.33  & 0.00 \\
		\multicolumn{1}{|c|}{} & 4     & 3.80  & 3.80  & 0.00  & 3.80  & 3.80  & 0.00  & 3.80  & 3.80  & 0.00  & 11    & 11.00 & 0.00  & 11.00 & 11.00 & 0.00  & 11.00 & 11.00 & 0.00  & 1.90  & 1.90  & 0.00  & 1.90  & 1.90  & 0.00  & 1.90  & 1.90  & 0.00  & 5.50  & 5.50  & 0.00  & 5.50  & 5.50  & 0.00  & 5.50  & 5.50  & 0.00 \\
		\multicolumn{1}{|c|}{} & 5     & 6.31  & 5.81  & -0.50 & 4.31  & 4.31  & 0.00  & 5.81  & 5.81  & 0.00  & 6     & 6.00  & 0.00  & 6.00  & 6.00  & 0.00  & 6.00  & 6.00  & 0.00  & 4.21  & 3.87  & -0.33 & 2.16  & 2.16  & 0.00  & 3.87  & 3.87  & 0.00  & 4.00  & 4.00  & 0.00  & 3.00  & 3.00  & 0.00  & 4.00  & 4.00  & 0.00 \\
		\multicolumn{1}{|c|}{} & 6     & 3.57  & 3.57  & 0.00  & 0.56  & 0.64  & 0.08  & 0.57  & 3.57  & 3.00  & 11    & 11.00 & 0.00  & 13.00 & 14.00 & 1.00  & 13.00 & 11.00 & -2.00 & 2.38  & 2.38  & 0.00  & 0.37  & 0.43  & 0.06  & 0.38  & 2.38  & 2.00  & 7.33  & 7.33  & 0.00  & 8.67  & 9.33  & 0.67  & 8.67  & 7.33  & -1.33 \\
		\multicolumn{1}{|c|}{} & 7     & 3.13  & 3.13  & 0.00  & 2.72  & 2.75  & 0.04  & 3.13  & 3.13  & 0.00  & 12    & 12.00 & 0.00  & 12.00 & 12.00 & 0.00  & 12.00 & 12.00 & 0.00  & 2.09  & 2.09  & 0.00  & 1.36  & 1.38  & 0.02  & 2.09  & 2.09  & 0.00  & 8.00  & 8.00  & 0.00  & 6.00  & 6.00  & 0.00  & 8.00  & 8.00  & 0.00 \\
		\multicolumn{1}{|c|}{} & 8     & 3.42  & 3.42  & 0.00  & 3.26  & 3.26  & 0.00  & 3.42  & 3.42  & 0.00  & 6     & 6.00  & 0.00  & 6.00  & 6.00  & 0.00  & 6.00  & 6.00  & 0.00  & 2.28  & 2.28  & 0.00  & 1.63  & 1.63  & 0.00  & 2.28  & 2.28  & 0.00  & 4.00  & 4.00  & 0.00  & 3.00  & 3.00  & 0.00  & 4.00  & 4.00  & 0.00 \\
		\multicolumn{1}{|c|}{} & 9     & 4.20  & 4.20  & 0.00  & 4.16  & 4.16  & 0.00  & 4.16  & 4.20  & 0.03  & 8     & 8.00  & 0.00  & 9.00  & 8.00  & -1.00 & 8.00  & 8.00  & 0.00  & 2.10  & 2.10  & 0.00  & 2.08  & 2.08  & 0.00  & 2.08  & 2.10  & 0.02  & 4.00  & 4.00  & 0.00  & 4.50  & 4.00  & -0.50 & 4.00  & 4.00  & 0.00 \\
		\multicolumn{1}{|c|}{} & 10    & 2.39  & 2.39  & 0.00  & 2.38  & 2.39  & 0.00  & 2.37  & 2.39  & 0.01  & 11    & 11.00 & 0.00  & 11.00 & 11.00 & 0.00  & 11.00 & 11.00 & 0.00  & 1.19  & 1.19  & 0.00  & 1.19  & 1.19  & 0.00  & 1.19  & 1.19  & 0.01  & 5.50  & 5.50  & 0.00  & 5.50  & 5.50  & 0.00  & 5.50  & 5.50  & 0.00 \\
		\midrule
		\multicolumn{1}{|c|}{\multirow{10}[2]{*}{Medium}} & 11    & 11.16 & 11.21 & 0.05  & 13.45 & 13.45 & 0.00  & 7.40  & 11.16 & 3.77  & 30    & 29.00 & -1.00 & 27.00 & 27.00 & 0.00  & 33.00 & 30.00 & -3.00 & 2.03  & 2.04  & 0.01  & 2.24  & 2.24  & 0.00  & 1.34  & 2.03  & 0.69  & 5.45  & 5.27  & -0.18 & 4.50  & 4.50  & 0.00  & 6.00  & 5.45  & -0.55 \\
		\multicolumn{1}{|c|}{} & 12    & 11.85 & 18.31 & 6.46  & 10.17 & 29.64 & 19.48 & 15.29 & 18.13 & 2.84  & 29    & 24.00 & -5.00 & 30.00 & 15.00 & -15.00 & 26.00 & 24.00 & -2.00 & 1.97  & 3.05  & 1.08  & 1.56  & 4.56  & 3.00  & 2.55  & 3.02  & 0.47  & 4.83  & 4.00  & -0.83 & 4.62  & 2.31  & -2.31 & 4.33  & 4.00  & -0.33 \\
		\multicolumn{1}{|c|}{} & 13    & 8.29  & 10.50 & 2.21  & 11.97 & 15.20 & 3.22  & 8.42  & 10.50 & 2.08  & 35    & 31.00 & -4.00 & 29.00 & 27.00 & -2.00 & 36.00 & 31.00 & -5.00 & 1.51  & 1.91  & 0.40  & 1.84  & 2.34  & 0.50  & 1.53  & 1.91  & 0.38  & 6.36  & 5.64  & -0.73 & 4.46  & 4.15  & -0.31 & 6.55  & 5.64  & -0.91 \\
		\multicolumn{1}{|c|}{} & 14    & 7.80  & 7.40  & -0.40 & 12.92 & 13.36 & 0.43  & 3.82  & 7.40  & 3.58  & 43    & 43.00 & 0.00  & 36.00 & 36.00 & 0.00  & 45.00 & 43.00 & -2.00 & 1.42  & 1.35  & -0.07 & 1.99  & 2.05  & 0.07  & 0.70  & 1.35  & 0.65  & 7.82  & 7.82  & 0.00  & 5.54  & 5.54  & 0.00  & 8.18  & 7.82  & -0.36 \\
		\multicolumn{1}{|c|}{} & 15    & 9.08  & 9.71  & 0.63  & 16.27 & 20.66 & 4.40  & 13.97 & 9.08  & -4.89 & 35    & 35.00 & 0.00  & 29.00 & 26.00 & -3.00 & 33.00 & 35.00 & 2.00  & 1.51  & 1.62  & 0.11  & 2.50  & 3.18  & 0.68  & 2.33  & 1.51  & -0.81 & 5.83  & 5.83  & 0.00  & 4.46  & 4.00  & -0.46 & 5.50  & 5.83  & 0.33 \\
		\multicolumn{1}{|c|}{} & 16    & 12.01 & 13.15 & 1.14  & 13.70 & 15.13 & 1.43  & 17.58 & 13.15 & -4.44 & 36    & 36.00 & 0.00  & 36.00 & 33.00 & -3.00 & 32.00 & 36.00 & 4.00  & 2.00  & 2.19  & 0.19  & 2.11  & 2.33  & 0.22  & 2.93  & 2.19  & -0.74 & 6.00  & 6.00  & 0.00  & 5.54  & 5.08  & -0.46 & 5.33  & 6.00  & 0.67 \\
		\multicolumn{1}{|c|}{} & 17    & 16.33 & 16.17 & -0.16 & 17.36 & 19.37 & 2.01  & 15.01 & 16.36 & 1.35  & 30    & 30.00 & 0.00  & 30.00 & 30.00 & 0.00  & 31.00 & 30.00 & -1.00 & 2.51  & 2.49  & -0.02 & 2.67  & 2.98  & 0.31  & 2.31  & 2.52  & 0.21  & 4.62  & 4.62  & 0.00  & 4.62  & 4.62  & 0.00  & 4.77  & 4.62  & -0.15 \\
		\multicolumn{1}{|c|}{} & 18    & 13.34 & 15.21 & 1.87  & 13.40 & 9.70  & -3.70 & 15.03 & 15.21 & 0.18  & 36    & 36.00 & 0.00  & 36.00 & 39.00 & 3.00  & 36.00 & 36.00 & 0.00  & 2.05  & 2.34  & 0.29  & 2.06  & 1.49  & -0.57 & 2.31  & 2.34  & 0.03  & 5.54  & 5.54  & 0.00  & 5.54  & 6.00  & 0.46  & 5.54  & 5.54  & 0.00 \\
		\multicolumn{1}{|c|}{} & 19    & 9.71  & 10.84 & 1.13  & 12.25 & 16.71 & 4.46  & 13.96 & 9.92  & -4.04 & 38    & 39.00 & 1.00  & 35.00 & 32.00 & -3.00 & 36.00 & 38.00 & 2.00  & 1.76  & 1.97  & 0.21  & 1.89  & 2.57  & 0.69  & 2.54  & 1.80  & -0.73 & 6.91  & 7.09  & 0.18  & 5.38  & 4.92  & -0.46 & 6.55  & 6.91  & 0.36 \\
		\multicolumn{1}{|c|}{} & 20    & 9.30  & 12.80 & 3.51  & 16.04 & 16.56 & 0.52  & 9.30  & 12.80 & 3.51  & 36    & 33.00 & -3.00 & 30.00 & 30.00 & 0.00  & 36.00 & 33.00 & -3.00 & 1.69  & 2.33  & 0.64  & 2.67  & 2.76  & 0.09  & 1.69  & 2.33  & 0.64  & 6.55  & 6.00  & -0.55 & 5.00  & 5.00  & 0.00  & 6.55  & 6.00  & -0.55 \\
		\midrule
		\multicolumn{1}{|c|}{\multirow{10}[2]{*}{Large}} & 21    & 42.84 & 29.79 & -13.05 & 42.88 & 34.00 & -8.89 & 28.56 & 29.79 & 1.23  & 45    & 54.00 & 9.00  & 44.00 & 51.00 & 7.00  & 56.00 & 54.00 & -2.00 & 4.08  & 2.84  & -1.24 & 3.73  & 2.96  & -0.77 & 2.72  & 2.84  & 0.12  & 4.29  & 5.14  & 0.86  & 3.83  & 4.43  & 0.61  & 5.33  & 5.14  & -0.19 \\
		\multicolumn{1}{|c|}{} & 22    & 25.48 & 24.70 & -0.78 & 43.69 & 53.94 & 10.25 & 21.47 & 24.70 & 3.23  & 60    & 58.00 & -2.00 & 41.00 & 35.00 & -6.00 & 64.00 & 58.00 & -6.00 & 2.68  & 2.47  & -0.21 & 3.80  & 4.69  & 0.89  & 2.26  & 2.47  & 0.21  & 6.32  & 5.80  & -0.52 & 3.57  & 3.04  & -0.52 & 6.74  & 5.80  & -0.94 \\
		\multicolumn{1}{|c|}{} & 23    & 18.86 & 31.95 & 13.09 & 44.08 & 55.57 & 11.49 & 23.93 & 31.96 & 8.03  & 61    & 49.00 & -12.00 & 39.00 & 30.00 & -9.00 & 55.00 & 49.00 & -6.00 & 1.98  & 3.36  & 1.38  & 3.83  & 4.83  & 1.00  & 2.52  & 3.36  & 0.84  & 6.42  & 5.16  & -1.26 & 3.39  & 2.61  & -0.78 & 5.79  & 5.16  & -0.63 \\
		\multicolumn{1}{|c|}{} & 24    & 50.00 & 50.49 & 0.49  & 53.17 & 44.14 & -9.03 & 34.91 & 50.49 & 15.58 & 30    & 32.00 & 2.00  & 27.00 & 33.00 & 6.00  & 42.00 & 32.00 & -10.00 & 5.00  & 5.05  & 0.05  & 4.83  & 4.01  & -0.82 & 3.67  & 5.05  & 1.37  & 3.00  & 3.20  & 0.20  & 2.45  & 3.00  & 0.55  & 4.42  & 3.20  & -1.22 \\
		\multicolumn{1}{|c|}{} & 25    & 35.36 & 37.92 & 2.55  & 47.21 & 39.44 & -7.77 & 27.00 & 33.85 & 6.84  & 54    & 51.00 & -3.00 & 42.00 & 48.00 & 6.00  & 57.00 & 53.00 & -4.00 & 3.54  & 3.61  & 0.07  & 4.11  & 3.43  & -0.68 & 2.70  & 3.22  & 0.52  & 5.40  & 4.86  & -0.54 & 3.65  & 4.17  & 0.52  & 5.70  & 5.05  & -0.65 \\
		\multicolumn{1}{|c|}{} & 26    & 45.73 & 26.68 & -19.05 & 27.84 & 49.21 & 21.37 & 18.93 & 26.68 & 7.75  & 44    & 58.00 & 14.00 & 57.00 & 42.00 & -15.00 & 63.00 & 58.00 & -5.00 & 4.36  & 2.54  & -1.81 & 2.42  & 4.28  & 1.86  & 1.80  & 2.54  & 0.74  & 4.19  & 5.52  & 1.33  & 4.96  & 3.65  & -1.30 & 6.00  & 5.52  & -0.48 \\
		\multicolumn{1}{|c|}{} & 27    & 20.76 & 37.69 & 16.92 & 37.51 & 26.03 & -11.48 & 26.19 & 37.69 & 11.50 & 51    & 38.00 & -13.00 & 37.00 & 48.00 & 11.00 & 48.00 & 38.00 & -10.00 & 2.44  & 3.77  & 1.33  & 3.75  & 2.60  & -1.15 & 3.08  & 3.77  & 0.69  & 6.00  & 3.80  & -2.20 & 3.70  & 4.80  & 1.10  & 5.65  & 3.80  & -1.85 \\
		\multicolumn{1}{|c|}{} & 28    & 40.03 & 34.76 & -5.27 & 45.86 & 32.30 & -13.56 & 26.05 & 34.76 & 8.71  & 45    & 47.00 & 2.00  & 40.00 & 48.00 & 8.00  & 53.00 & 47.00 & -6.00 & 4.00  & 3.31  & -0.69 & 3.99  & 2.81  & -1.18 & 2.61  & 3.31  & 0.71  & 4.50  & 4.48  & -0.02 & 3.48  & 4.17  & 0.70  & 5.30  & 4.48  & -0.82 \\
		\multicolumn{1}{|c|}{} & 29    & 49.24 & 27.91 & -21.32 & 43.33 & 31.47 & -11.87 & 36.98 & 27.91 & -9.07 & 41    & 56.00 & 15.00 & 45.00 & 54.00 & 9.00  & 50.00 & 56.00 & 6.00  & 4.48  & 2.54  & -1.94 & 3.77  & 2.74  & -1.03 & 3.36  & 2.54  & -0.82 & 3.73  & 5.09  & 1.36  & 3.91  & 4.70  & 0.78  & 4.55  & 5.09  & 0.55 \\
		\multicolumn{1}{|c|}{} & 30    & 38.54 & 27.84 & -10.70 & 44.74 & 31.58 & -13.16 & 31.33 & 27.84 & -3.49 & 57    & 65.00 & 8.00  & 54.00 & 62.00 & 8.00  & 63.00 & 65.00 & 2.00  & 3.50  & 2.53  & -0.97 & 3.89  & 2.75  & -1.14 & 2.85  & 2.53  & -0.32 & 5.18  & 5.91  & 0.73  & 4.70  & 5.39  & 0.70  & 5.73  & 5.91  & 0.18 \\
		\midrule
		\multicolumn{2}{|c|}{Average} & 17.18 & 16.47 & -0.71 & 20.20 & 20.20 & -0.01 & 14.37 & 16.28 & 1.92  & 30.80 & 31.07 & 0.27  & 27.77 & 27.80 & 0.03  & 32.83 & 31.13 & -1.70 & 2.64  & 2.59  & -0.05 & 2.66  & 2.73  & 0.07  & 2.33  & 2.57  & 0.23  & 5.42  & 5.35  & -0.07 & 4.52  & 4.46  & -0.06 & 5.65  & 5.35  & -0.30 \\
		\midrule
		\multicolumn{2}{|p{10.29em}|}{P-value from \newline{}paired t-test} & \multicolumn{3}{c|}{0.6122} & \multicolumn{3}{c|}{0.9951} & \multicolumn{3}{c|}{0.0468} & \multicolumn{3}{c|}{0.8004} & \multicolumn{3}{c|}{0.9761} & \multicolumn{3}{c|}{0.0163} & \multicolumn{3}{c|}{0.7034} & \multicolumn{3}{c|}{0.6791} & \multicolumn{3}{c|}{0.0481} & \multicolumn{3}{c|}{0.5640} & \multicolumn{3}{c|}{0.6546} & \multicolumn{3}{c|}{0.0081} \\
		\midrule
		\multicolumn{2}{|p{10.29em}|}{Significant difference} & \multicolumn{3}{c|}{} & \multicolumn{3}{c|}{} & \multicolumn{3}{c|}{\checkmark} & \multicolumn{3}{c|}{} & \multicolumn{3}{c|}{} & \multicolumn{3}{c|}{\checkmark} & \multicolumn{3}{c|}{} & \multicolumn{3}{c|}{} & \multicolumn{3}{c|}{\checkmark} & \multicolumn{3}{c|}{} & \multicolumn{3}{c|}{} & \multicolumn{3}{c|}{\checkmark} \\
		\bottomrule
	\end{tabular}%
	}
	\label{tbl_sensitivity_output2}%
\end{table}%

\end{landscape}

\end{appendices}

\newpage
\noindent \textbf{\Large{References}} \\

\noindent Adegbohun, F., A. von Jouanne, E. Agamloh and A. Yokochi (2023). Geographical Modeling of Charging Infrastructure Requirements for Heavy-Duty Electric Autonomous Truck Operations. \textit{Energies (Basel)} 16(10), 4161. \\[-5pt]

\noindent Aredah, A., J. Du, M. Hegazi, G. List and H.A. Rakha (2024). Comparative Analysis of Alternative Powertrain Technologies in Freight Trains: A Numerical Examination Towards Sustainable Rail Transport. \textit{Applied Energy} 356, 122411.    \\[-5pt]

\noindent Bai, T., Y. Li, K.H. Johansson and J. Mårtensson (2023). Rollout-Based Charging Strategy for Electric Trucks With Hours-of-Service Regulations. \textit{IEEE Control Systems Letters} 7, 2167–2172. \\[-5pt]

\noindent Barbosa, F.C. (2024). Battery Only Electric Traction for Freight Trains - A Technical and Operational Assessment. \textit{Proceedings of the Institution of Mechanical Engineers. Part F, Journal of Rail and Rapid Transit} 238(3), 322-337. \\[-5pt]

\noindent Biedenbach, F. and  K. Strunz (2024). Multi-Use Optimization of a Depot for Battery-Electric Heavy-Duty Trucks. \textit{World Electric Vehicle Journal} 15, 84. \\[-5pt]

\noindent Cech, L. (2024). DB Cargo’s Dual-Mode Vectrons Undergo Intensive Trials in Germany. \textit{Rail Market} June 19, 2024. \\[-5pt]

\noindent D'Ambrosio, C., A. Lodi and S. Martello (2010). Piecewise Linear Approximation of Functions of Two Variables in MILP Models. \textit{Operations Research Letters} 28, 29-46.  \\[-5pt]

\noindent FreightWaves (2025). FRA Study Sees New Locomotive Tech as Gateway to Electric Freight Trains. https://www.freightwaves.com/news/fra-study-sees-new-locomotive-tech-as-gateway-to-electric-freight-trains? [Accessed 22 June 2025]. \\[-5pt]

\noindent Hughes, G.M. (2023). Varamis Rail Links up with Carousel Logistics to Boost Pan-European Operations. \textit{Rail Advent} September 13, 2023. \\[-5pt]

\noindent Karlsson, J. and A. Grauers (2023). Agent-based Investigation of Charger Queues and Utilization of Public Chargers for Electric Long-haul Trucks. \textit{Energies} 16(12), 4704. \\[-5pt] 

\noindent Lu, A. and J.G. Allen (2024). Intermittent Electrification with Battery Locomotives and the Post-diesel Future of North American Freight Railroads. \textit{Transportation Research Record} 2678(11), 1691-1718.   \\[-5pt]

\noindent Mazzone, A., M. Schönbacher and X. Larrea (2018). Future Freight Locomotives in Shift2Rail – Development of Full Electric Last Mile Propulsion System. \textit{Proceedings of $7^{\text{th}}$ Transport Research Arena TRA}, April 16-19, 2018, Vienna, Austria. \\[-5pt]

\noindent Speth, D., P. Plötz and M. Wietschel (2025). An Optimal Capacity-constrained Fast Charging Network for Battery Electric Trucks in Germany. \textit{Transportation Research Part A: Policy and Practice} 193, 104383. \\[-5pt]

\noindent Wang, F. Y., Z. Chen and Z. Hu (2024). Comprehensive Optimization of Electrical Heavy-duty Truck Battery Swap Stations with a SOC-dependent Charge Scheduling Method. \textit{Energy} 308, 132773. \\[-5pt]

\noindent Wang, Q., X. Liu, J. Du and F. Kong (2016). Smart Charging for Electric Vehicles: A Survey From the Algorithmic Perspective. \textit{arXiv.org}. DOI: https://doi.org/10.48550/arxiv.1607.07298. \\[-5pt]

\noindent Whitehead, J., J. Whitehead, M. Kane and Z. Zheng (2022). Exploring Public Charging Infrastructure Requirements for Short-haul Electric Trucks. \textit{International Journal of Sustainable Transportation} 16(9), 775–791. \\[-5pt]

\noindent Ye, Z., N. Yu and R. Wei (2024). Joint Planning of Charging Stations and Power Systems for Heavy-duty Drayage Trucks. \textit{Transportation Research Part D} 134, 104320. \\[-5pt]

\noindent Zähringer, O. Teichert, G. Balke, J. Schneider and M. Lienkamp (2024). Optimizing the Journey: Dynamic Charging Strategies for Battery Electric Trucks in Long-Haul Transport. \textit{Energies} 17, 973.  \\[-5pt]
\end{document}